\documentclass[a4paper,12pt]{article}
\usepackage{amsmath}
\usepackage{amscd}
\usepackage{amssymb}
\usepackage{cite}
\renewcommand{\theequation}{\arabic {section}.\arabic{equation}}

\allowdisplaybreaks

\begin{document}


\begin{titlepage}
\begin{flushright}
EPHOU-05-001 \\
February, 2005
\end{flushright}

\vspace{5cm}

\begin{center}
{\Large $N=4$ Twisted Superspace from Dirac-K\"ahler Twist } \\ 
\vspace{0.3cm}
{\Large and }\\
\vspace{0.3cm}
{\Large Off-shell SUSY Invariant Actions in Four Dimensions} \\ 

\vspace{1cm}

{\scshape Junji Kato}\footnote{jkato@particle.sci.hokudai.ac.jp}, 
{\scshape Noboru Kawamoto}\footnote{kawamoto@particle.sci.hokudai.ac.jp},
{\scshape Akiko Miyake}\footnote{miyake@particle.sci.hokudai.ac.jp}\\

{\textit{ Department of Physics, Hokkaido University }}\\
{\textit{ Sapporo, 060-0810, Japan}}\\
\end{center}

\vspace{2cm}

\begin{abstract}
We propose $N=4$ twisted superspace formalism in four dimensions by 
introducing Dirac-K\"ahler twist. In addition to the BRST charge as a 
scalar counter part of twisted supercharge we find vector and tensor 
twisted supercharges. By introducing twisted chiral superfield we 
explicitly construct off-shell twisted $N=4$ SUSY invariant action. 
We can propose variety of supergauge invariant actions by introducing 
twisted vector superfield. We may, however, need to find further constraints 
to identify twisted $N=4$ super Yang-Mills action. 
We propose a superconnection formalism of twisted superspace where constraints 
play a crucial role. It turns out that $N=4$ superalgebra of Dirac-K\"ahler 
twist can be decomposed into $N=2$ sectors. We can then construct twisted 
$N=2$ super Yang-Mills actions by the superconnection formalism of twisted 
superspace in two and four dimensions. 
\end{abstract}

\end{titlepage}

\newpage
\renewcommand{\theequation}{\arabic {section}.\arabic{equation}}
\section{Introduction}

One of the most fundamental questions in modern particle physics is to 
understand the origin of the supersymmetry(SUSY) between boson and fermion. 
It was pointed out that quantized version of four-dimensional 
topological Yang-Mills 
action with instanton gauge fixing led to $N=2$ twisted super 
Yang-Mills action\cite{W,bs,bms,lp,BRT1}. The twisting procedure 
generates matter fermions from the ghost-related fermions by 
relating the spin of matter fermion and internal $R$ symmetry of 
$N=2$ SUSY algebra. 
Even though this is a very special example, we may recognize that 
SUSY appears from the quantization procedure of bosonic 
topolotical theory. 
Many related works appeared to find better understandings of the 
connection between the SUSY and the topological field theory 
\cite{BG,Kanno,BK,EY,BBT,BBRT}\cite{GM,LSSTV}.

One of the important characteristics of the Witten's twisting procedure 
is that the BRST charge of the quantization of topolotical field theories 
is responsible for the twisted SUSY as a supercharge. 
In the investigations of the quantization of topological field theories of 
Schwarz type; Chern-Simons action and BF actions, a new type of vector 
SUSY was 
discovered\cite{BRT,BR,DGS,DLPS}. It was recognized that this vector 
SUSY belongs to a twisted version of an extended SUSY of 
$N=2$ or $N=4$. The origin of the vector SUSY was recognized 
in some particular examples to be related to the fact that the energy-momentum 
tensor can be expressed as a pure BRST variation\cite{MS,SSVV,Pi}. 
It was later stressed that a (pseudo) scalar SUSY 
was also accompanied together with BRST and vector SUSY\cite{CLS}. 
Then there appeared many related works
\cite{GMS,BM,MSore,LPS,LSZ,BSSV,GGPS,GGNPS,FTVVSS}
\cite{BT,Soda,GMR,Pa}. 
The connection of the extended SUSY and the quantization procedure of 
anti-field formalism by Batalin and Vilkovisky was also 
investigated\cite{DR,BDL}.

One of the authors (N.K.) and Tsukioka pointed out that the two-dimensional version of topological Yang-Mills action obtained from generalized 
gauge theory \cite{KW,KW2,KOS,KSTU} with instanton gauge fixing led to 
two-dimensional version of $N=2$ twisted super Yang-Mills action\cite{KT}. 
It was found in this investigation that the twisting procedure to relate 
between the ghost-related fermions and matter fermions is essentially 
Dirac-K\"ahler fermion mechanism 
\cite{IL,Kahler,G,BJ,Rabin,BDH,BennT,Bull}
and the "flavor" degrees of freedom of 
the Dirac-K\"ahler fermion can be interpreted as that of the extended 
SUSY \cite{KT}. 
It turned out that this Dirac-K\"ahler twisting mechanism works universally 
in the quantization of topological field theory and an extended SUSY 
is generated. Here the BRST charge in the quantization is equivalent 
to the scalar component of twisted supercharge\cite{KKU}. 
It was then recognized that the twisted superspace formulation is hidden 
behind the formulation. 
In the previous paper two of the authors (J.K. and N.K.) with Uchida 
proposed twisted superspace formalism for $N=2$ twisted SUSY 
in two dimensions and derived off-shell SUSY invariant 
BF, Wess-Zumino and super Yang-Mills actions \cite{KKU}. 
The quantized action of Yang-Mills type in two dimensions has the close 
connection with $N=2$ super Yang-Mills 
action obtained from the different context\cite{DVF,SchapT}.
In this paper we propose four-dimensional $N=4$ twisted superspace 
formalism as a natural extension of the $N=D=2$ twisted superspace formalism.
Related works with the similar contexts with our formulation was given 
by Labastida and collaborators for $N=2$ twisted SUSY by spinor formulation in 
two and four dimensions\cite{LL,AL}, while our formulation is based 
on the scalar-vector-tensor formulation due to the Dirac-K\"ahler twisting 
procedure. 

Concerning to the twisting procedure of $N=4$ SUSY there are several 
twisting procedure\cite{Yam,Mac,VW,DM,LCL}. The twisting procedure we propose 
in this paper has close connection with that of Marcus\cite{Mac}. 
Off-shell $N=4$ SUSY invariant actions were proposed and the corresponding superspace 
formulation was investigated by several authors but so far it is not successful
\cite{GSW,Sohnius,SSW}. 

One of the other important motivations of current investigation comes from 
the recent lattice SUSY investigation\cite{DKKN}. It is well known 
that the Dirac-K\"ahler fermion mechanism is fundamentally related 
to the lattice formulation
\cite{KSuss,Suss,KS}\cite{Gliozzi:1982ib,Kluberg-Stern:1983dg}. 
In fact recently $N=2$ twisted superspace 
in two dimensions has been successfully formulated on a lattice with 
an introduction of mild non-commutability 
\cite{Woronowicz:1989rt,Dimakis:1992pk,Aschieri:1993wg,Dai:2000vf,Vaz1,KK} 
for lattice difference operator and twisted supercharges\cite{DKKN}. 
It is strongly suggested that $N=4$ twisted superspace formalism 
in four dimensions is important to formulate four-dimensional 
SUSY on a lattice. In particular $N=4$ twisted super Yang-Mills 
action is needed to formulate supergauge invariant action on a 
four-dimensional lattice. 

The recent AdS/CFT correspondence\cite{JMM,AGMOO} from superstring formulation 
also suggests 
that superspace formulation of $N=4$ super Yang-Mills action will help 
to understand the fundamental structure of Yang-Mills theory based on 
the brane dynamics. 

This paper is organized as follows: We first give the general formulation 
of $N=4$ twisted SUSY algebra based on Dirac-K\"ahler twist in section 2. 
Then we formulate twisted superspace and superfield for $N=D=2$ and 
$N=D=4$ in section 3. We introduce twisted chiral and 
vector superfields and propose off-shell twisted SUSY invariant actions. 
In section 4 we propose superconnection formalism to formulate twisted $N=2$ 
SUSY invariant super Yang-Mills actions in two and four 
dimensions.  We summarize the results in section 5. We provide several 
appendixes to give the details of tensor kinematics of twisted algebra 
and full expression of twisted $N=4$ SUSY invariant action.

\renewcommand{\theequation}{\arabic {section}.\arabic{equation}}
\section{Twisted SUSY from Dirac-K\"ahler twist}

The twisting procedure was first proposed by Witten in the derivation of 
the twisted version of $N=2$ super Yang-Mills action\cite{W}. It was soon 
recognized that the super Yang-Mills action can be derived by the 
quantization of topological Yang-Mills action with instanton gauge 
fixing, where the ghost-related fields turned into matter fermions 
via twisting mechanism\cite{bs,bms,lp,BRT1}. It was also shown that 
$N=D=2$ twisted 
super Yang-Mills action can be derived from instanton gauge fixing 
of the two-dimensional version of generalized topological Yang-Mills 
theory in the similar way as the four-dimensional case\cite{KT}. 
It was then found in the $N=D=2$ twisted SUSY formulation that 
the twisting mechanism is essentially related to the Dirac-K\"{a}hler 
fermion mechanism\cite{KT}. Then it has led to the proposal of the 
$N=2$ twisted superspace formalism in two dimensions\cite{KKU}.
Here we propose twisting mechanism of $N=4$ 
twisted super symmetry in four dimensions. We may call this twisting 
procedure as Dirac-K\"{a}hler twist. In order to show that a general 
formulation of the Dirac-K\"ahler twist of twisted SUSY for 
$N=2$ in two dimensions 
and $N=4$ in four dimensions has an intimate similarity, we construct the four-
dimensional formulation parallel to the two-dimensional case. 

The supercharges of extended supersymmetry algebra satisfy the following relations:
\begin{eqnarray}
\{ Q_{\alpha i} , \overline{Q}_{j \beta} \} &=& 2\delta_{ij} P_\mu \gamma^\mu
 _{\alpha\beta} ,
\label{eq:44cal}
\end{eqnarray}
where the indices $\{\alpha,\beta\}$ and the indices $\{ i,j\}$ are Lorentz spinor 
and internal $R$-symmetry suffix of an extended SUSY, respectively. 
For $N=2$ extended SUSY in two dimensions we take $i,j =1,2 $, while we take 
$i,j =1,2,3,4 $ for $N=4$ extended SUSY in four dimensions. 
Since we introduce $\gamma$-matrix in the right hand side of (\ref{eq:44cal}) 
the conjugate supercharge $\overline{Q}_{i \alpha}$ should be related 
to $Q_{\alpha i}$. 
$\gamma^\mu$ and $P_\mu$ are the corresponding $\gamma$-matix and momentum 
generator in two and four 
dimensions, respectively, and the explicit representation of four-dimensional 
$\gamma^\mu$ is given in Appendix A. Throughout this paper we consider 
Euclidean spacetime. 

\subsection{$N=2$ twisted SUSY in two dimensions}

In defining Dirac-K\"ahler twist we identify the right index of the supercharge 
$Q_{\alpha i}$ as spinor suffix, then we can decompose the charge into the 
following scalar, vector and pseudo-scalar components which we call twisted 
supercharges:
\begin{align}
 Q_{\alpha i}= \left(\mathbf{1} s + \gamma^\mu s_\mu + \gamma^5
\tilde{s}\right)_{\alpha i}.
\end{align}
We introduce the conjugate supercharge as 
$\overline{Q}_{i \alpha}  = (C^{-1}Q^TC)_{i\alpha}$ where we can take 
$C=\mathbf{1}$ in two-dimensional Euclidean spacetime and thus 
$\overline{Q}_{i \alpha} = Q_{\alpha i}$. The details of the notation 
in two dimensions can be found in \cite{KKU}. 

The relations (\ref{eq:44cal}) can now be rewritten by the twisted
generators as: 
\begin{equation}
\begin{split}
\{s,s_\mu\}&=P_\mu,\ \ \ 
\{\tilde{s},s_\mu\}=-\epsilon_{\mu\nu}P^\nu, \\ 
s^2=\tilde{s}^2&=\{s,\tilde{s}\}=\{s_\mu,s_\nu\}=0. \\
\end{split}
\label{eq:N=2 T-algebra1}
\end{equation}
This is the twisted $N=D=2$ SUSY algebra obtained from Dirac-K\"ahler twist. 

Similar to the supercharges we can introduce twisted superparameters as 
\begin{align}
 \theta_{\alpha i}= \frac{1}{2}\left(\mathbf{1} \theta + 
\gamma^\mu \theta_\mu + \gamma^5 \tilde{\theta}\right)_{\alpha i},
\end{align}
then we can define $N=2$ supersymmetry transformation as
\begin{equation}
\delta_\theta \ = \ \theta_{\alpha i} Q_{\alpha i} \ = \  
\theta s+\theta^\mu s_\mu+\tilde{\theta}\tilde{s}.
\end{equation}

In general $N=2$ SUSY in two dimensions includes super-Poincar\'{e} symmetry 
together with $R$-symmetry as follows: 
\begin{equation}
\begin{split}
[P_\mu,Q_{\alpha i}]&=0,\\
[J,Q_{\alpha i}]&=\tfrac{i}{2}(\gamma^5)_\alpha{^\beta}Q_{\beta i},\\
[R,Q_{\alpha i}]&=\tfrac{i}{2}(\gamma^5)_i{^j}Q_{\alpha j},\\
[J,P_\mu]&=i\epsilon_\mu{^\nu}P_\nu,\\
[P_\mu,P_\nu]&=[P_\mu,R]=[J,R]=0.
\end{split}
\label{eq:N=2 algebra}
\end{equation}
 $J$ and $R$ are generators of $SO(2)$ Lorentz and $SO(2)_I$ internal rotation 
of extended SUSY called $R$ symmetry, respectively, while $\gamma^5$ 
can be identified as the rotation generator of spinor suffix for those 
rotations.

The essential meaning of the Dirac-K\"ahler twist is to identify the extended 
SUSY indices as the spinor ones. Then the internal extended SUSY 
should transform as spinor under the Lorentz transformation. This will lead to a 
redefinition of the energy-momentum tensor and the Lorentz rotation generator.

We can redefine the energy-momentum tensor $T_{\mu\nu}$ as the following
relation without breaking the conservation law:
\begin{align}
T^\prime_{\mu\nu}
=T_{\mu\nu}+\epsilon_{\mu\rho}\partial^\rho R_\nu
+\epsilon_{\nu\rho}\partial^\rho R_\mu,
\end{align}
where $R_\mu$ is the conserved current associated with 
$R$ symmetry\cite{W,EY,LL}. This
modification leads to a redefinition of the Lorentz rotation generator,
\begin{align}
J^\prime=J+R.
\end{align}
This new rotation group is interpreted as the diagonal subgroup of $SO(2)\times
SO(2)_I$.

The twisted version of the super-Poiancar\'{e} and $R$-symmetry algebra can be 
obtained as follows:
\begin{equation}
\begin{split}
[J^\prime,s]&=[J^\prime,\tilde{s}]=0,\
[J^\prime,s_\mu]=i\epsilon_{\mu\nu}s^\nu,\\
[R,s]&=\tfrac{i}{2}\tilde{s},\
[R,s_\mu]=\tfrac{i}{2}\epsilon_{\mu\nu}s^\nu,\
[R,\tilde{s}]=-\tfrac{i}{2}s,\ \\
[J^\prime,P_\mu]&=i\epsilon_{\mu\nu}P^\nu,\ \\
[P_\mu,P_\nu]&=[P_\mu,R]=[J^\prime,R]=0.
\end{split}
\label{eq:N=2 T-algebra}
\end{equation}
This is the twisted $D=N=2$ SUSY algebra.

\subsection{$N=4$ twisted SUSY in four dimensions}

In this subsection we consider an extension of the twisting procedure 
using Dirac-K\"ahler twist into four dimensions. 
The Dirac-K\"ahler fermion mechanism is formulated in any dimensions
\cite{IL,Kahler,G,BJ,Rabin,BDH,BennT,Bull}. 
It is particularly convenient to formulate in even dimensions since the 
corresponding Clifford algebra is unambiguously defined with a chiral generator.
Similar to the two-dimensional case, we introduce the Dirac-K\"ahler twisting 
procedure in four dimensions. In this case 
the situation is more involved since the charge conjugation matrix
$C$ cannot be taken as unit matrix any more. 
Throughout this paper we consider the four-dimensional Euclidean
spacetime. 

As in the two-dimensional case we identify 
the extended SUSY suffix $\{i\}$ as spinor suffix of a diagonal subgroup 
$SO(4) \times SO(4)_I$. Analogous to the Dirac-K\"ahler mechanism, we can expand 
$Q_{\alpha i}$ as:
\begin{eqnarray}
Q_{\alpha i} &=& \frac{1}{\sqrt{2}}\bigg( {\bf 1} s + \gamma^{\mu} s_\mu+ 
\frac{1}{2}\gamma^{\mu\nu} s_{\mu\nu}+ \frac{1}{3!}\gamma^{\mu\nu\rho} s_{\mu\nu\rho}+ 
\frac{1}{4!}\gamma^{\mu\nu\rho\sigma} s_{\mu\nu\rho\sigma} \bigg)_{\alpha i}
\nonumber  \\
&=& \frac{1}{\sqrt{2}}\Big( {\bf 1} s + \gamma^{\mu} s_\mu+ 
\frac{1}{2}\gamma^{\mu\nu} s_{\mu\nu}+ \tilde{\gamma}^\mu \tilde{s}_\mu + 
\gamma_5 \tilde{s}\Big)_{\alpha i},
\label{eq:DKf}
\end{eqnarray}
where\ $\tilde{s}_\mu \equiv
\frac{1}{3!}{\epsilon_{\mu\nu\rho\sigma}s^{\nu\rho\sigma}}$ and $
\tilde{s} \equiv \frac{1}{4!}
\epsilon^{\mu\nu\rho\sigma} s_{\mu\nu\rho\sigma}$. 
We define the following conjugate supercharge:
\begin{eqnarray}
\overline{Q}_{i \alpha } = (C^{-1} Q^T C)_{i \alpha },
\end{eqnarray}
where $C$ is the charge conjugation matrix in four-dimensional Euclidean 
spacetime and satisfies 
\begin{equation}
\gamma_\mu{}^T = C \gamma_\mu C^{-1}, \ \ \ C^T = -C.
\end{equation}
We give explicit representation of the $\gamma$-matrix and the 
charge conjugation matrix in four dimensions and useful relations in Appendix A. 

The algebra (\ref{eq:44cal}) may be equivalently rewritten as 
\begin{eqnarray}
\{Q_{\alpha i}, Q_{\beta j}\}= 2C_{j i}(\gamma^\mu C^{-1})_{\alpha\beta} P_\mu.
\end{eqnarray}
The twisted supercharges $\{s_I\}$ are related to the supercharges $Q_{\alpha i}$ 
of $N=4$ extended SUSY as: 
\begin{eqnarray}
s &=& \frac{1}{2\sqrt{2}} \text{Tr} Q, \nonumber \\
s_\mu &=& \frac{1}{2\sqrt{2}} \text{Tr} (Q \gamma_\mu), \nonumber \\
s_{\mu\nu} &=& -\frac{1}{2\sqrt{2}} \text{Tr}(Q\gamma_{\mu\nu}), \nonumber \\
\tilde{s}_{\mu}&=& -\frac{1}{2\sqrt{2}} \text{Tr} (Q \tilde{\gamma}_\mu ), \nonumber \\
\tilde{s}&=& \frac{1}{2\sqrt{2}} \text{Tr} (Q \gamma_5).
\end{eqnarray}
We can then explicitly calculate the commutation relations of supercharges $\{s_I\}$
\begin{eqnarray}
\{ s, s_\mu\} &=& P_\mu, \qquad \{ s_\mu , s_{\rho\sigma} \} =
-(\delta_{\mu\rho}P_\sigma - \delta_{\mu\sigma} P_{\rho}), \nonumber \\
\{ \tilde{s}, \tilde{s}_\mu\} &=& P_\mu, \qquad 
\{ \tilde{s}_\mu, s_{\rho\sigma}  \}=\epsilon_{\mu\rho\sigma\nu} P^\nu, 
\nonumber \\
\{s,s_{\mu\nu}\}&=&\{s, \tilde{s}_\mu\} =\{s, \tilde{s} \}=
\{ s_\mu, \tilde{s}_\nu\}=\{ s_\mu \tilde{s}\}=\{s_{\mu\nu}, \tilde{s}\}=0, 
\nonumber \\
\{s,s\} &=& \{s_\mu ,s_\nu \}=\{s_{\mu\nu} , s_{\rho\sigma} \}=
\{\tilde{s}_\mu ,\tilde{s}_\nu\}= \{ \tilde{s}, \tilde{s}\}=0. 
\label{N=4_T-algebra}
\end{eqnarray}

Here we introduce Grassmann odd twisted SUSY parameters corresponding to the 
supercharges (\ref{eq:DKf})
\begin{eqnarray}
\theta_{\alpha i} 
&=& \frac{1}{2\sqrt{2}}\Big( {\bf 1} \theta + \gamma^{\mu} \theta_\mu+ 
\frac{1}{2}\gamma^{\mu\nu} 
\theta_{\mu\nu}+ \tilde{\gamma}^\mu \tilde{\theta}_\mu + 
\gamma_5 \tilde{\theta}\Big)_{\alpha i}.
\label{N4twisted-parameter}
\end{eqnarray}
Then the explicit form of $N=4$ twisted SUSY generator can be given by 
\begin{equation}
\delta_\theta \ = \ \overline{\theta}_{i \alpha} Q_{\alpha i} \ = \  
\theta s + \theta^{\mu} s_\mu+ \frac{1}{2}\theta^{\mu\nu} s_{\mu\nu}+ 
\tilde{\theta}^\mu \tilde{s}_\mu + \tilde{\theta} \tilde{s}, 
\end{equation}
where $\overline{\theta}$ has the same relation as (\ref{eq:DKf}) with 
$\theta$: 
$\overline{\theta}_{i \alpha}=(C^{-1}\theta^T C)_{i \alpha}$.

Next we consider super-Poincar\'{e} and $R$-rotation of $N=4$ SUSY algebra, 
\begin{eqnarray}
{[} P_\mu , P_\nu{]}&=&0, \nonumber \\
{[} Q_{\alpha i} ,P_\mu {]}&=&0,\nonumber \\
{[}M_{\mu\nu}, M_{\rho\sigma}{]}&=& -i(\delta_{\mu\rho} M_{\nu\sigma} -
\delta_{\nu\rho} M_{\mu\sigma}-\delta_{\mu\sigma} M_{\nu\rho} +
\delta_{\nu\sigma} M_{\mu\rho}),\nonumber \\
{[} P_\mu , M_{\rho\sigma}{]}&=&i(\delta_{\mu\rho}P_\sigma-
\delta_{\mu\sigma}P_\rho), \nonumber \\
{[}  Q_{\alpha i} , M_{\mu\nu} {]}&=& 
\frac{i}{2} (\gamma_{\mu\nu})_{\alpha\beta} Q_{\beta i}, 
\label{twisted-algebra1} \\
{[}R_{\mu\nu}, R_{\rho\sigma}{]} &=& -i(\delta_{\mu\rho} R_{\nu\sigma} -
\delta_{\nu\rho} R_{\mu\sigma}-\delta_{\mu\sigma} R_{\nu\rho} +
\delta_{\nu\sigma} R_{\mu\rho}),\nonumber \\
{[}  Q_{\alpha i} , R_{\mu\nu} {]}&=& 
\frac{i}{2} (\Gamma_{\mu\nu})_{ij} Q_{\alpha j}, \nonumber \\
{[}R_{\mu\nu} , P_{\rho} {]}&=&0,\nonumber \\
{[}R_{\mu\nu} , M_{\rho\sigma} {]}&=&0,\nonumber 
\end{eqnarray}
where $M_{\mu\nu}$ is $SO(4)$ Lorentz generator and $R_{\mu\nu}$ is $SO(4)_I$ internal 
space rotation generator called R-symmetry. $\Gamma_\mu$ satisfies the same
Clifford algebra as $\gamma_\mu$ and can be identified as the rotation generators 
of spinor suffix of $SO(4)_I$.

We introduce the Dirac-K\"ahler twisting procedure where we identify the internal 
SUSY suffix $\{i\}$ as spinor suffix of $SO(4)_I$. We can then identify 
$\Gamma^\mu_{ij}$ as $(\gamma^\mu)_{\alpha\beta}$ in particular 
$\Gamma^\mu =(\gamma^\mu)^T$. In this case the twisted algebras 
do not have the Lorentz invariance of original Lorentz group any more. For the 
$R$-rotation we obtain 
\begin{eqnarray}
& &[Q_{\alpha i}, R_{\mu\nu}]=\frac{i}{2}(\Gamma_{\mu\nu})_{ij} Q_{\alpha j} 
\nonumber \\
&\to& [Q_{\alpha i}, R_{\mu\nu}]=-\frac{i}{2}(\gamma_{\mu\nu})^T_{i j} Q_{\alpha j} 
\nonumber \\
& &\hspace{20mm}=-\frac{i}{2}Q_{\alpha j} (\gamma_{\mu\nu})_{j i}.
\end{eqnarray}
Substituting $Q_{\alpha i}$ into (\ref{eq:DKf}), we obtain the following algebraic 
relations: 
\begin{eqnarray}
{[} s, R_{\mu\nu}{]} &=& \frac{i}{2}s_{\mu\nu}, \nonumber \\
{[} s_\rho, R_{\mu\nu}{]}&=& \frac{i}{2}( \delta_{\mu\rho} s_{\nu}-
\delta_{\nu\rho}s_{\mu}) -\frac{i}{2}\epsilon_{\mu\nu\rho\sigma} \tilde{s}^\sigma, 
\nonumber \\
{[} s_{\rho\sigma }, R_{\mu\nu}{]}&=& 
-\frac{i}{2} (\delta_{\mu\rho} \delta_{\nu\sigma} -\delta_{\nu\rho}\delta_{\mu\sigma})s
-\frac{i}{2} (\delta_{\mu\rho} s_{\sigma\nu} -\delta_{\mu\sigma} s_{\rho\nu}- 
\delta_{\nu\rho} s_{\sigma\mu}+\delta_{\nu\sigma} s_{\rho\mu})  \nonumber \\
& & +\frac{i}{2} \epsilon_{\mu\nu\rho\sigma} \tilde{s} \nonumber, \\
{[}\tilde{s}_{\rho} ,R_{\mu\nu} {]} &=&
-\frac{i}{2} \epsilon_{\mu\nu\rho\sigma} s^\sigma +
\frac{i}{2} (\delta_{\mu\rho} \tilde{s}_\nu -\delta_{\nu\rho} \tilde{s}_\mu)
\nonumber, \\
{[} \tilde{s} , R_{\mu\nu} {]} &=& -\frac{i}{4} \epsilon_{\mu\nu\rho\sigma}s^{\rho\sigma}.
\end{eqnarray}
Other algebraic relations can be obtained in the similar way. It should be noted that 
$\{s_I\}$ are not tensors under the original Lorentz transformation. They are, however, 
tensors of newly defined Lorentz generators,
\begin{eqnarray}
M'_{\mu\nu} = M_{\mu\nu} +R_{\mu\nu}.
\label{eq:RM}
\end{eqnarray}
Using the new Lorentz generator $M'_{\mu\nu}$, we can represent the super-Poincar\'{e} and 
$R$-symmetry part of $N=4$ twisted SUSY algebras as follows: 
\begin{allowdisplaybreaks}
\begin{eqnarray}
{[} P_\mu , P_\nu{]}&=&0, \nonumber \\
{[}s,P_\rho {]}&=&{[}s_\mu,P_\rho {]}={[}s_{\mu\nu},P_\rho {]}=
{[}\tilde{s}_\mu,P_\rho {]}={[}\tilde{s},P_\rho {]}=0,\nonumber \\
{[}M'_{\mu\nu}, M'_{\rho\sigma}{]}&=& -i(\delta_{\mu\rho} M'_{\nu\sigma} -
\delta_{\nu\rho} M'_{\mu\sigma}-\delta_{\mu\sigma} M'_{\nu\rho} +
\delta_{\nu\sigma} M'_{\mu\rho}),\nonumber \\
{[} P_\mu , M'_{\rho\sigma}{]}&=&i(\delta_{\mu\rho}P_\sigma-
\delta_{\mu\sigma}P_\rho), \nonumber \\
{[}s,M'_{\rho\sigma} {]}&=&{[}\tilde{s},M'_{\rho\sigma} {]}=0,\nonumber \\
{[}s_\mu , M'_{\rho\sigma}{]} &=& 
i(\delta_{\mu\rho}s_\sigma-\delta_{\mu\sigma} s_\rho ), \qquad 
{[}\tilde{s}_\mu , M'_{\rho\sigma}{]} = i(\delta_{\mu\rho}\tilde{s}_\sigma-
\delta_{\mu\sigma} \tilde{s}_\rho ), \nonumber \nonumber \\
{[} s_{\rho\sigma} , M'_{\mu\nu} {]}&=&-i (\delta_{\mu\rho} s_{\sigma\nu} -
\delta_{\mu\sigma} s_{\rho\nu}- \delta_{\nu\rho} s_{\sigma\mu}+
\delta_{\nu\sigma} s_{\rho\mu}),\nonumber \\
{[}R_{\mu\nu}, R_{\rho\sigma}{]} &=& -i(\delta_{\mu\rho} R_{\nu\sigma} -
\delta_{\nu\rho} R_{\mu\sigma}-\delta_{\mu\sigma} R_{\nu\rho} +
\delta_{\nu\sigma} R_{\mu\rho}), \nonumber \\
{[} s, R_{\mu\nu}{]} &=& \frac{i}{2}s_{\mu\nu}, \nonumber \\
{[} s_\rho, R_{\mu\nu}{]}&=& \frac{i}{2}( \delta_{\mu\rho} s_{\nu}-
\delta_{\nu\rho}s_{\mu}) -\frac{i}{2}\epsilon_{\mu\nu\rho\sigma} \tilde{s}^\sigma, 
\nonumber \\
{[} s_{\rho\sigma }, R_{\mu\nu}{]}&=& 
-\frac{i}{2} (\delta_{\mu\rho} \delta_{\nu\sigma} -
\delta_{\nu\rho}\delta_{\mu\sigma})s
-\frac{i}{2} (\delta_{\mu\rho} s_{\sigma\nu} -
\delta_{\mu\sigma} s_{\rho\nu}- \delta_{\nu\rho} s_{\sigma\mu}+
\delta_{\nu\sigma} s_{\rho\mu})  \nonumber \\
& & +\frac{i}{2} \epsilon_{\mu\nu\rho\sigma} \tilde{s}, \nonumber \\
{[}\tilde{s}_{\rho} ,R_{\mu\nu} {]} &=&
-\frac{i}{2} \epsilon_{\mu\nu\rho\sigma} s^\sigma +
\frac{i}{2} (\delta_{\mu\rho} \tilde{s}_\nu -\delta_{\nu\rho} \tilde{s}_\mu), 
\nonumber \\
{[} \tilde{s} , R_{\mu\nu} {]} &=& 
-\frac{i}{4} \epsilon_{\mu\nu\rho\sigma}s^{\rho\sigma},\nonumber \\
{[}R_{\mu\nu} , P_{\rho} {]}&=&0,\nonumber \\
{[}R_{\mu\nu} , M'_{\rho\sigma} {]}&=&-i(\delta_{\mu\rho} R_{\nu\sigma} -
\delta_{\nu\rho} R_{\mu\sigma}-\delta_{\mu\sigma} R_{\nu\rho} +
\delta_{\nu\sigma} R_{\mu\rho}).
\end{eqnarray}
\end{allowdisplaybreaks}
As we can see in this algebra, $s,s_\mu,s_{\mu\nu},\tilde{s}_\nu,\tilde{s}$ transform 
as scalar, vector, tensor, pseudo-vector and pseudo-scalar, respectively under the 
new Lorentz generator $M'_{\mu\nu}$.

Next we consider how the relation (\ref{eq:RM}) has an effect on the
transformation law of component fields. We define the Dirac-K\"ahler field as 
follows\cite{IL,Kahler,G,BJ,Rabin,BDH,BennT,Bull}, 
\begin{eqnarray}
\Psi_{\alpha i} = \frac{1}{\sqrt{2}}\Big( {\bf 1}\psi + \gamma^{\mu} \psi_\mu+ \frac{1}{2}\gamma^{\mu\nu} \psi_{\mu\nu}+ \tilde{\gamma}^\mu  \tilde{\psi}_\mu + \gamma_5 \tilde{\psi}\Big)_{\alpha i},
\label{def-DK-fields}
\end{eqnarray}
where $\Psi_{\alpha i} $ appears in the $N=4$ extended supersymmetric theory while 
$\{\psi ,\ \psi_\mu ,\ \psi_{\mu\nu} , \\ 
  \tilde{\psi}_\mu ,\ \tilde{\psi}\}$ appear 
in the twisted $N=4$ supersymmetric theory.
As $M'$ is the Lorentz generator in the twisted theory, we can define
the transformation laws of $\{ \psi,\psi_\mu, \psi_{\mu\nu}, \\ 
\tilde{\psi}_\mu, \tilde{\psi} \}$ as 
\begin{eqnarray}
\delta_{M'} \psi &=& 0, \nonumber  \\
\delta_{M'} \psi_\mu &=& 2ik_{\mu\nu} \psi^\nu, \nonumber  \\
\delta_{M'} \psi_{\mu\nu} &=& -2i(k_{\mu} { }^\rho \psi_{\nu\rho}-k_{\nu} {
}^\rho \psi_{\mu\rho} ), \nonumber  \\
\delta_{M'} \tilde{\psi}_\mu &=& 2ik_{\mu\nu} \tilde{\psi}^\nu \nonumber,  \\
\delta_{M'} \tilde{\psi} &=& 0,  
\end{eqnarray}
where $k_{\mu\nu}$ is bosonic anti-symmetric tensor. Therefore the Dirac-K\"ahler
field transforms in the following form:
\begin{eqnarray}
\delta_{M'} \Psi = \frac{i}{2}[k_{\mu\nu} \gamma^{\mu\nu}, \Psi].
\end{eqnarray}
We can also define the R-symmetry transformation law of theses fields, 
\begin{eqnarray}
\delta_R \psi &=& \frac{i}{2} k^{\mu\nu} \psi_{\mu\nu}, \nonumber \\
\delta_R \psi_\mu &=& ik_{\mu\nu}\psi^\nu - \frac{i}{2}k^{\rho\sigma}
\epsilon_{\rho\sigma\mu\nu} \tilde{\psi}^\nu, \nonumber \\
\delta_R \psi_{\mu\nu} &=& -ik_{\mu\nu} \psi  -i(k_\mu { }^\rho
\psi_{\nu\rho}-k_\nu { }^\rho \psi_{\mu\rho} )
+\frac{i}{2}k^{\rho\sigma} \epsilon_{\rho\sigma\mu\nu} \tilde{\psi}, 
\nonumber \\
\delta_R \tilde{\psi}_\mu &=& -\frac{i}{2} k^{\rho\sigma}
\epsilon_{\rho\sigma\mu\nu} \psi^\nu +ik_{\mu\nu}\tilde{\psi}^\nu,  \nonumber \\
\delta_R \tilde{\psi} &=& -\frac{i}{4} k^{\rho\sigma}\epsilon_{\mu\nu\rho\sigma} \psi^{\mu\nu}.  \nonumber \\
\end{eqnarray}
The Dirac-K\"ahler field transforms under the R-symmetry transformation in the following:
\begin{eqnarray}
\delta_R \Psi = -\frac{i}{2} \Psi k_{\rho\sigma} \gamma^{\rho\sigma}.
\end{eqnarray}
Here we may replace $\gamma_{\mu\nu}$ by $\Gamma_{\mu\nu}$ and then 
\begin{eqnarray}
\delta_R \Psi_{\alpha i} = \frac{i}{2} k_{\rho\sigma} (\Gamma^{\rho\sigma})_{ij} \Psi_{\alpha j}.
\end{eqnarray}
This transformation is the R-symmetry transformation of spinor filed with
$SO(4)_I$ internal symmetry. 
On the other hand the Lorentz transformation induced by $M=M'- R$ is
\begin{eqnarray}
\delta_{M} \Psi &=& \delta_{M'} \Psi -  \delta_{R}\Psi \nonumber \\
&=& \frac{i}{2}[k_{\mu\nu} \gamma^{\mu\nu}, \Psi]+\frac{i}{2}\Psi k_{\rho\sigma} \gamma^{\rho\sigma} \nonumber \\
&=& \frac{i}{2} k_{\mu\nu} \gamma^{\mu\nu} \Psi,
\end{eqnarray}
which precisely coincides with the Lorentz transformation of a spinor
field. Thus the R-symmetry generator $R$ plays the role of shifting the
integer spin ghost-related fermions into the half integer spin matter fermions. 
This shows that Dirac-K\"ahler fermion mechanism is essentially related to the 
twisting procedure of $N=4$ extended SUSY in four dimensions just like 
the two-dimensional case\cite{KT,KKU}.

\section{Twisted superspace and superfield }

In the Dirac-K\"ahler twisting procedure we have introduced the twisted supercharges 
in the following form: 
\begin{eqnarray}
 Q_{\alpha i} = a_D \left(\gamma^I s_I \right)_{\alpha i},        
\end{eqnarray}
where the normalization constant $a_D$ has a dimension dependence in our notation:
\begin{equation}
\left\{ 
\begin{array}{ll}
          a_2 = 1 & (\mbox{$N=2$ in two dimensions}),\\
          a_4 = \frac{1}{\sqrt{2}} & (\mbox{$N=4$ in four dimensions}).
\end{array}
\right. 
\end{equation}
We then introduce the corresponding twisted superparameters as:
\begin{align}
 \theta_{\alpha i}= \frac{a_D}{2}\left(\gamma^I \theta_I \right)_{\alpha i}. 
\end{align} 
Then we can define a twisted SUSY transformation as 
\begin{equation}
\delta_\theta \ = \ \overline{\theta}_{i \alpha} Q_{\alpha i} \ = \  
\overline{\theta}_I s_I.
\label{def_delta_theta}
\end{equation}
We can then define the following supergroup element acting on a twisted superspace:
\begin{align}
G(x^\mu,\theta_A)
 =e^{i(-x^\mu P_\mu + \delta_\theta)}. 
\end{align}
$N=D=2$ and $N=D=4$ twisted superspace of extended SUSY is defined in the 
parameter space of $(x^\mu,\theta_I)$.

By using the relations (\ref{eq:N=2 T-algebra1}) or (\ref{N=4_T-algebra}) and 
Baker-Hausdorff formula, we can obtain the following relation:
\begin{align}
G(0,\xi_I) \ G(x^\mu,\theta_I) = G(x^\mu+b^\mu,\theta_I+\xi_I),
\end{align}
where 
\begin{equation}
\begin{split}
b^\mu=&\frac{i}{2}\xi\theta^\mu+\tfrac{i}{2}\xi^\mu\theta
 +\tfrac{i}{2}\epsilon^\mu{_\nu}\xi^\nu\tilde{\theta}
 +\tfrac{i}{2}\epsilon^\mu{_\nu}\tilde{\xi}\theta^\nu ~~~\\
&(\hbox{$N=2$ in two dimensions}),\\
b^\mu=&\frac{i}{2} \xi \theta^\mu + \frac{i}{2} \xi^\mu \theta - 
\frac{i}{2} \xi_\nu \theta^{\nu\mu}- \frac{i}{2} \xi^{\nu\mu} \theta_\nu - 
\frac{i}{4} \epsilon^{\mu\nu\rho\sigma} \tilde{\xi}_\nu \theta_{\rho\sigma} 
- \frac{i}{4} \epsilon^{\mu\nu\rho\sigma} \xi_{\rho\sigma} \tilde{\theta}_\nu 
+\frac{i}{2} \tilde{\xi}^\mu \tilde{\theta} + 
\frac{i}{2} \tilde{\xi} \tilde{\theta}^\mu \\
&(\hbox{$N=4$ in four dimensions}).  
\end{split}
\label{N=4-a-parameter}
\end{equation}
This multiplication induces a shift transformation in superspace
$(x^\mu,\{\theta_I\})$:
\begin{equation}
(x^\mu,\{\theta_I\}) \rightarrow (x^\mu+b^\mu,\{\theta_I+\xi_I\}),
\label{parameter-shift1}
\end{equation}
which is in general generated by supercharge differential operators $\mathcal{Q}_A$. 
Accordingly supercharge differential operators are introduced to satisfy the following 
parameter shifts: 
\begin{align}
\delta_\xi
\begin{pmatrix}
x^\mu\\
\theta_I\\
\end{pmatrix}
=(\xi_I \mathcal{Q}_I)
\begin{pmatrix}
x^\mu\\
\theta_I\\
\end{pmatrix}
=
\begin{pmatrix}
a^\mu\\
\xi_I\\
\end{pmatrix}. \label{parameter-shift2}
\end{align}
We introduce a general superfield $\Upsilon(x,\{\theta_I\})$,
\begin{eqnarray}
\Upsilon(x,\{\theta_I\}) &=& \phi + \theta_I \phi_I + 
\frac{1}{2}\theta_I \theta_J \phi_{IJ} + \cdots . 
\label{eq:gsf}
\end{eqnarray}
We then define twisted SUSY transformation of component fields as follows:
\begin{eqnarray}
\delta_\xi \Upsilon(x,\{\theta_I\}) &=&\delta_\xi \phi + \theta_I \delta_\xi \phi_I 
+ \frac{1}{2}\theta_I \theta_J \delta_\xi \phi_{IJ} + \cdots \nonumber \\
&\equiv & (\xi_I \mathcal{Q}_I) \Upsilon(x,\{\theta_I\}),
\end{eqnarray}
where $\delta_\xi = \overline{\xi}_I s_I $.
It should be noted that this operator applies not on the superfield but on component fields 
as well. We thus obtain the full SUSY transformation law of
component fields in principle. In general superfield is, however, highly reducible 
and thus we need to introduce a constraint on the general superfield. 
 Once we know the SUSY transformation of the component fields by any means, 
we can construct the superfield itself as follows:
\begin{eqnarray}
\Upsilon(x,\{\theta_I\}) &=& e^{\delta_\theta}\phi \nonumber \\
                     &=& \phi + \delta_\theta \phi + 
\frac{1}{2}(\delta_\theta)^2 \phi  + \cdots ,
\label{superfield-expansion}
\end{eqnarray} 
where $\phi$ is the parent field to generate other component fields in the same 
twisted supermultiplet.

\subsection{$N=2$ twisted superspace in two dimensions}

The twisted $N=2$ supercharge differential operators generating the 
parameter shift (\ref{parameter-shift1}) and (\ref{parameter-shift2}) are given by 
\begin{equation}
\begin{split}
\mathcal{Q}
 &=\frac{\partial}{\partial \theta}
  +\frac{i}{2}\theta^\mu\partial_\mu,\\
\mathcal{Q}_\mu
 &=\frac{\partial}{\partial \theta^\mu}
  +\frac{i}{2}\theta\partial_\mu
  -\frac{i}{2}\tilde{\theta}\epsilon_{\mu\nu}\partial^\nu,\\
\tilde{\mathcal{Q}}
 &=\frac{\partial}{\partial \tilde{\theta}}
  -\frac{i}{2}\theta^\mu\epsilon_{\mu\nu}\partial^\nu,~~~~ 
\end{split}
\label{eq:Q-operator}
\end{equation}
which satisfy the following $N=2$ twisted superalgebra: 
\begin{eqnarray}
\{ \mathcal{Q} \ , \mathcal{Q}_{\mu} \} &=& i \partial_{\mu},  \nonumber \\
 \{ \tilde{\mathcal{Q}} \ , \mathcal{Q}_{\mu}\} &=& -i\epsilon_{\mu\nu} \partial^{\nu},  \nonumber \\
 \{\mathcal{Q}_\mu ,\mathcal{Q}_\nu\}&=&\{\mathcal{Q},\tilde{\mathcal{Q}}\}=\mathcal{Q}^2=\tilde{\mathcal{Q}}^2 =0.
\end{eqnarray} 
It should be note that the sign of spacetime derivative is reversed with respect to the 
original algebra (\ref{eq:N=2 T-algebra1}) due to the reverse order 
in operation. We introduce 
superderivative differential operators $\{\mathfrak{D}_I\}$ 
which anticommute with $\{ \mathcal{Q}_I \}$,
\begin{align}
\{\mathcal{Q}^I,\mathfrak{D}^J\}=0.
\end{align}
We find
\begin{eqnarray}
\mathfrak{D} &=& \frac{\partial}{\partial \theta} -\frac{i}{2} \theta^\mu \partial_\mu, \nonumber \\
\mathfrak{D}_\mu &=& \frac{\partial}{\partial \theta^\mu} -\frac{i}{2}\theta \partial_\mu
+\frac{i}{2} \tilde{\theta} \epsilon_{\mu \nu} \partial^\nu, \nonumber \\
\tilde{\mathfrak{D}} &=& \frac{\partial}{\partial \tilde{\theta}} +\frac{i}{2} \theta^{\mu}\epsilon_{\mu \nu} \partial^\nu.
\label{def_2d-super-derivative}
\end{eqnarray}
We can see that these operators satisfy the following algebra:
\begin{eqnarray}
\{ \mathfrak{D} \ , \mathfrak{D}_{\mu} \} &=& -i \partial_{\mu},  \nonumber \\
 \{ \tilde{\mathfrak{D}} \ , \mathfrak{D}_{\mu}\} &=& i\epsilon_{\mu\nu} \partial^{\nu},  \nonumber \\
 \{\mathfrak{D}_\mu ,\mathfrak{D}_\nu\}&=&\{\mathfrak{D},\tilde{\mathfrak{D}}\}=\mathfrak{D}^2=\tilde{\mathfrak{D}}^2 =0,
\label{2d-super-derivative-algebara}
\end{eqnarray} 
which has the same algebraic structure as (\ref{eq:N=2 T-algebra1}).
Similarly we can derive the twisted angular momentum and $R$-symmetry 
differential operators which satisfy the $N=2$ twisted SUSY algebra 
(\ref{eq:N=2 T-algebra}): 
\begin{eqnarray}
\cal{J}^{\prime} &=& i\epsilon^{\mu\nu} x_\mu \partial_\nu + 
i\epsilon^{\mu\nu}\theta_\mu \frac{\partial}{\partial\theta^\nu}, \nonumber \\
\cal{R}&=&- \frac{i}{2} \tilde{\theta}\frac{\partial}{\partial\theta}+
\frac{i}{2}\theta \frac{\partial}{\partial\tilde{\theta}}+
\frac{i}{2}\epsilon^{\mu\nu}\theta_\mu \frac{\partial}{\partial\theta^\nu},
\end{eqnarray}
where 
\begin{equation}
\cal{J}^{\prime} = \cal{J} + \cal{R}.
\end{equation}

\subsection{$N=4$ twisted superspace in four dimensions}

In the previous section we have derived $N=4$ twisted SUSY algebra based on 
the Dirac-K\"{a}hler twisting procedure. In this section we will construct a $N=4$ 
twisted superspace formalism. The $N=4$ twisted supercharges satisfy the 
algebra (\ref{N=4_T-algebra}):
\begin{eqnarray}
\{ s, s_\mu\} &=& -i\partial_\mu, \qquad \{ s_\mu , s_{\rho\sigma} \} =i(\delta_{\mu\rho}\partial_\sigma - \delta_{\mu\sigma} \partial_{\rho})\equiv i\mathcal{D}_{\mu,\rho\sigma}, \nonumber \\
\{ \tilde{s}, \tilde{s}_\mu\} &=& -i\partial_\mu, \qquad \{ \tilde{s}_\mu, s_{\rho\sigma}  \}=-i\epsilon_{\rho\sigma\mu\nu} \partial^\nu, \nonumber \\
\{\hbox{others}\}&=& 0, 
\label{eq:44al}
\end{eqnarray}
where we have introduced an explicit representation of momentum $P_\mu=-i\partial_\mu$.
We construct superspace formalism based on this $N=4$ twisted SUSY algebra.

The twisted supercharge differential operators generating the parameter shifts 
(\ref{parameter-shift1}) and (\ref{parameter-shift2}) for the $N=4$ twisted superspace 
are given by 
\begin{eqnarray}
\mathcal{Q} &=& \frac{\partial}{\partial\theta} +\frac{i}{2} \theta^\mu\partial_\mu, \nonumber \\
\mathcal{Q}_\mu &=& \frac{\partial}{\partial\theta^\mu} +\frac{i}{2} \theta \partial_\mu -\frac{i}{2} \theta_{\mu\nu} \partial^\nu, \nonumber \\
\mathcal{Q}_{\rho\sigma} &=& \frac{\partial}{\partial\theta^{\rho\sigma}} -\frac{i}{2} (\theta_\rho \partial_\sigma - \theta_\sigma \partial_\rho)+\frac{i}{2}\tilde{\theta}^\mu \epsilon_{\mu\nu\rho\sigma} \partial^\nu, \nonumber \\
\tilde{\mathcal{Q}}_\mu  &=& \frac{\partial}{\partial \tilde{\theta}^\mu}+\frac{i}{4} \theta^{\rho\sigma} \epsilon_{\rho\sigma\mu\nu} \partial^{\nu} +\frac{i}{2} \tilde{\theta} \partial_\mu, \nonumber \\
\tilde{\mathcal{Q}} &=& \frac{\partial}{\partial\tilde{\theta}} + \frac{i}{2}\tilde{\theta}^\mu \partial_\mu.
\end{eqnarray}
These differential operators satisfy the following relations:
\begin{eqnarray}
\{ \mathcal{Q}, \mathcal{Q}_\mu\} &=& i\partial_\mu, \qquad \{ \mathcal{Q}_\mu , \mathcal{Q}_{\rho\sigma} \} = -i\mathcal{D}_{\mu,\rho\sigma}, \nonumber \\
\{ \tilde{\mathcal{Q}}, \tilde{\mathcal{Q}}_\mu\} &=& i\partial_\mu, \qquad \{ \tilde{\mathcal{Q}}_\mu, \mathcal{Q}_{\rho\sigma}  \}=i\epsilon_{\rho\sigma\mu\nu} \partial^\nu, \nonumber \\
 \{\hbox{others}\} &=& 0 
\end{eqnarray}
where we have introduced the following notations:
\begin{eqnarray}
\mathcal{D}_{\mu,\rho\sigma} &\equiv& \delta_{\mu\rho}\partial_\sigma-
\delta_{\mu\sigma}\partial_\rho = \delta_{\mu\nu,\rho\sigma}\partial^\nu, \nonumber \\
\delta_{\mu\nu,\rho\sigma} &\equiv& \delta_{\mu\rho}\delta_{\nu\sigma} -
\delta_{\mu\sigma}\delta_{\nu\rho}, \nonumber \\
\frac{\partial \theta^{\mu\nu}}{\partial \theta^{\rho\sigma}} &=& 
\delta^\mu _\rho \delta^\nu _\sigma  -\delta^\mu _\sigma \delta^\nu _\rho =
\delta^{\mu\nu} { }_{\rho\sigma}.
\end{eqnarray} 
Next we introduce superderivative differential operators $\{\mathfrak{D}_I\}$ which anticommute with all differential operators $\{\mathcal{Q}_I\}$:
\begin{eqnarray}
\mathfrak{D} &=& \frac{\partial}{\partial\theta} -\frac{i}{2} \theta^\mu\partial_\mu \nonumber, \\
\mathfrak{D}_\mu &=& \frac{\partial}{\partial\theta^\mu} -\frac{i}{2} \theta \partial_\mu +\frac{i}{2} \theta_{\mu\nu} \partial^\mu, \nonumber \\
\mathfrak{D}_{\rho\sigma} &=& \frac{\partial}{\partial\theta^{\rho\sigma}} +\frac{i}{2} (\theta_\rho \partial_\sigma + \theta_\sigma \partial_\rho)-\frac{i}{2}\tilde{\theta}^\mu \epsilon_{\mu\nu\rho\sigma} \partial^\nu, \nonumber \\
\tilde{\mathfrak{D}}_\mu  &=& \frac{\partial}{\partial \tilde{\theta}^\mu}-\frac{i}{4} \theta^{\rho\sigma} \epsilon_{\rho\sigma\mu\nu} \partial^{\nu} -\frac{i}{2} \tilde{\theta} \partial_\mu, \nonumber \\
 \tilde{\mathfrak{D}} &=& \frac{\partial}{\partial\tilde{\theta}} - \frac{i}{2}\tilde{\theta}^\mu \partial_\mu.
\end{eqnarray}
We use these operators $\{\mathfrak{D}_I\}$ to impose chiral and anti-chiral conditions 
and to reduce the unnecessary degrees of freedom in a twisted superfield.
The differential operator of the twisted Lorentz generator $M_{\rho\sigma}'$ and
the R-symmetry generator $R_{\rho\sigma}$ are given by
\begin{eqnarray}
\cal{M}_{\rho\sigma}'&=& -i\delta_{\rho\sigma,\alpha\beta}x^\alpha\partial^\beta
-i\delta_{\rho\sigma,\alpha\beta}\theta^\alpha \frac{\partial}{\partial\theta_\beta}
-i\delta_{\rho\sigma,\alpha\beta}\tilde{\theta}^\alpha 
\frac{\partial}{\partial\tilde{\theta}_\beta}
-i\delta_{\rho\sigma,\alpha\beta}\theta^{\alpha\gamma} 
\frac{\partial}{\partial\theta_\beta { }^\gamma}, \nonumber \\
\cal{R}_{\rho\sigma} &=&- 
\frac{i}{2}\delta_{\rho\sigma,\alpha\beta}\theta^\alpha 
\frac{\partial}{\partial\theta_\beta}
-\frac{i}{2}\delta_{\rho\sigma,\alpha\beta}\tilde{\theta}^\alpha 
\frac{\partial}{\partial\tilde{\theta}_\beta}
-\frac{i}{2}\delta_{\rho\sigma,\alpha\beta}\theta^{\alpha\gamma} 
\frac{\partial}{\partial\theta_\beta { }^\gamma} 
-\frac{i}{2}\theta \frac{\partial}{\partial\theta^{\rho\sigma}} \nonumber \\
& &+\frac{i}{4}\epsilon_{\rho\sigma\alpha\beta}\tilde{\theta}
\frac{\partial}{\partial\theta_{\alpha\beta}}
+\frac{i}{2}\epsilon_{\rho\sigma\alpha\beta}\theta^\alpha 
\frac{\partial}{\partial\tilde{\theta}_{\beta}}
+\frac{i}{2}\epsilon_{\rho\sigma\alpha\beta}\tilde{\theta}^\alpha 
\frac{\partial}{\partial\theta_{\beta}}
+\frac{i}{2}\theta_{\rho\sigma}\frac{\partial}{\partial\theta} \nonumber \\
& &- \frac{i}{4}\epsilon_{\rho\sigma\alpha\beta}\theta^{\alpha\beta} 
\frac{\partial}{\partial\tilde{\theta}},  
\end{eqnarray}
where the newly defined Lorentz generators $\cal{M}_{\rho\sigma}'$ 
and the original Lorentz generators $\cal{M}_{\rho\sigma}$ 
have the same relation as (\ref{eq:RM}),
\begin{equation}
\cal{M}_{\rho\sigma}'= \cal{M}_{\rho\sigma} + \cal{R}_{\rho\sigma}.
\end{equation}
\subsection{Chiral and anti-chiral superfields}

We consider chiral superfield characterized by the following condition:
\begin{eqnarray}
\mathfrak{D}\Psi(x^\mu,\{\theta_I\}) = \mathfrak{D}_{\rho\sigma}\Psi(x^\mu,\{\theta_I\}) 
= \tilde{\mathfrak{D}} \Psi(x^\mu,\{\theta_I\}) =0,
\label{eq:4ccs1}
\end{eqnarray}
where $\Psi$ has all the 
$\{\theta_I\} = \{\theta,\theta^\mu,\theta^{\mu\nu},\tilde{\theta}^\mu,\tilde{\theta}\}$ 
dependence at this stage. The details of the twisted chiral 
superfield formulation for $N=D=2$ can be found in \cite{KKU,DKKN}. 
The reason why we call these condition as the chiral condition stems from the 
Dirac-K\"ahler fermion formulation on a lattice. The chiral transformation 
on a lattice is interpreted as an interchange between the even sites 
and odd sites which in turn can be translated into the interchange between 
even and odd differential forms. We may then identify the conditions in 
(\ref{eq:4ccs1}) as the chiral sector of even forms. 
There are several 
possible treatments to solve this type of differential equations. 
It is convenient to rewrite the chiral and anti-chiral conditions by using the operator 
which satisfy the following relations for the differential operators:
\begin{align}
U\mathfrak{D}U^{-1} \ &= \ \frac{\partial}{\partial \theta} , \ \ 
U\mathfrak{D}_{\rho\sigma}U^{-1} \ = \ \frac{\partial}{\partial \theta^{\rho\sigma}},
\ \ \ 
U\tilde{\mathfrak{D}}U^{-1} \ = \ \frac{\partial}{\partial \tilde{\theta}}, \nonumber \\
\label{diagonalization1}
\end{align}
and 
\begin{align}
U^{-1} \mathfrak{D}_\mu U \ &= \ \frac{\partial}{\partial \theta^\mu}, \ \ \ 
U^{-1} \tilde{\mathfrak{D}}_\mu U \ = \ \frac{\partial}{\partial \tilde{\theta}^\mu}, 
\label{diagonalization2}
\end{align}
where 
\begin{eqnarray}
U&=&e^{-a^\mu \partial_\mu}, \nonumber \\
a^\mu &=&  \frac{i}{2} \theta \theta^\mu  +\frac{i}{2} \theta_\nu \theta^{\nu\mu}+ 
\frac{i}{4} \epsilon^{\mu\nu\rho\sigma} \tilde{\theta}_\nu \theta_{\rho\sigma} 
-\frac{i}{2} \tilde{\theta}^\mu \tilde{\theta}.
\end{eqnarray}
Then the chiral conditions (\ref{eq:4ccs1}) can be transformed into 
\begin{align}
\frac{\partial}{\partial \theta} \ 
U \Psi(x^\mu,\{\theta_I\})  \ = \ 0, \ \ 
\frac{\partial}{\partial \theta^{\rho\sigma}} \ 
U \Psi(x^\mu,\{\theta_I\})  \ = \ 0, \ \ 
\frac{\partial}{\partial \tilde{\theta}} \ 
U \Psi(x^\mu,\{\theta_I\}) \ = \ 0, 
\label{scalar_condition_2}
\end{align}
which leads 
\begin{align}
U \Psi(x^\mu,\{\theta_I\})  \ = \ 
\Psi^{\prime} (x^\mu,\theta^\mu,\tilde{\theta}^\mu).
\label{solution_of_scalar_condition_2}
\end{align}
Then the solution for the original chiral condition 
(\ref{eq:4ccs1}) is obtained as 
\begin{align}
\Psi(x^\mu,\{\theta_I\})  \ = \ 
U^{-1} \Psi^{\prime} (x^\mu,\theta^\mu,\tilde{\theta}^\mu)  \ = \ 
\Psi^\prime (y^\mu,\theta^\mu,\tilde{\theta}^\mu), 
\label{solution_of_scalar_condition_3}
\end{align}
where 
\begin{equation}
y^\mu=x^\mu+a^\mu. 
\label{def_of_y}
\end{equation}
This solution can be expanded as follows:
\begin{eqnarray}
\Psi^\prime (y^\mu ,\theta^\mu , \tilde{\theta}^\mu) 
=\Psi^0(y^\mu,\tilde{\theta}^\mu) + 
\theta^m \Psi^1 _m(y^\mu,\tilde{\theta}^\mu) +\frac{1}{2} \theta^m
\theta^n\Psi^2 _{mn}(y^\mu,\tilde{\theta}^\mu) \nonumber \\
+(\theta^3)^m \Psi^3{}_{m}(y^\mu,\tilde{\theta}^\mu) 
+\theta^4 \Psi^4(y^\mu,\tilde{\theta}^\mu), 
\label{eq:4csf}
\end{eqnarray}
where $(\theta^3)^m=\frac{1}{3!} \theta^\mu
\theta^\nu\ \theta^\rho \epsilon_{\mu\nu\rho} {}^{m}$ and $\theta^4 =
\frac{1}{4!}\epsilon_{\mu\nu\rho\sigma}\theta^\mu\theta^\nu\ \theta^\rho
\theta^\sigma$ with 
$\{ \Psi^0,\ \Psi^1_m,\ \Psi^2_{mn},\ \Psi^3_m,\ \Psi^4 \}$ 
defined by 
\begin{eqnarray}
\Psi^0(y^\mu,\tilde{\theta}^\mu) &=&
\psi^0 + 
\tilde{\theta}^\mu \psi^0 _{,\mu} 
+\frac{1}{2} \tilde{\theta}^\mu \tilde{\theta}^\nu \psi^0 _{,\mu\nu} 
+(\tilde{\theta}^3)^\sigma \tilde{\psi^0} {}_{,\sigma} 
+\tilde{\theta}^4 \tilde{\psi^0}, \nonumber \\
\Psi^1 {}_{m}(y^\mu,\tilde{\theta}^\mu) &=&
\psi^1 _{m,}+ 
\tilde{\theta}^\mu \psi^1 _{m,\mu} 
+\frac{1}{2} \tilde{\theta}^\mu \tilde{\theta}^\nu \psi^1 _{m,\mu\nu} 
+(\tilde{\theta}^3)^\sigma \tilde{\psi^1} {}_{m,\sigma} 
+\tilde{\theta}^4 \tilde{\psi^1} _{m,}, \nonumber \\
\Psi^2 {}_{mn}(y^\mu,\tilde{\theta}^\mu) &=&
\psi^2 _{mn,}+ 
\tilde{\theta}^\mu \psi^2 _{mn,\mu} 
+\frac{1}{2} \tilde{\theta}^\mu \tilde{\theta}^\nu \psi^2 _{mn,\mu\nu} 
+(\tilde{\theta}^3)^\sigma \tilde{\psi^2} {}_{mn,\sigma} 
+ \tilde{\theta}^4 \tilde{\psi^2} _{mn,}, \nonumber \\
\Psi^3 {}_{m}(y^\mu,\tilde{\theta}^\mu) &=&
\psi^3_{m,} + 
\tilde{\theta}^\mu \psi^3 _{m,\mu} 
+\frac{1}{2} \tilde{\theta}^\mu \tilde{\theta}^\nu \psi^3 _{m,\mu\nu} 
+(\tilde{\theta}^3)^\sigma \tilde{\psi^3} {}_{m,\sigma} 
+ \tilde{\theta}^4 \tilde{\psi^3} _{m,}, \nonumber  \\
\Psi^4(y^\mu,\tilde{\theta}^\mu) &=&
\psi^4 + 
\tilde{\theta}^\mu \psi^4 _{,\mu} 
+\frac{1}{2} \tilde{\theta}^\mu \tilde{\theta}^\nu \psi^4 _{,\mu\nu} 
+(\tilde{\theta}^3)^\sigma \tilde{\psi^4} {}_{,\sigma} 
+ \tilde{\theta}^4 \tilde{\psi^4}, \nonumber 
\end{eqnarray}
where $(\tilde{\theta}^3)^\sigma=\frac{1}{3!} \tilde{\theta}^\mu
\tilde{\theta}^\nu\ \tilde{\theta}^\rho \epsilon_{\mu\nu\rho} {}^{\sigma}$ and 
$\tilde{\theta}^4 =
\frac{1}{4!}\epsilon_{\mu\nu\rho\sigma}\tilde{\theta}^\mu
\tilde{\theta}^\nu\ \tilde{\theta}^\rho\tilde{\theta}^\sigma$.

The relation given in (\ref{solution_of_scalar_condition_3}) can be understood 
also from the following relation:
\begin{eqnarray}
\Psi(x^\mu,\{ \theta \}) &=& e^{\delta_\theta} \psi^0(x) = e^{(\theta^\mu s_\mu+\tilde{\theta}^\mu \tilde{s}_\mu )}e^{(\theta s +\frac{1}{4} \theta^{\mu\nu} s_{\mu\nu} + \tilde{\theta} \tilde{s})} e^{a^\mu \partial_\mu} \psi^0(x)  \nonumber \\
&=& e^{(\theta^\mu s_\mu+\tilde{\theta}^\mu \tilde{s}_\mu )}e^{(\theta s +\frac{1}{4} \theta^{\mu\nu} s_{\mu\nu} + \tilde{\theta} \tilde{s})} \psi^0(y)  \nonumber \\
&=& e^{(\theta^\mu s_\mu+\tilde{\theta}^\mu \tilde{s}_\mu )} \psi^0(y)\nonumber \\
&=& \Psi^\prime (y^\mu ,\theta^\mu , \tilde{\theta}^\mu), 
\label{parameter_change1}
\end{eqnarray}
where we can recognize that the all the component fields can be identified as 
the fields obtained by operating the twisted supercharge $s_\mu$ and 
$\tilde{s}_\mu$ to the parent field $\psi^0(y)$.

Anti-chiral conditions for a superfield are given by 
\begin{eqnarray}
\mathfrak{D}_\mu \Phi(x^\mu,\{\theta_I\}) = \tilde{\mathfrak{D}}_\mu \Phi(x^\mu,\{\theta_I\}) 
= 0.
\label{eq:4accs}
\end{eqnarray}
Similar to the chiral condition (\ref{scalar_condition_2}), 
we can transform the original anti-chiral condition 
(\ref{eq:4accs}) into the following form: 
\begin{eqnarray}
U^{-1} \mathfrak{D}_\mu U 
U^{-1} \Phi(x^\mu,\{\theta_I\}) \ &=& \ 
\frac{\partial}{\partial \theta^\mu} 
U^{-1} \Phi(x^\mu,\{\theta_I\}) \ = \ 0,  \nonumber \\
U^{-1} \tilde{\mathfrak{D}}_\mu U 
U^{-1} \Phi(x^\mu,\{\theta_I\}) \ &=& \ 
\frac{\partial}{\partial \tilde{\theta}^\mu} 
U^{-1} \Phi(x^\mu,\{\theta_I\}) \ = \ 0,
\label{anti-chiral-condition_2}
\end{eqnarray}
which leads 
\begin{align}
U^{-1} \Phi(x^\mu,\{\theta_I\}) \ = \ 
\Phi^\prime (x^\mu,\theta,\theta^{\rho\sigma},\tilde{\theta}).
\label{solution_of_anti-chiral_condition_2}
\end{align}
Then the solution for the original anti-chiral condition 
(\ref{eq:4accs}) is obtained as 
\begin{align}
\Phi(x^\mu,\{\theta_I\}) \ = \ 
U \Phi^\prime (x^\mu,\theta,\theta^{\rho\sigma},\tilde{\theta})  \ = \ 
\Phi^\prime (w^\mu,\theta,\theta^{\rho\sigma},\tilde{\theta}), 
\label{solution_of_anti-chiral_condition_3}
\end{align}
where 
\begin{equation}
w^\mu=x^\mu-a^\mu.
\end{equation}
This solution can be expanded as follows:
\begin{eqnarray}
\Phi^\prime(w^\mu,\theta,\tilde{\theta},\theta^{\mu\nu})
=\Phi^0(w^\mu,\theta^{\mu\nu}) + 
\theta \Phi^1(w^\mu,\theta^{\mu\nu}) +
\tilde{\theta} \Phi^2(w^\mu,\theta^{\mu\nu}) + 
\theta \tilde{\theta} \Phi^3(w^\mu,\theta^{\mu\nu}), 
\nonumber \\
\label{N=4_anti-chiral_SF}
\end{eqnarray}
with 
\begin{eqnarray}
\Phi^m(w^\mu,\theta^{\mu\nu}) &=& \phi^m+\frac{1}{2}\theta^{ab}\phi^m _{ab}
+\frac{1}{2!2^2}\theta^{ab} \theta^{cd}\phi^m _{ab,bd}
+\frac{1}{3! 2^3}\theta^{ab} \theta^{cd} \theta^{ef}\phi^m _{ab,bd,ef} \nonumber  \\
& &+\frac{1}{4! 2^7}\theta^{ab} \theta^{cd} \theta^{ef} \theta^{gh}
\Gamma_{ab,cd,ef,gh,ij,kl} \tilde{\phi^m}^{ij,kl} \nonumber  \\
& &+\frac{1}{5! 2^6}\theta^{ab} \theta^{cd} \theta^{ef} \theta^{gh} \theta^{ij}
\Gamma_{ab,cd,ef,gh,ij,kl} \tilde{\phi^m}^{kl} \nonumber  \\
& &+\frac{1}{6! 2^6}\theta^{ab} \theta^{cd} \theta^{ef} \theta^{gh}
\theta^{ij} \theta^{kl}
\Gamma_{ab,cd,ef,gh,ij,kl} \tilde{\phi}^m,
\label{eq:4acsf}
\end{eqnarray}
where $(m=0,1,2,3)$, $\Gamma_{ab,cd,ef,gh,ij,kl}$ is 
like a six-dimensional totally anti-symmetric tensor. 
Independent degrees of freedom of the tensor is six because of the 
anti-symmetric nature of the suffix $\{ab\}$.
We give the definition of $\Gamma_{ab,cd,ef,gh,ij,kl}$ and useful relations 
in the Appendix B.  

Similar to the relation (\ref{parameter_change1}) the corresponding relation for 
anti-chiral field (\ref{solution_of_anti-chiral_condition_3}) can be understood 
from the following relation as well:
\begin{eqnarray}
\Phi(x^\mu,\{ \theta \}) &=& e^{\delta_\theta} \phi^0(x) = 
e^{(\theta s +\frac{1}{4} \theta^{\mu\nu} s_{\mu\nu} + \tilde{\theta} \tilde{s})} 
e^{(\theta^\mu s_\mu+\tilde{\theta}^\mu \tilde{s}_\mu )}e^{a^\mu \partial_\mu} \phi^0(x)  
\nonumber \\
&=& e^{(\theta s +\frac{1}{4} \theta^{\mu\nu} s_{\mu\nu} + \tilde{\theta} \tilde{s})} 
e^{(\theta^\mu s_\mu+\tilde{\theta}^\mu \tilde{s}_\mu )} \phi^0(w)  \nonumber \\
&=& e^{(\theta s+ \frac{1}{4}\theta^{\mu\nu} s_{\mu\nu} +
\tilde{\theta} \tilde{s} )} \phi^0(w) \nonumber \\
&=& \Phi^\prime (w^\mu,\theta,\theta^{\rho\sigma},\tilde{\theta}),
\end{eqnarray}  
where we can again recognize that the all the component fields can be identified 
as the fields obtained by operating the twisted supercharge $s,\tilde{s}$ and 
$s_{\mu\nu}$to the parent field $\phi^0(w)$.

The SUSY transformations of the chiral and anti-chiral 
superfields are given by 
\begin{align}
s_I \Psi(x^\mu,\{\theta_I\}) \ &= \ 
\mathcal{Q}_I \Psi(x^\mu,\{\theta_I\}), \nonumber \\  
s_I \Phi(x^\mu,\{\theta_I\}) \ &= \ 
\mathcal{Q}_I \Phi(x^\mu,\{\theta_I\}), 
\label{susy_trasformation_1}
\end{align}
where $\{s_I\} =\{s,s_\mu,s_{\mu\nu},\tilde{s}_\mu,\tilde{s}\}$ and 
$ \{\mathcal{Q}_I\} = \{\mathcal{Q}, \mathcal{Q}_\mu,
\mathcal{Q}_{\mu\nu},\tilde{\mathcal{Q}}_\mu, \tilde{\mathcal{Q}} \}$.

These SUSY transformation can be transformed into the 
following form by using the operator $U$: 
\begin{align}
s_I U \Psi(x^\mu,\{\theta_I\}) \ &= \ 
U\mathcal{Q}_IU^{-1} U\Psi(x^\mu,\{\theta_I\}),\nonumber \\  
s_I U^{-1}\Phi(x^\mu,\{\theta_I\}) \ &= \ 
U^{-1}\mathcal{Q}_IU U^{-1}\Phi(x^\mu,\{\theta_I\}), 
\label{susy_trasformation_2}
\end{align}
which can be equivalently written as 
\begin{align}
s_I \Psi^{\prime} (x^\mu,\theta^\mu , \tilde{\theta}^\mu) \ &= \ 
\mathcal{Q}^{\prime}_I \Psi^{\prime} (x^\mu,\theta^\mu , \tilde{\theta}^\mu) ,\nonumber \\  
s_I \Phi^\prime (x^\mu,\theta,\theta^{\rho\sigma},\tilde{\theta}) \ &= \ 
\mathcal{Q}^{\prime\prime}_I \Phi^\prime (x^\mu,\theta,\theta^{\rho\sigma},\tilde{\theta}), 
\label{susy_trasformation_3}
\end{align}
where $\mathcal{Q}^{\prime}_I=U\mathcal{Q}_IU^{-1}$ and 
$\mathcal{Q}^{\prime \prime}_I=U^{-1}\mathcal{Q}_IU$ 
are given by 
\begin{eqnarray}
\mathcal{Q}' &=& \frac{\partial}{\partial\theta} +i \theta^\mu \partial_\mu,\qquad   \tilde{\mathcal{Q}'} = \frac{\partial}{\partial\tilde{\theta}} + i\tilde{\theta}^\mu \partial_\mu ,\nonumber \\
\mathcal{Q}'_\mu &=& \frac{\partial}{\partial\theta^\mu},\hspace{24mm} \tilde{\mathcal{Q}'}_\mu  = \frac{\partial}{\partial \tilde{\theta}^\mu},  \nonumber \\
\mathcal{Q}'_{\rho\sigma} &=& \frac{\partial}{\partial\theta^{\rho\sigma}} -i (\theta_\rho \partial_\sigma - \theta_\sigma \partial_\rho)+ i \tilde{\theta}^\mu \epsilon_{\mu\nu\rho\sigma} \partial^\nu,
\label{s-charge-diff-op-prime}
\end{eqnarray}
\begin{eqnarray}
\mathcal{Q}'' &=& \frac{\partial}{\partial\theta},\qquad  \tilde{\mathcal{Q}}'' = \frac{\partial}{\partial\tilde{\theta}},\qquad \mathcal{Q}''_{\rho\sigma} = \frac{\partial}{\partial\theta^{\rho\sigma}}, \nonumber \\
\mathcal{Q}''_\mu &=& \frac{\partial}{\partial\theta^\mu} +i \theta \partial_\mu -i \theta_{\mu\nu} \partial^\nu, \nonumber \\
\tilde{\mathcal{Q}}_\mu''   &=& \frac{\partial}{\partial \tilde{\theta}^\mu}+\frac{i}{2} \theta^{\rho\sigma} \epsilon_{\rho\sigma\mu\nu} \partial^\nu +i \tilde{\theta} \partial^\mu.
\label{s-charge-diff-op-doubleprime}
\end{eqnarray}
The explicit form of the $N=4$ twisted SUSY transformation of the component fields 
for chiral and anti-chiral superfield are given in the Appendix C.  

\setcounter{equation}{0}
\section{Twisted SUSY invariant actions}
Based on the twisted superspace formalism given in the previous section and 
further introduction of twisted vector superfield we derive variety of SUSY 
invariant actions.

\subsection{Off-shell $N=4$ twisted SUSY invariant action}

We first show that a naive extension of the $N=D=2$ twisted SUSY invariant 
action to $N=4$ in four dimensions leads to an action with a lengthy 
expression. This is, however, a first nontrivial example of an action 
which has off-shell $N=4$ twisted SUSY invariance. 

The chiral and anti-chiral conditions for $N=2$ twisted superfield in two dimensions 
can, respectively, be given by the following conditions: 
\begin{align}
\mathfrak{D}\Psi(x^\mu,\theta,\tilde{\theta},\theta^\mu) \ = \ 0 , \ \ 
\tilde{\mathfrak{D}}\Psi(x^\mu,\theta,\tilde{\theta},\theta^\mu) \ = \ 0  
\label{scalar_condition_1}
\end{align}
and 
\begin{equation}
\mathfrak{D}_\mu\overline{\Psi}(x^\mu,\theta,\tilde{\theta},\theta^\mu) \ = \ 0.
\label{anti-chiral-condition_1}
\end{equation}
Parallel to the formulation given in the previous subsections, we can obtain fermionic 
chiral and anti-chiral superfields as
\begin{equation}
\begin{split}
\Psi(x^\mu,\theta,\tilde{\theta},\theta^\mu)  \ &= \ 
\Psi^\prime (y^\mu,\theta^\mu)
 =ie^{\theta^\mu s_\mu}c(y) 
 =ic(y)+\theta^\mu \omega_\mu(y) +i\theta^2 \lambda(y), \\
\overline{\Psi}(x^\mu,\theta,\tilde{\theta},\theta^\mu) \ &= \ 
\overline{\Psi}^\prime (w^\mu,\theta,\tilde{\theta})
  =ie^{\theta s + \tilde{\theta} \tilde{s}}\overline{c}(w)
  =i\overline{c}(w)+\theta b(w)
  +\tilde{\theta}\phi(w)
  -i\theta \tilde{\theta}\rho(w),
\label{expansion_psi_psibar}  
\end{split}
\end{equation}
where 
\begin{eqnarray}
y^\mu&=&x^\mu+\frac{i}{2}\theta\theta^\mu-\frac{i}{2}\epsilon^\mu{_\nu}\theta
^\nu\tilde{\theta}, \nonumber \\
w^\mu&=&x^\mu-\frac{i}{2}\theta\theta^\mu+
\frac{i}{2}\epsilon^\mu{_\nu}\theta^\nu\tilde{\theta}. 
\label{def_of_z_and_tilde_z}
\end{eqnarray}
Here the parent fields $c(y)$ and $\overline{c}(w)$ of chiral 
and anti-chiral superfields are Grassmann odd fields. 
The two-dimensional counterparts of the supercharge operators 
$\mathcal{Q}'$ in (\ref{s-charge-diff-op-prime}) and $\mathcal{Q}''$ in 
(\ref{s-charge-diff-op-doubleprime}) are given by 
\begin{align}
\mathcal{Q}^{\prime} \ &= \ \frac{\partial}{\partial \theta}
  + i \theta^\mu\partial_\mu, \ 
&\tilde{\mathcal{Q}}^{\prime}& \ = \ \frac{\partial}{\partial \tilde{\theta}}
  - i \theta^\mu\epsilon_{\mu\nu}\partial^\nu, \ 
&\mathcal{Q}^{\prime}_\mu& \ = \ \frac{\partial}{\partial \theta^\mu}, 
\label{susy_trasformation_4}\\
\mathcal{Q}^{\prime\prime} \ &= \ \frac{\partial}{\partial \theta}, \ 
&\tilde{\mathcal{Q}}^{\prime\prime}& \ = \ \frac{\partial}{\partial \tilde{\theta}}, \ 
&\mathcal{Q}^{\prime\prime}_\mu& \ = \ \frac{\partial}{\partial \theta^\mu} 
 + i \theta\partial_\mu - i \tilde{\theta}\epsilon_{\mu\nu}\partial^\nu. 
\label{susy_trasformation_5}
\end{align}
We can then obtain $N=2$ twisted SUSY transformation of the component fields of 
the (anti-)chiral field which we show in Table 1.
\begin{table}[htbp]
\[
\begin{array}{|c||c|c|c|}
\hline
& s &  s_{\mu} &  \tilde{s}  \\
\hline
c & 0 & -i\omega_{\mu} & 0 \\
\omega_{\nu} & \partial_{\nu}c  & -i\epsilon_{\mu\nu}\lambda &
-\epsilon_{\nu\rho} \partial^{\rho} c \\
\lambda & \epsilon^{\mu\nu}\partial_{\mu} \omega_{\nu} & 0
& -\partial^{\mu}\omega_{\mu} \\
\hline
\bar{c} & -ib & 0 & -i\phi \\
b & 0 & \partial_{\mu} \bar{c} & -i\rho \\
\phi & i\rho & -\epsilon_{\mu\nu} \partial^\nu \bar{c} & 0 \\
\rho & 0 & -\partial_{\mu} \phi - \epsilon_{\mu\nu} \partial^{\nu} b& 0 \\
\hline
\end{array}
\]
\label{SUSY-transformation1}
\caption{$N=2$ twisted SUSY transformation.}
\end{table}
The details of the twisted chiral superfield formulation for $N=D=2$ can be 
found in \cite{KKU,DKKN}.

Then the bi-linear product of the twisted chiral and anti-chiral superfields 
leads to an off-shell $N=2$ twisted SUSY invariant action
\begin{align}
S_f&= \int d^2x \int d^4\theta \  
 ( i \overline{\Psi} \Psi )
 = \int d^2x \int d^4\theta \ e^{\delta_\theta}(-i\overline{c}c)
 \nonumber \\
 &=\int d^2x \ 
s\tilde{s}\frac{1}{2}\epsilon^{\mu\nu}s_\mu s_\nu (-i\bar{c}c)\nonumber\\
 &=\int d^2x
\Bigl(
\phi\epsilon^{\mu\nu}\partial_\mu\omega_\nu
+b \partial^\mu\omega_\mu
+i\partial_\mu\overline{c}\partial^\mu c
+i\rho\lambda
\Bigr),
\label{eq:WZ 3}
\end{align}
where $\delta_\theta$ is defined in (\ref{def_delta_theta}).
This action can be identified as the quantized BF action 
with auxiliary fields and has off-shell $N=2$ twisted SUSY invariance 
up to the surface terms by construction. 

As we have seen from the formulation of Dirac-K\"ahler twist, the two-dimensional 
and four-dimensional formulation of the twisted superspace have close similarity. 
It is then natural to expect that we can obtain the four-dimensional counterpart 
of the action (\ref{eq:WZ 3}) which has off-shell twisted $N=4$ SYSY invariance. 
The bi-linear product of the 
chiral superfield $\Psi$ in (\ref{eq:4csf}) and the anti-chiral superfield 
$\Phi$ in (\ref{N=4_anti-chiral_SF}) leads to the following action: 
\begin{eqnarray}
S &=& \int d^4 x d^{16} \theta\  \Phi(x)\ \Psi(x) \\
&=& \int d^4 x \{  -\tilde{\phi}^3 \tilde{\psi}^4 \nonumber \\
& &-i \tilde{\phi}^2 \partial^\mu \tilde{\psi}^3 { }_{\mu,}
 +i\tilde{\phi}^1 \partial^a \tilde{\psi}^4 { }_{,a}
 +\frac{i}{2}\mathcal{D}_{\sigma,ab}\tilde{\phi}^{3,ab}\tilde{\psi}^{3,\sigma}
 -\frac{i}{2} \epsilon_{abcd} \partial^d \tilde{\phi}^{3,bc}\tilde{\psi}^{4,a} \nonumber \\
& &+\frac{1}{4}\mathcal{D}_{\sigma,ab}\tilde{\phi}^{2,ab}\epsilon^{\sigma \alpha \beta \gamma } \partial_\alpha \tilde{\psi}^2 {}_{\beta\gamma}
+\frac{1}{2}\mathcal{D}_{\sigma,ab}\tilde{\phi}^{1,ab}\partial_c\tilde{\psi}^{3,\sigma,c} \nonumber \\
& &-\frac{1}{4^2}\epsilon^{\mu\nu\rho\sigma} \mathcal{D}_{\mu,ab} \mathcal{D}_{\nu,cd}\tilde{\phi}^{3,ab,cd}\tilde{\psi}^2 {}_{\rho\sigma}
-\frac{1}{2} \epsilon_{abcd} \partial^d \tilde{\phi}^{2,bc}\partial^\delta \tilde{\psi}^3 {}_{\delta,a}\nonumber \\
& &+\frac{1}{4} \epsilon_{abcd} \partial^d \tilde{\phi}^{1,bc}\epsilon^{a\alpha\beta\gamma}\partial_\alpha \psi^{4} { }_{,\beta\gamma}+\frac{1}{4}\epsilon_{abcd}\partial^d \mathcal{D}_{\mu,ef} \tilde{\phi}^{3,bc,ef}\tilde{\psi}^{3,\mu,a}\nonumber \\
& &+\tilde{\phi}^0 \partial^\mu \partial^a \tilde{\psi}^3 {}_{\mu,a} +\frac{1}{4}\epsilon_{abcd}\partial^d \partial^f \tilde{\phi}^{3,ea,bc}\psi^4 {}_{,ef} + \cdots \} ,
\label{eq:44action}
\end{eqnarray}
where $\Phi$ and $ \Psi$ are Grassmann odd superfields. Here we display 
the action up to the second order of derivative. We show the complete expression of 
the action in the Appendix D. As we can see, the full expression of the action is 
very lengthy and includes up to the 8-th order of derivatives. In two-dimensional 
case the action has an interpretation that it is the quantized version of 
two-dimensional BF theory\cite{KKU}. This four-dimensional action does not have an obvious 
correspondence with a known action. It is, however, interesting to recognize that 
this is the first example of the action which is derived from $N=4$ twisted 
superfields and has the exact off-shell $N=4$ 
twisted SUSY in four dimensions. To keep the exact off-shell $N=4$ 
twisted SUSY we need 8-th order derivative terms, which is very non-trivial 
and can never be found unless we have twisted superspace formulation.

\subsection{$N=2$ decomposition of $N=4$ twisted SUSY algebra 
in four dimensions} 

In this subsection we introduce a (anti-)self-dual decomposition of the $N=4$ twisted 
SUSY into two $N=2$ twisted SUSY sectors. We decompose twisted 
supercharges as follows:
\[
\left\{
	\begin{array}{l}
	s^{\pm} \equiv \frac{1}{\sqrt{2}} (s \pm \tilde{s}),\\
        s^{\pm}_\mu \equiv \frac{1}{\sqrt{2}} (s_\mu \pm \tilde{s}_\mu),\\
        s^{\pm}_{\mu\nu}\equiv  \frac{1}{\sqrt{2}}(s_{\mu\nu} \mp \frac{1}{2}\epsilon_{\mu\nu\rho\sigma}s^{\rho\sigma}),
	\end{array}
\right.
\]
where the second rank tensor twisted supercharge satisfies (anti-)self-dual condition  
\begin{eqnarray}
\frac{1}{2}\epsilon_{\mu\nu\rho\sigma}s^{\pm\rho\sigma}=\mp s^{\pm} _{\mu\nu}.
\end{eqnarray}
We may identify this decomposition as self-dual and anti-self-dual 
decomposition. 
These supercharges satisfy the following two $N=2$ factorized algebras:
\begin{eqnarray}
\{ s^\pm,s^\pm _\mu\} &=& -i\partial_\mu, \nonumber \\
\{ s^\pm _{\mu\nu} , s^\pm _\rho \}&=& i(\delta_{\mu\nu , 
\rho\sigma}\partial^\sigma \mp \epsilon_{\mu\nu\rho\sigma} \partial^\sigma) \nonumber \\
&\equiv& i\mathcal{D}^\pm _{\rho,\mu\nu},\nonumber  \\
\mbox{\{others\}} &=& 0. 
\label{eq:4al+-}
\end{eqnarray}
The supercharges $\{s^+_I\}\equiv \{s^+,s_\mu^+,s_{\mu\nu}^+\}$ are the dual 
partner of $\{s^-_I\}\equiv \{s^-,s_\mu^-,s_{\mu\nu}^-\}$ and anticommute with 
each other. It is interesting to note that the algebras of self-dual and 
anti-self-dual supercharges close 
by themselves and construct twisted $N=2$ superalgebra in four dimensions. In other words 
the $N=4$ superalgebra generated by Dirac-K\"ahler twist can be decomposed into 
two dual pairs of $N=2$ twisted superalgebras which we use to formulate $N=2$ 
twisted superspace formalism in the later subsection 5.2. 
Corresponding to the factorization of the supercharges, we can decompose the 
superparameters in the similar way by the following relation:
\begin{eqnarray}
\delta_\theta &\equiv&\theta s +\theta^\mu s_\mu+ 
\frac{1}{2}\theta^{\mu\nu} s_{\mu\nu} + \tilde{\theta}^{\mu}\tilde{s}_\mu+\tilde{\theta} 
\tilde{s}  \nonumber \\
&\equiv& \delta_{\theta^+} + \delta_{\theta^-},
\end{eqnarray}
where
\begin{eqnarray}
\delta_{\theta^+} &\equiv& \theta^+s^+ +\theta^{+\mu} s^+ _{\mu} +
\frac{1}{4}\theta^{+\mu\nu} s^+ _{\mu\nu}, \nonumber \\
\delta_{\theta^-}&\equiv& \theta^- s^-+\theta^{-\mu} s^- _{\mu}  + 
\frac{1}{4}\theta^{-\mu\nu} s^- _{\mu\nu}, 
\end{eqnarray}
with 
\[
\left\{
	\begin{array}{l}
	\theta^{\pm} \equiv \frac{1}{\sqrt{2}} (\theta \pm \tilde{\theta}),\\
        \theta^{\pm}_\mu \equiv \frac{1}{\sqrt{2}} (\theta_\mu \pm \tilde{\theta}_\mu),\\
        \theta^{\pm}_{\mu\nu}\equiv  \frac{1}{\sqrt{2}}(\theta_{\mu\nu} \mp \frac{1}{2}\epsilon_{\mu\nu\rho\sigma}\theta^{\rho\sigma}).
	\end{array}
\right.
\]
By using the above dual decomposed superparameters we can find supercharge 
differential operators: 
\begin{eqnarray}
\mathcal{Q}^\pm &=& \frac{\partial}{\partial \theta^\pm}+
\frac{i}{2} \theta^{\pm\mu}\partial_\mu, \nonumber \\
\mathcal{Q}^\pm_\mu &=& \frac{\partial}{\partial \theta^{\pm \mu}}+
\frac{i}{2} \theta^{\pm} \partial_\mu -
\frac{i}{2} \theta^{\pm} { }_{\mu\nu}\partial^\nu, \nonumber \\
\mathcal{Q}^\pm _{\mu\nu} &=& \frac{\partial}{\partial \theta^{\pm \mu\nu}}-
\frac{i}{2}\theta^{\pm \rho}\mathcal{D}^{\pm} _{\rho,\mu\nu},  
\label{t-s-c-d-operator}
\end{eqnarray}
where we introduce the following notations: 
\begin{eqnarray}
& &\mathcal{D}^{\pm} _{\rho,\mu\nu} = \delta^{\pm}_{\mu\nu,\rho\sigma} 
\partial^\sigma, \\
& &\delta^\pm {}_{\mu\nu,\rho\sigma} \equiv 
\frac{\partial}{\partial \theta^{\pm \mu\nu}}\theta^{\pm}_{\rho\sigma} = 
\delta_{\mu\rho} \delta_{\nu\sigma} -
\delta_{\mu\sigma} \delta_{\nu\rho} \mp \epsilon_{\mu\nu\rho\sigma}. \nonumber
\end{eqnarray}
These $N=2$ supercharge differential operators satisfy the following relations:
\begin{eqnarray}
\{ \mathcal{Q}^\pm,\mathcal{Q}^\pm _\mu\} &=& i\partial_\mu, \nonumber \\
\{ \mathcal{Q}^\pm _{\mu\nu} , \mathcal{Q}^\pm _\rho \}&=& -
i\mathcal{D}^\pm _{\rho,\mu\nu},\nonumber  \\
\text{\{others\}} &=& 0, 
\label{t-s-c-d-op-algebra}
\end{eqnarray}
where the sign of the spacetime derivative is reversed with respect to the algebra 
(\ref{eq:4al+-}).
Similar to the $N=4$ full twisted algebra, we introduce superdifferential operators 
which anticommute with all the supercharge differential operators 
$\{\mathcal{Q}^\pm_I\}$,
\begin{eqnarray}
\mathfrak{D}^\pm &=& \frac{\partial}{\partial \theta^\pm}-\frac{i}{2} \theta^{\pm\mu}\partial_\mu, \nonumber \\
\mathfrak{D}^\pm_\mu &=& \frac{\partial}{\partial \theta^{\pm \mu}}-\frac{i}{2} \theta^{\pm} \partial_\mu +\frac{i}{2} \theta^{\pm} { }_{\mu\nu}\partial^\nu, \nonumber \\
\mathfrak{D}^\pm _{\mu\nu} &=& \frac{\partial}{\partial \theta^{\pm \mu\nu}}+\frac{i}{2}\theta^{\pm \rho}\mathcal{D}^{\pm} _{\rho,\mu\nu},
\label{def-chirally-decomposed-superderivatives}
\end{eqnarray}
which satisfy the same algebra as (\ref{eq:4al+-}) with replacements: 
$s^{\pm}_I \rightarrow \mathfrak{D}^\pm_I$.

We now consider only the positive part of supercharges 
$\{s^+_I\}$, the corresponding supercharge differential operators 
$\{\mathcal{Q}^+_I\}$ and superdifferential operators 
$\{\mathfrak{D}^+_I\}$, which fulfill $N=2$ twisted SUSY algebra 
in four dimensions. 
We can impose chiral and anti-chiral conditions by using these 
four-dimensional $N=2$ positive superdifferential operators: 
\begin{eqnarray}
\mathfrak{D}^+ \Phi &=& \mathfrak{D}^+ _{\mu\nu} \Phi=0, \label{eq:42chi1} \\
\mathfrak{D}^+ _\mu \Psi&=& 0 \label{eq:42achi2} .
\end{eqnarray} 
The following combination of the coordinates satisfy the chiral condition 
(\ref{eq:42chi1}): 
\[
 \left\{
\begin{array}{l}
y^\mu = x^\mu +\frac{i}{2}\theta^+\theta^{+\mu}-\frac{i}{2}\theta^+
_{\nu} \theta^{+\mu\nu}, \\
\theta^{+\mu}.
\end{array}
 \right.
\] 
Thus the general solution of chiral superfield $\Phi$ can be given by
\begin{eqnarray}
\Phi(y^\mu,\theta^{+\mu}) =e^{\theta^{+\mu}s^+_\mu}\phi =
\phi + \theta^{+\mu} C_\mu +\frac{1}{2}\theta^{+\mu}\theta^{+\nu}\phi_{\mu\nu}+ \theta^{3+} _{\mu}\tilde{\psi}^\mu +\theta^{4+} \tilde{\phi}. \label{eq:42csf}
\end{eqnarray}
where $ \theta^{3+} _{\mu} \equiv
\frac{1}{3!}\epsilon_{\mu\nu\rho\sigma}\theta^{+\nu}\theta^{+\rho}\theta^{+\sigma}, \quad
\theta^{4+}\equiv
\frac{1}{4!}\epsilon_{\mu\nu\rho\sigma}\theta^{+\mu}\theta^{+\nu}\theta^{+\rho}\theta^{+\sigma}
$.
We show the twisted $N=2$ SUSY transformation of the Abelian chiral multiplets 
of the superfield in Table 2.
\begin{table}[htpb]
\[
 \begin{array}{|c||c|c|c|}
\hline
 & s^+ &s^+ _\mu & s^+ _A \\
\hline
\phi & 0 & C_\mu & 0 \\ 
C_{\nu} & -i\partial_\nu \phi & -\phi_{\mu\nu} & i\mathcal{D}_{\nu,A} \phi \\
\phi_{\rho\sigma} & i(\partial_\rho C_\sigma-\partial_\sigma C_\rho ) & -\epsilon_{\mu\rho\sigma\nu} \tilde{\psi}^\nu & -i( \mathcal{D}_{\rho,A} C_\sigma - \mathcal{D}_{\sigma,A} C_\rho) \\
\tilde{\psi}_\nu & -\frac{i}{2} \epsilon_{\nu\rho\sigma\mu}\partial^\mu \phi^{\rho\sigma} & -\delta_{\mu\nu} \tilde{\phi} & \frac{i}{2}\mathcal{D}_{\mu,A}\epsilon_{\nu} { }^{\mu\rho\sigma}\phi_{\rho\sigma} \\
\tilde{\phi} & i\partial^\mu \tilde{\psi}_\mu & 0 & -i\mathcal{D}_{\mu,A}\tilde{\psi}^\mu \\
\hline
 \end{array}
\]
\label{4d-N2-chiral-ab-transfs1}
\caption{Twisted $N=2$ SUSY transformation of Abelian chiral multiplets.}
\end{table}
We can now define the following action:
\begin{eqnarray}
S &=& \int d^4y d^4 \theta\ (\Phi(y,\theta^{+\mu}))^2 \nonumber \\
&=&\int d^4y \frac{1}{4!} \epsilon^{\mu\nu\rho\sigma}
s^+_\mu s^+_\nu s^+_\rho s^+_\sigma\  (\phi^2) \nonumber \\
&=& 2\int   d^4x \{\phi \tilde{\phi}- C_\mu \tilde{\psi}^\mu 
+\frac{1}{8}\epsilon^{\mu\nu\rho\sigma}\phi_{\mu\nu}\phi_{\rho\sigma} \}, 
\label{4d-ab-N2-action-chiral1}
\end{eqnarray}
where the coordinate shift 
$y^\mu \rightarrow x^\mu$ is allowed since this action includes only 
chiral superfield which has only $y^\mu$ dependence. 
Due to this possible coordinate shift the action does not include any 
derivative terms. 
It should be compared with the action of previous subsection which has a 
bilinear form of chiral and anti-chiral superfields and each of the superfield 
has different chiral coordinate dependence, which is the origin of 
generating derivatives. 

Next we consider the anti-chiral condition (\ref{eq:42achi2}). We find that 
the following combination of coordinate satisfies the condition: 
\[
 \left\{
\begin{array}{l}
w^\mu = x^\mu -\frac{i}{2}\theta^+\theta^{+\mu} +\frac{i}{2}\theta^+
_{\nu} \theta^{+\mu\nu}, \\
\theta^+ ,\quad \theta^{+A}.
\end{array}
 \right.
\]
Then the general solution for anti-chiral superfield can be given by
\begin{eqnarray}
\Psi(w,\theta^+,\theta^{+A})&=& 
e^{\theta^+s^++\frac{1}{4}\theta^{\mu\nu}s^+_{\mu\nu}} \overline{\phi} \nonumber \\
&=&\overline{\phi} + \frac{1}{4}\theta^{+A}\chi^+ _A + 
\frac{1}{4}(\theta^{2+})^{A}\tilde{M}^+ _A +(\theta^{3+})\tilde{\chi} 
\nonumber  \\
& &+\theta^+ (\chi + \frac{1}{4}\theta^{+A}N^+ _A + 
\frac{1}{4}(\theta^{2+})^{A}\tilde{\chi}^+ _A +(\theta^{3+})L ), 
\label{def-antichiral-Psi}
\end{eqnarray}
where $(\theta^{2+})_{A} \equiv \frac{1}{2\cdot4^2}\Gamma^+
_{ABC}\theta^{+B}\theta^{+C}$ and $ (\theta^{3+}) \equiv  
\frac{1}{3!\cdot4^3}\Gamma^+
_{ABC}\theta^{+A}\theta^{+B}\theta^{+C} $. 
Here the suffixes $A,B,C$
denote tensor suffix, for example $\phi^+ _A \psi^{+A} $ stands for  
$\phi^+ _{\mu\nu} \psi^{+\mu\nu}$. $\Gamma^{\pm ABC} $ is defined as follows: 
\begin{eqnarray}
\Gamma^{\pm \mu \alpha , \nu \beta , \rho \gamma}
&=& \delta^{\alpha \nu} \delta^{\beta \rho} \delta^{\gamma \mu} 
+ \delta^{\mu \nu} \delta^{\beta \gamma } \delta^{\rho \alpha} 
+ \delta^{\alpha \beta} \delta^{\nu \gamma} \delta^{\rho \mu} 
+ \delta^{\mu \beta} \delta^{\nu \rho} \delta^{\gamma \alpha} \nonumber \\
& &-( \delta^{\alpha \nu} \delta^{\beta \gamma} \delta^{\rho \mu} 
+ \delta^{\mu \nu} \delta^{\beta \rho } \delta^{\gamma \alpha} 
+ \delta^{\alpha \beta} \delta^{\nu \rho} \delta^{\gamma \mu} 
+ \delta^{\mu \beta} \delta^{\nu \gamma } \delta^{\rho \alpha} ) \nonumber \\
& & \mp \varepsilon^{\mu \alpha \beta \gamma } \delta^{\nu \rho}
\mp \varepsilon^{\mu \alpha \nu \rho } \delta^{\beta \gamma}
\pm \varepsilon^{\mu \alpha \nu \gamma } \delta^{\beta \rho}
\pm \varepsilon^{\mu \alpha \beta \rho } \delta^{\nu \gamma}. 
\label{def-Gamma-pm}
\end{eqnarray}
We summarize details of tensor notations and useful formulae in Appendix E.
We show twisted $N=2$ SUSY transformation of Abelian anti-chiral multiplets 
in Table \ref{tb:42actrans}. 

\begin{table}[htbp]
\[
 \begin{array}{|c||c|c|c|}
\hline
 & s^+ &s^+ _\mu & s^+ _A \\
\hline
\overline{\phi} & \chi & 0 & \chi^+ _A \\ 
\chi^+_{B} & -N^+ _B & i\mathcal{D}_{\mu,B}\overline{\phi} & -\frac{1}{4}\Gamma^+ _{ABC} M^{+C} \\
M^+ _B & \tilde{\chi}^+ _B  &-\frac{i}{4}\Gamma^+ _{B\mu\nu C} \partial^\nu \chi^{+C}    & \delta^+ _{AB}\tilde{\chi} \\
\tilde{\chi} & -L & i\partial^\nu M^+_{\mu\nu} & 0  \\
\hline
\chi &0& -i\partial_\mu \overline{\phi} & N^+ _A \\
N^+ _B & 0 & i(\partial_\mu \chi^+ _B + \mathcal{D}^+ _{\mu,B} \chi ) & -\frac{1}{4}\Gamma^+ _{ABC}\tilde{\chi}^{+C} \\ 
\tilde{\chi}^+ _B&0 &-i(\partial_\mu M^+ _B +\frac{i}{4}\Gamma^+ _{B\mu\nu C}\partial^\nu N^{+C}) & \delta^+ _{AB} L\\
L &0& i(\partial_\mu \tilde{\chi}+ \partial^\nu \tilde{\chi}^+ _{\mu\nu})&0 \\
\hline
 \end{array}
\]
\caption{Twisted $N=2$ SUSY transformation of Abelian anti-chiral multiplets}
\label{tb:42actrans}
\end{table}
A twisted $N=2$ SUSY invariant action can be also obtained by anti-chiral 
superfield as 
\begin{eqnarray}
S &=& \int d^4w d^4 \theta (\Psi(w,\theta^+,\theta^+ _A))^2 \nonumber \\
&=&
\int d^4 w \ s^+\frac{1}{3!4^3}\Gamma^{+A,B,C} s^+_A s^+_B s^+_C 
(\overline{\phi}^2)
\nonumber \\
&=&\int d^4x \{\overline{\phi}L + \frac{1}{4}\chi^+ _A \tilde{\chi}^{+A} + \frac{1}{4}M^+ _A N^{+A} + \tilde{\chi}\chi \},
\label{4d-ab-N2-action-anti-chiral1}
\end{eqnarray}
where we make the coordinate shift $w^\mu \rightarrow x^\mu$ again. 
By the same reason as the previous action there is no derivative in 
the action. The action has $N=2$ twisted SUSY invariance where the 
supertransformation of the anti-chiral superfield is given in Table 3.
In the following subsection we introduce superconnection formalism where 
several constraints will be introduced. The constraints relate the chiral 
and anti-chiral superfields and generate derivatives and eventually 
super Yang-Mills actions can be generated.

\subsection{Vector superfield and gauge symmetry 
for $N=4$ twisted SUSY invariant actions} 
In this subsection we propose to formulate a twisted version of vector superfield 
to derive supergauge invariant actions. 
Using the operators $\{\mathfrak{D}^\pm_I\}$ defined in 
(\ref{def-chirally-decomposed-superderivatives}), we may impose two types of 
chiral conditions. The first type is given as follows: 
\begin{eqnarray}
\mathfrak{D}^+ \Psi &=& \mathfrak{D}^- \Psi = \mathfrak{D}^+ _{\mu\nu}\Psi = 
\mathfrak{D}^- _{\mu\nu} \Psi =0, \label{eq:44cons1}\\
\mathfrak{D}^+ _\mu \Phi &=& \mathfrak{D}^- _\mu \Phi=0, \label{eq:44cons2}
\end{eqnarray}
which are essentially equivalent to the previous chiral conditions 
(\ref{eq:4ccs1}) and (\ref{eq:4accs}), respectively. 
Here we propose to impose other type of chiral conditions
\begin{eqnarray}
\mathfrak{D}^+ \Phi^+ &=& \mathfrak{D}^+ _{\mu\nu} \Phi^+ = 
\mathfrak{D}^- _\mu \Phi^+ =0, \label{eq:44cons3} \\
\mathfrak{D}^- \Phi^- &=& \mathfrak{D}^- _{\mu\nu} \Phi^- = 
\mathfrak{D}^+ _\mu \Phi^- =0 \label{eq:44cons4}, 
\end{eqnarray}
where we find that these conditions are mutually transformed into each other 
by the replacement: $+ \leftrightarrow -$. 
We may loosely abuse the words, chiral and anti-chiral in this section, 
instead of self-dual and anti-self-dual of the previous subsection, and call 
(\ref{eq:44cons3}) as a chiral condition for a chiral superfield $\Phi^+$ 
and (\ref{eq:44cons4}) as an anti-chiral condition for an anti-chiral 
superfield $\Phi^-$.

The simplest way to find a general solution of the chiral condition 
(\ref{eq:44cons3}) is to recognize that the following parameters are solution of 
the chiral condition:
\[
\left\{
	\begin{array}{l}
	y^{+-\mu}=x^\mu + a^{+\mu} - a^{-\mu}, \\
        \theta^- ,\quad \theta^- _{\mu\nu} ,\quad \theta^+ _\mu, \\
	\end{array}
\right.
\]
where 
\begin{equation}
a^{\pm\mu} = \frac{i}{2}\theta^{\pm}\theta^{\pm\mu} +
\frac{i}{2}\theta^{\pm} _{\rho}\theta^{\pm\rho\mu}.
\end{equation}
Therefore a general solution of the chiral condition (\ref{eq:44cons3}) is given by 
an arbitrary function of these parameters:
\begin{eqnarray}
\Phi^+ = \Phi^+ (y^{+-\mu},\theta^- , \theta^- _{\mu\nu}, \theta^+ _\mu).
\end{eqnarray}
Similarly we find that the following parameters satisfy the anti-chiral condition 
(\ref{eq:44cons4}),
\[
\left\{
	\begin{array}{l}
	y^{-+\mu}=x^\mu - a^{+\mu} + a^{-\mu}, \\
        \theta^+ ,\quad \theta^+ _{\mu\nu} ,\quad \theta^- _\mu. \\
	\end{array}
\right.
\]
Thus a general solution can be given by
\begin{eqnarray}
\Phi^- = \Phi^- (y^{-+\mu},\theta^+ , \theta^+ _{\mu\nu}, \theta^- _\mu).
\end{eqnarray}
Supertransformation of the component fields of these chiral superfields 
can be obtained by the similar procedure as in the subsection 3.3. We, 
however, don't bother to write them explicitly here.   

Let us now introduce an operator $P$ which interchanges the chirality of the 
superparameters and coordinate, $+ \leftrightarrow -$: 
\begin{eqnarray}
P\{y^{\pm\mp\mu},\theta^{\mp},\theta^{\mp} _{\mu\nu},\theta^{\pm} _{\mu}\} = 
\{y^{\mp\pm\mu},\theta^{\pm},\theta^{\pm} _{\mu\nu},\theta^{\mp} _\mu\}. 
\end{eqnarray}

We define a chiral and anti-chiral superfields; $\Lambda^+ $ and $ \Lambda^-$, 
which satisfy the following conditions:
\begin{eqnarray}
\mathfrak{D}^+ \Lambda^+ &=& \mathfrak{D}^+ _{\mu\nu} \Lambda^+ = 
\mathfrak{D}^- _\mu \Lambda^+ =0, 
\label{chiral_cond_lambda}\\
\mathfrak{D}^- \Lambda^- &=& \mathfrak{D}^- _{\mu\nu} \Lambda^- = 
\mathfrak{D}^+ _\mu \Lambda^- =0 \label{anti-chiral_cond_lambda}.
\end{eqnarray}
A chiral pair of superfields $\Lambda^+ $ and $ \Lambda^-$ which 
satisfy these chirally conjugate conditions can be related by the following 
conjugate relation:
\begin{eqnarray}
P\Lambda^{\pm}(y^{\pm\mp\mu},\theta^{\mp},\theta^{\mp} _{\mu\nu},\theta^{\pm} _{\mu}) 
= \Lambda^{\mp}(y^{\mp\pm\mu},\theta^{\pm},\theta^{\pm} _{\mu\nu},\theta^{\mp} _\mu).
\end{eqnarray}
These chiral and anti-chiral superfields can be expanded into component 
fields as follows:
\begin{eqnarray}
\Lambda^\pm(y^{\pm\mp\mu},\theta^{\mp},\theta^{\mp} _{\mu\nu},\theta^{\pm} _{\mu}) 
&=& \lambda_0^\pm(x) + \theta^{\pm\mu} \lambda_{1\mu}^\pm(x) 
+\frac{1}{4}\theta^{\mp A}\lambda_{1A}^\pm(x)
+ \theta^\mp \lambda_1^\pm(x) \nonumber \\
& & +\frac{1}{2} \theta^{\pm\mu}\theta^{\pm\nu}\lambda_{2\mu\nu}^\pm(x) 
+\frac{1}{4} \theta^{\mp A}\theta^{\pm\mu}\lambda_{2A,\mu}^\pm(x) 
+\frac{1}{4}\theta^{\mp A}\theta^{\mp B} \lambda_{2AB}^\pm(x) \nonumber \\
& & +\theta^\mp\theta^{\pm\mu}\lambda_{2\mu}^\pm(x) 
+\frac{1}{4} \theta^\mp\theta^{\mp A}\lambda_{2A}^\pm(x) \nonumber \\
& & +\frac{i}{2}\theta^\pm\theta^{\pm\mu}\partial_\mu\lambda_0^\pm(x)  
+\frac{i}{2}\theta^\pm_\rho\theta^{\pm\rho\mu}\partial_\mu\lambda_0^\pm(x) 
-\frac{i}{2}\theta^\mp\theta^{\mp\mu}\partial_\mu\lambda_0^\pm(x) \nonumber \\ 
& & -\frac{i}{2}\theta^\mp_\rho\theta^{\mp\rho\mu}\partial_\mu\lambda_0^\pm(x) 
+\cdots,
\nonumber
\end{eqnarray}
where we have expanded up to the second order of twisted superparameters.

We define the vector superfield which satisfies the following condition:
\begin{eqnarray}
PV(x^\mu , \theta^\pm, \theta^\pm _{\mu}, \theta^{\pm} _{\mu\nu}) = 
V(x^\mu , \theta^\pm, \theta^\pm _{\mu}, \theta^{\pm} _{\mu\nu}).
\end{eqnarray}
Here we can introduce the following chiral supergauge transformation for the vector 
superfield: 
\begin{eqnarray}
e^V &\to& e^{-\Lambda^-}e^V e^{-\Lambda^+} \nonumber \\
e^{-V} &\to& e^{\Lambda^+}e^{-V} e^{\Lambda^-},
\end{eqnarray}
where we may consider non-Abelian vector superfield and thus the two chiral 
gauge superparameters $\Lambda^+, \Lambda^-$ may carry non-Abelian nature 
and should satisfy the chiral and anti-chiral conditions 
(\ref{chiral_cond_lambda}) and (\ref{anti-chiral_cond_lambda}), respectively.
Then the small supergauge transformation can be given by 
\begin{equation}
\delta V = -\Lambda^+ - \Lambda^- - \frac{1}{2} \ [V,\Lambda^+ - \Lambda^-]
\label{small_super_gauge_tr}+ \cdots.
\end{equation}
The vector superfield can be expanded into component fields as follows:
\begin{eqnarray}
V = A_0(x) &+& \theta^{+\mu}A_{1\mu}^+ + \frac{1}{4}\theta^{-A} A_{1A}^- + \theta^-A_1^- 
\nonumber \\  
&+& \theta^{-\mu}A_{1\mu}^- + \frac{1}{4}\theta^{+A} A_{1A}^+ + \theta^+A_1^+  
\nonumber \\
&+& \theta^+\theta^{+\mu} A_{2\mu}^+ + \theta^-\theta^{-\mu} A_{2\mu}^- 
\nonumber \\
&+& \theta_\rho^+\theta^{+\rho\mu}B_{2\mu}^+ + 
\theta_\rho^-\theta^{-\rho\mu}B_{2\mu}^- \cdots,  
\end{eqnarray}
where we keep only some of the component fields in the second order superparameter 
terms. 

As we can see from the small supergauge transformation (\ref{small_super_gauge_tr}) 
we can gauge away the component fields 
$(A_0, A_{1\mu}^\pm, A_{1A}^\pm, A_1^\pm, \cdots)$ by the corresponding gauge 
parameters 
$(-\lambda_0^+-\lambda_0^-, \lambda_{1\mu}^\pm, \lambda_{1A}^\mp, \lambda_1^\mp, 
\cdots)$, respectively. 
In fact we can recognize that all the component fields of the vector superfield, 
which are coefficients of any combination of the product 
$(\theta^-,\theta^-_{\mu\nu},\theta^+_\mu)$ or 
$(\theta^+,\theta^+_{\mu\nu},\theta^-_\mu)$, can be gauged away. 
We identify this gauging away procedure as Wess-Zumino 
gauge choice. 
Then we can find gauge fields $A_{2\mu}^\pm$ and $B_{2\mu}^\pm$ which have the 
following gauge transformation: 
\begin{eqnarray}
\delta A_{2\mu}^\pm(x) &=& -\frac{i}{2}\partial_\mu(\lambda_0^\pm(x) - \lambda_0^\mp(x)) 
- \frac{i}{2}[A_{2\mu}^\pm(x),\lambda_0^\pm(x) - \lambda_0^\mp(x)], \nonumber \\
\delta B_{2\mu}^\pm(x) &=& -\frac{i}{2}\partial_\mu(\lambda_0^\pm(x) - \lambda_0^\mp(x)) 
- \frac{i}{2}[B_{2\mu}^\pm(x),\lambda_0^\pm(x) - \lambda_0^\mp(x)]. 
\label{super_gauge_transformation_2}
\end{eqnarray}
It is interesting to note that there is only one chiral pair of gauge parameters 
$-\frac{i}{2}(\lambda_0^\pm(x) - \lambda_0^\mp(x))$ for two chiral pairs 
of gauge fields $A_{2\mu}^\pm$ and $B_{2\mu}^\pm$.  

Next we consider the gauge invariant quantity which satisfies the chiral 
constraint (\ref{eq:44cons3}). We first define the following differential operator:
\begin{eqnarray}
\mathfrak{D}^{8+} \equiv \mathfrak{D}^+ \frac{1}{3!4^3} \Gamma^{+ABC} \mathfrak{D}^+ _A \mathfrak{D}^+ _B \mathfrak{D}^+ _C \frac{1}{4!} \epsilon^{\mu\nu\rho\sigma} \mathfrak{D}^- _\mu \mathfrak{D}^- _\nu \mathfrak{D}^- _\rho \mathfrak{D}^- _\sigma,
\end{eqnarray}
where the suffixes $A,B,C$ denote tensor suffixes similar as in 
(\ref{def-antichiral-Psi}) and $\Gamma^{+ ABC}$ is defined in (\ref{def-Gamma-pm}). 
Since the chiral and anti-chiral supergauge parameters $\Lambda^+$ and 
$\Lambda^-$ satisfy the chiral and anti-chiral conditions, they satisfy 
the following relations:
\begin{eqnarray}
e^{\mp\Lambda^\pm}(\mathfrak{D}^\pm, \mathfrak{D}^\pm _{\mu\nu}, 
\mathfrak{D}^\mp _\mu) &=& 
(\mathfrak{D}^\pm, \mathfrak{D}^\pm_{\mu\nu}, \mathfrak{D}^\mp_\mu) 
e^{\mp\Lambda^\pm}.
\end{eqnarray}

When the differential operator $\mathfrak{D}^{8+}$ is applied to a general
superfield $\Phi$, $\mathfrak{D}^{8+}\Phi$ satisfies the chiral condition
(\ref{eq:44cons3}) up to possible total divergence terms. It is then possible 
to find all the possible chiral invariant combinations of the 
differential operators 
$\{\mathfrak{D}^-, \mathfrak{D}^- _A , \mathfrak{D}^+ _\mu\}$ acting on a 
vector superfield $e^V$:
\begin{eqnarray}
W^{1-} = \mathfrak{D}^{8+} e^{-V}\mathfrak{D}^- e^{V}, \ \ \ 
W^{1-}_A = \mathfrak{D}^{8+} e^{-V}\mathfrak{D}^-_A e^{V}, \ \ \ 
W^{1+}_\mu = \mathfrak{D}^{8+} e^{-V}\mathfrak{D}^+_\mu e^{V} ,\nonumber
\end{eqnarray}
\begin{eqnarray}
W^{2-} _A = \mathfrak{D}^{8+} e^{-V}(\mathfrak{D}^{2-})_A  e^{V}, \ \ \
W^{2+} _{\mu\nu} = \mathfrak{D}^{8+} e^{-V}(\mathfrak{D}^{2+})_{\mu\nu} e^{V}, \ \ \
W^{2-}_\mu = \mathfrak{D}^{8+} e^{-V}\mathfrak{D}^- \mathfrak{D}^- _A e^{V} ,
\nonumber
\end{eqnarray}
\begin{eqnarray}
W^{3-}  = \mathfrak{D}^{8+} e^{-V}(\mathfrak{D}^{3-}) e^{V}, \ \ \ 
W^{3+}_\mu = \mathfrak{D}^{8+} e^{-V}(\mathfrak{D}^{3+})_\mu e^{V}, \ \ \ 
W^{3-}_A = \mathfrak{D}^{8+} e^{-V}\mathfrak{D}^- (\mathfrak{D}^{2-})_A e^{V} ,
\nonumber
\end{eqnarray}
\begin{eqnarray}
W^{4-}  = \mathfrak{D}^{8+} e^{-V}\mathfrak{D}^- (\mathfrak{D}^{3-})e^{V}, \ \ \ 
\tilde{W}^{4+} = \mathfrak{D}^{8+} e^{-V}(\mathfrak{D}^{4+}) e^{V} ,
\end{eqnarray}
where 
\begin{eqnarray}
(\mathfrak{D}^{2-})_A &=& \frac{1}{2!\cdot 4^3} \Gamma^-
_{ABC} \mathfrak{D}^{-B} \mathfrak{D}^{-C}, \ \ \ 
(\mathfrak{D}^{2+})_{\mu\nu}=
\frac{1}{2}\epsilon_{\mu\nu\rho\sigma}\mathfrak{D}^{+\rho}
\mathfrak{D}^{+\sigma},  \nonumber \\ 
(\mathfrak{D}^{3-}) &=& \frac{1}{3!\cdot 4^2} \Gamma^-
_{ABC} \mathfrak{D}^{-A} \mathfrak{D}^{-B} \mathfrak{D}^{-C}, \ \ \  
(\mathfrak{D}^{3+})_\mu
=\frac{1}{3!}\epsilon_{\mu\nu\rho\sigma}\mathfrak{D}^{+\nu} \mathfrak{D}^{+\rho} 
\mathfrak{D}^{+\sigma}, \nonumber \\
(\mathfrak{D}^{4+})
&=&\frac{1}{4!}\epsilon_{\mu\nu\rho\sigma} \mathfrak{D}^{+\mu} \mathfrak{D}^{+\nu} 
\mathfrak{D}^{+\rho} \mathfrak{D}^{+\sigma}.
\end{eqnarray}
These \lq\lq supercurvature" terms $\{W_I\}$ transform adjointly under a supergauge transformation and satisfy the chiral condition (\ref{eq:44cons3}). 
Opposite chiral combinations of the \lq\lq supercurvature" terms can be obtained 
simply by the interchange of the chirality for the above terms: $+ \leftrightarrow -$.

We can then formally construct a variety of supergauge 
invariant actions which have off-shell $N=4$ twisted SUSY and non-Abelian 
gauge symmetry; 
\begin{eqnarray}
S_1 &=& \int d^4x \ d^{8+}\theta \ \mbox{Tr}(W^{1-} W^{1-}), 
\label{action1}\\
S_2 &=& \int d^4x \ d^{8+}\theta \ \mbox{Tr}(W^{4-} \tilde{W}^{4+}),
\nonumber \\
    & & \cdots.
\label{susy-gauge-iv-action}
\end{eqnarray}
where 
\begin{equation}
d^{8+}\theta = d\theta^-\frac{1}{3!\cdot 4^3} \Gamma^-
_{ABC} d\theta^{-A} d\theta^{-B} d\theta^{-C} 
\frac{1}{4!}\epsilon_{\mu\nu\rho\sigma} d\theta^{+\mu} d\theta^{+\nu} 
d\theta^{+\rho} d\theta^{+\sigma}.
\end{equation}
It should be noted that the chiral superfields $\{W_I\}$ given above are function 
of ($y^{+-\mu},\theta^-,\theta^-_{A},\theta^+_\mu$) and thus the coordinate change 
$y^{+-\mu} \rightarrow x^\mu$ is allowed since the action includes only the 
chiral superfields and thus $d^{8+}\theta$ include only chiral superparameters. 

These actions are highly reducible in the sense that they may include too many 
superfluous component fields even if we take the Wess-Zumino gauge. 
As we have shown in the previous subsection 4.1 and in the Appendix D, the $N=4$ 
twisted supersymmetric action constructed from chiral and anti-chiral pair of 
superfields have included too many terms. We expect that these gauge invariant 
actions given above include too many fields as well when they are 
expressed by the component fields. 

It is, however, not unreasonable to expect that these actions may include 
twisted $N=4$ super Yang-Mills terms in the action since they include gauge fields as 
we have shown in (\ref{super_gauge_transformation_2}). 
If we try to find how the suspected gauge fields, $A_{2\mu}^\pm$ and $B_{2\mu}^\pm$ 
are included in the action, we find too many spacetime derivatives. 
For example in the Abelian gauge we can consider how the gauge field 
$A_{2\mu}^+$ survives after the operation of $\mathfrak{D}^{8+} \mathfrak{D}^-$ 
and before the integration $d\theta^{8+}$ in 
\begin{eqnarray}
W^{1-} &=& \mathfrak{D}^{8+} \mathfrak{D}^- V 
= \mathfrak{D}^{8+} \mathfrak{D}^- (\cdots + \theta^+\theta^{+\mu} A_{2\mu}^+ + \cdots) 
\nonumber \\
&=&-\frac{1}{3!\cdot 4^3 4!}(\frac{i}{2})^7 \Gamma^{-ABC} \theta^{+\alpha} 
\theta^{+\beta} \theta^{+\gamma} \mathcal{D}^{+} _{\alpha,A}\mathcal{D}^{+} _{\beta,B}
\mathcal{D}^{+} _{\gamma,C}  
\nonumber \\ 
& & \epsilon^{\mu\nu\rho\sigma} 
(\theta^-\partial_\nu - \theta^-_{\nu\nu'}\partial^{\nu'} )
(\theta^-\partial_\rho - \theta^-_{\rho\rho'}\partial^{\rho'} )
(\theta^-\partial_\sigma - \theta^-_{\sigma\sigma'}\partial^{\sigma'} ) 
\theta^{+\tau} \partial_\mu A^+_{2\tau} + \cdots. 
\nonumber
\end{eqnarray}
As we can see that this term includes 7-th power of spacetime derivative. 
Other component gauge fields include higher derivatives as well. 
Thus the action (\ref{action1}) and (\ref{susy-gauge-iv-action}) include the 
gauge terms but 
higher derivatives are accompanied and then Yang-Mills term is missing. 
We may further need to impose constraints in addition to the chiral 
conditions to find super Yang-Mills actions in the vector superfield formulation. 
In the next section we introduce superconnection to the corresponding superderivatives and impose constraints to find super Yang-Mills action.

\setcounter{equation}{0}
\section{Superconnection formalism}

\subsection{$N=2$ superconnection formalism in two dimensions}

In the previous subsection we have given a general formulation to obtain 
off-shell $N=4$ twisted SUSY invariant and gauge invariant actions by introducing 
vector superfield. The supergauge invariance was introduced by the transformation 
of the vector superfield. It turned out that the proposed supergauge 
invariant actions include too higher derivatives and thus we expect 
that it is necessary to impose further constraints to find twisted 
$N=4$ SUSY invariant Yang-Mills action. 
Here we propose an alternative approach, twisted superconnection formalism, 
to find supergauge invariant actions. A similar 
type of formulation was proposed by Labastida and his collaborators 
by the spinor notations \cite{LL,AL}. Here the formulation is based on the 
Dirac-K\"ahler twist to generate twisted superspace. 
Since the four-dimensional twisted $N=4$ SUSY invariant formulation is 
expected to be obtained quite parallel to the two-dimensional $N=2$ version of 
the formulation, we first show the simpler twisted $N=D=2$ formulation of 
superconnection formalism. 

We first introduce fermionic superconnection $\{\Gamma_I\}$ which allows us to 
define the following gauge superspace covariant derivatives: 
\begin{eqnarray}
 \nabla &\equiv& \mathfrak{D}-i\Gamma, \nonumber \\
\tilde{\nabla}&\equiv& \tilde{\mathfrak{D}}-i\tilde{\Gamma}, \nonumber \\
\nabla_\mu &\equiv& \mathfrak{D}_\mu-i\Gamma_\mu, 
\label{def_2d-super-covariant-derivative}
\end{eqnarray}
where $\mathfrak{D},\tilde{\mathfrak{D}}$ and $\mathfrak{D}_\mu$ are 
two-dimensional superderivatives defined in (\ref{def_2d-super-derivative}) 
and satisfy the $N=2$ twisted SUSY algebra (\ref{2d-super-derivative-algebara}). 
We further introduce the following bosonic covariant derivative: 
\begin{equation}
\nabla_{\underline{\mu}} \equiv \partial_{\underline{\mu}}-i\Gamma_{\underline{\mu}},
\end{equation} 
where the lowest $\theta$ independent component of $\Gamma_{\underline{\mu}}$ 
is identified as the usual gauge field: 
$\Gamma_{\underline{\mu}}|_{\theta=0}=\omega_\mu$. We call them as 
supercovariant derivatives. It should be noted that these superconnections
$\{\Gamma_I\}$ are superfields. 

Corresponding to the superconnections $\{\Gamma_I\}$, we can introduce the following 
supergauge transformation: 
\begin{eqnarray}
\delta_{gauge} \Gamma_I = \nabla_I K \equiv \mathfrak{D}_I K - i [\Gamma_I,K],
\end{eqnarray}
where $K$ is an arbitrary superfield. Since the superconnections include many 
superfluous component fields, we gauge away them by the supergauge 
freedom of the superfield $K$, by taking Wess-Zumino gauge, and keep the usual 
gauge degrees of freedom. 

Even after the Wess-Zumino gauge fixing, we still have many component fields. 
In the superconnection formalism we impose reasonable constraints to reduce the 
unnecessary component fields in the superfields. 
In general the anticommutators of those supercovariant derivatives 
$\{\nabla_I\}= \{\nabla,\tilde{\nabla}, \nabla_\mu\} $ have the following form: 
\begin{eqnarray}
\{\nabla_I ,\nabla_J \} = T_{IJ} {}^{\underline{\mu}} \nabla_{\underline{\mu}} 
-i \mathcal{F}_{IJ},  
\end{eqnarray}
where we identify $\mathcal{F}_{IJ}$ as supercurvatures and 
$T_{IJ} {}^{\underline{\mu}}$ as 
superspace torsions with a possible bosonic covariant derivative 
$\nabla_{\underline{\mu}}$. 
We also introduce fermionic supercurvature 
\begin{eqnarray}
[\nabla_I ,\nabla_{\underline{\nu}} ] =  -i \mathcal{F}_{I{\underline{\nu}}},  
\end{eqnarray}
and a bosonic curvature which includes usual curvature term:
\begin{eqnarray}
[\nabla_{\underline{\mu}} ,\nabla_{\underline{\nu}} ] =  
-i \mathcal{F}_{{\underline{\mu}}{\underline{\nu}}},  
\end{eqnarray} 
where torsion terms are suppressed. 
The lowest ($\theta$-independent) component of 
$\mathcal{F}_{{\underline{\mu}}{\underline{\nu}}}$ coincides with the usual 
curvature: 
\begin{equation}
\mathcal{F}_{{\underline{\mu}}{\underline{\nu}}}|_{\theta=0} \equiv 
F_{\mu\nu} = \partial_\mu \omega_\nu- \partial_\nu \omega_\mu 
-i[\omega_\mu,\omega_\nu].
\end{equation}
We summarize the commutators and anticommutators of these supercovariant 
derivatives in Table \ref{def-supercurvature} where torsion terms are 
supecified as in the Table.
\begin{table}[htbp]
\[
\begin{array}{|c|c|c|c||c|}
\hline
& \nabla &  \tilde{\nabla} & \nabla_\nu  & \nabla_{\underline{\nu}}   \\
\hline
\nabla & -i\mathcal{W}_1 & -i\mathcal{W}_3 & -i\nabla_{\underline{\nu}} -i \mathcal{F}_\nu & -i\mathcal{F}_{\underline{\nu}}\\ 
\tilde{\nabla} && -i\mathcal{W}_2 &i\epsilon_{\nu\rho} \nabla^{\underline{\rho}}-i \tilde{\mathcal{F}}_\nu &-i\tilde{\mathcal{F}}_{\underline{\nu}} \\
\nabla_\mu &&&-i\mathcal{F}_{\mu\nu} &-i\mathcal{F}_{\mu\underline{\nu}} \\
\hline 
\hline
\nabla_{\underline{\mu}} &&&& -i\mathcal{F}_{\underline{\mu}\underline{\nu}}\\
\hline
\end{array}
\]
\caption{Definition of supercurvature.}
\label{def-supercurvature}
\end{table}

In order to suppress superfluous component fields we propose to impose the following 
constraints on the supercurvatures:
\begin{eqnarray}
\mathcal{W}_1=\mathcal{W}_2 \equiv \mathcal{W} \ ,\qquad \mathcal{W}_3 =0 \ ,\qquad  \mathcal{F}_{\mu\nu} \equiv \delta_{\mu\nu} F ,\qquad \mathcal{F}_\mu= \tilde{\mathcal{F}}_\mu =0.
\label{eq:curvature con}
\end{eqnarray}
We eventually find out that these constraints are necessary conditions to derive 
twisted $N=2$ SUSY invariant Yang-Mills actions and can be interpreted as 
suppression conditions of higher spin fields from the superspace. 
In the Appendix F we explicitly show in the details how the component fields 
of the superconnection can be gauged away, by taking Wess-Zumino gauge, 
to be consistent with these constraints. 

We can then obtain series of nontrivial relations via Jacobi identity 
of these supercovariant derivatives. 
For example, 
\begin{eqnarray}
& &{[} \nabla, \{\tilde{\nabla},\nabla_\mu  \} {]}+{[}\tilde{\nabla}, 
\{\nabla,\nabla_\mu  \}  {]}+ {[} \nabla_\mu, \{\nabla,\tilde{\nabla} \} {]}=0, 
\nonumber \\
&\to &{[}\nabla ,i\epsilon_{\mu\nu} \nabla^{\underline{\nu}} {]}+{[}\tilde{\nabla},
-i\nabla_{\underline{\mu}} {]}+{[}\nabla_\mu, 0 {]}=0,\nonumber \\
&\to& \epsilon_{\mu\nu} \mathcal{F}^{\underline{\nu}}
-\tilde{\mathcal{F}}_{\underline{\mu}}=0.
\end{eqnarray}
We can similarly obtain other identities, 
\begin{eqnarray}
& &\nabla \mathcal{W} \equiv \mathfrak{D}\mathcal{W} -i[\Gamma,\mathcal{W}]= 0, 
\label{eqn:g1}\\
& &\tilde{\nabla}\mathcal{W}  \equiv \tilde{\mathfrak{D}}\mathcal{W}
-i[\tilde{\Gamma},\mathcal{W}]= 0 ,\label{eqn:g2}\\
& &\nabla_\mu \mathcal{F} \equiv \mathfrak{D}_\mu \mathcal{F} 
-i [\Gamma_\mu,\mathcal{F}]=0 ,\label{eq:f} \\
& &\mathcal{F}_{\underline{\mu}} = -\frac{i}{2} \nabla_\mu \mathcal{W}, \\
& &\tilde{\mathcal{F}}_{\underline{\mu}} = -\frac{i}{2}\epsilon_{\mu\nu} 
\nabla^\nu \mathcal{W},\\
& &\mathcal{F}_{\mu \underline{\nu}} =  -\frac{i}{2} \delta_{\mu\nu} \nabla \mathcal{F} 
+\frac{i}{2}\epsilon_{\mu\nu} \tilde{\nabla} \mathcal{F}, \\ 
& &\mathcal{F}_{\underline{\mu} \underline{\nu}}=
-\frac{1}{2} \epsilon_{\mu\nu}\nabla\tilde{\nabla}\mathcal{F} 
+\frac{i}{4}\epsilon_{\mu\nu}\epsilon^{\rho\sigma}
\nabla_\rho \nabla_\sigma \mathcal{W},
\label{eq.f5}
\end{eqnarray}
where we assume that the gauge algebra is non-Abelian. 

If the gauge algebra is Abelian the first three conditions 
(\ref{eqn:g1}),(\ref{eqn:g2}) and (\ref{eq:f}) are simplified and 
lead to chiral and anti-chiral conditions,
\begin{eqnarray}
& &\mathfrak{D}\mathcal{W}=\tilde{\mathfrak{D}}\mathcal{W} =0, \label{eq:chiral}\\
& &\mathfrak{D}_\mu \mathcal{F} =0. \label{eq:anti-chiral}
\end{eqnarray}
Following to the same procedure given in subsection 4.1, we find the 
following component expansion of chiral and anti-chiral superfields: 
\begin{eqnarray}
\mathcal{W}(y) &=& A + \theta^\mu \lambda_\mu +\theta^2 D, \\
\mathcal{F}(w) &=& B+\theta \rho +\tilde{\theta}\tilde{\rho}+ \theta \tilde{\theta} E.
\end{eqnarray}
The lowest ($\theta=0$) component of the constraint (\ref{eq.f5}) in the 
Abelian case leads to the following relation: 
\begin{eqnarray}
-D+E=\epsilon^{\mu\nu}F_{\mu\nu},
\label{DEF-relation}
\end{eqnarray}
where $F_{\mu\nu} \equiv \partial_\mu\omega_\nu-\partial_\nu\omega_\mu$.
Since the explicit form of the chiral and anti-chiral superfields is given, 
the $N=2$ twisted SUSY transformation of the component fields can be obtained 
by the same procedure of subsection 4.1. 

In the case of non-Abelian gauge group the constraints (\ref{eqn:g1})
(\ref{eqn:g2}) and (\ref{eq:f}) are not (anti-)chiral condition anymore, 
we cannot obtain the component field expansion of (anti-)chiral superfield 
easily like in the Abelian case. We thus proceed differently to obtain the 
component field expansion of (anti-)chiral superfield. 
We first derive twisted SUSY transformation of the all component fields 
of the (anti-)chiral superfield by using the constraints obtained from the 
Jacobi identities.   

We first identify the lowest component of the superfields as follows:
\begin{eqnarray}
& & \mathcal{W}|=A , \qquad \nabla_\mu \mathcal{W}| = 
\lambda_\mu, \qquad \frac{1}{2} \epsilon^{\mu\nu}\nabla_\mu \nabla_\nu 
\mathcal{W}| = -D, \\
& & \mathcal{F}| = B , \qquad  \nabla \mathcal{F}| = 
\rho , \qquad  \tilde{\nabla}\mathcal{F}| = \tilde{\rho} , 
\qquad \nabla \tilde{\nabla}\mathcal{F}| = -E,
\end{eqnarray}
where $|=|_{\theta=0}$. 

For example the SUSY transformation of the field $A$ can be derived by the 
constraint (\ref{eqn:g1}),
\begin{eqnarray}
sA=s\mathcal{W}|=\nabla \mathcal{W}| =[\nabla,i\{\nabla,\nabla\}]|=0,
\end{eqnarray}
where we can replace $s$ with $\nabla$ since the $\theta$-independent terms 
coincide due to the choice of Wess-Zumino gauge.  
We show another example in the following:
\begin{eqnarray}
s\rho &=& s\nabla \mathcal{F} | = \nabla \nabla \mathcal{F}| \nonumber \\
&=& \nabla (\mathfrak{D}\mathcal{F} -i[\Gamma,\mathcal{F}])| \nonumber \\
&=& \mathfrak{D}(\mathfrak{D}\mathcal{F} -i[\Gamma,\mathcal{F}])|
-i\{\Gamma,\mathfrak{D}\mathcal{F} -i[\Gamma,\mathcal{F}] \}|  \nonumber \\
&=& -i[\mathfrak{D}\Gamma , \mathcal{F}]| 
-\{\Gamma ,[\Gamma, \mathcal{F}] \}| \nonumber \\
&=& -i[\mathfrak{D}\Gamma-i \Gamma^2 , \mathcal{F}]|,
\end{eqnarray}
where
\begin{eqnarray}
-i\mathcal{W} =\{\nabla , \nabla\} =\{\mathfrak{D} -i\Gamma,\mathfrak{D}-i\Gamma\}=-2i\mathfrak{D}\Gamma-2\Gamma^2 \nonumber.
\end{eqnarray}
We then find
\begin{eqnarray}
s\rho &=& -\frac{i}{2}[\mathcal{W},\mathcal{F}] | = -\frac{i}{2} [A,B].
\end{eqnarray}
Similarly we can find twisted SUSY transformation of all component fields, which we show in Table \ref{tb:2dabelian}, where we introduce the notation of usual 
covariant derivative: $D_\mu=\partial_\mu -i[\omega_\mu,\cdots]$.
\begin{table}[htbp]
\[
\begin{array}{|c||c|c|c|}
\hline
& s &  s_{\mu} &  \tilde{s}  \\
\hline
A & 0 & \lambda_{\mu} & 0 \\
\lambda_{\nu} & -iD_{\nu}A  & -\epsilon_{\mu\nu} D+\frac{i}{2}\delta_{\mu\nu}[A,B] &
i\epsilon_{\nu\rho} D^{\rho}A \\
D & i\epsilon^{\mu\nu} D_{\mu} \lambda_{\nu}+\frac{i}{2}[A,\tilde{\rho}] & 
\frac{i}{2}[B,\epsilon_{\mu\nu}\lambda^\nu] & -iD^{\mu}\lambda_{\mu} 
-\frac{i}{2}[A,\rho]\\
\hline
B & \rho & 0 & \tilde{\rho} \\
\rho & -\frac{i}{2}[A,B] & -iD_{\mu}B  & E  \\
\tilde{\rho} & -E & i\epsilon_{\mu\nu}D^{\nu}B & -\frac{i}{2}[A,B]\\
E & \frac{i}{2}[A,\tilde{\rho}] & iD_\mu \tilde{\rho} +i\epsilon_{\mu\nu}D^\nu\rho 
+\frac{i}{2}[B,\epsilon_{\mu\nu}\lambda^\nu]&-\frac{i}{2}[A,\rho] \\
\hline
G & i\epsilon^{\mu\nu} D_{\mu} \lambda_{\nu}+ i[A,\tilde{\rho}] & 
iD_\mu \tilde{\rho} +i\epsilon_{\mu\nu}D^\nu\rho +i[B,\epsilon_{\mu\nu}\lambda^\nu] &
-iD^{\mu}\lambda_{\mu} - i[A,\rho] \\
\omega_\nu &- \frac{i}{2}\lambda_\nu & \frac{i}{2}(\epsilon_{\mu\nu}\tilde{\rho}
- \delta_{\mu\nu} \rho) &  -\frac{i}{2}\epsilon_{\nu\rho}\lambda^\rho \\
\hline
\end{array}
\]
\caption{Twisted $N=2$ SUSY transformation of the component fields of 
non-Abelian super Yang-Mills theory 
with $D_\mu=\partial_\mu -i[\omega_\mu,\cdots]$.}
\label{tb:2dabelian}
\end{table}

The SUSY transformation of the gauge field can be derived as follows.  
The lowest components of the constraint (\ref{eq.f5}), 
\begin{eqnarray}
\mathcal{F}_{\underline{\mu}\underline{\nu}}|=
-\frac{1}{2} \epsilon_{\mu\nu}\nabla\tilde{\nabla}\mathcal{F}| 
+\frac{i}{4}\epsilon_{\mu\nu}\epsilon^{\rho\sigma}\nabla_\rho 
\nabla_\sigma \mathcal{W} |,
\end{eqnarray}
leads 
\begin{eqnarray}
F_{\mu\nu} &=& \frac{1}{2}\epsilon_{\mu\nu}(E-D),
\label{curvature-E-D}
\end{eqnarray}
where $F_{\mu\nu} \equiv \partial_\mu \omega_\nu- \partial_\nu \omega_\mu 
-i[\omega_\mu,\omega_\nu]$. 
Here we define a new field 
\begin{eqnarray}
G = E+D,
\label{def-G}
\end{eqnarray}
and then the degrees of freedom of $E$ and $D$ can be replaced by 
$\epsilon^{\mu\nu}F_{\mu\nu}$ and $G$. 

Operating $s$ charge on (\ref{curvature-E-D}), we obtain the following 
relation:
\begin{eqnarray}
& &sE-sD =-i\epsilon^{\mu\nu} D_\mu \lambda_\nu \nonumber \\
&=& 2(\epsilon^{\mu\nu}\partial_\mu s \omega_\nu 
-i \epsilon^{\mu\nu}[s\omega_\mu,\omega_\nu])=
2\epsilon^{\mu\nu}D_\mu s\omega_\nu,
\label{SUSYtr-ex2}
\end{eqnarray}
where we have used the $s$ transformation of $E$ and $D$ from the Table 5. 
Thus the $s$ transformation law of the gauge filed can be identified up to 
the gauge transformation as follows: 
\begin{eqnarray}
s\omega_\mu =- \frac{i}{2}\lambda_\mu.
\end{eqnarray}
We can derive other twisted supercharge transformations of the gauge field 
in the similar way. 
It should be understood in the Table 5 that the transformation of the fields 
$E$ and $D$ can be replaced by that of $G$ and the gauge field $\omega_\mu$ 
due to the relations (\ref{curvature-E-D}) and (\ref{def-G}). 

As we can see in the identification of SUSY transformation in 
(\ref{SUSYtr-ex2}), the twisted SUSY algebra is expected to be valid up to 
some gauge transformations. We can confirm that the twisted SUSY algebra 
is in fact valid up to the gauge transformation with the following specific 
choice of gauge parameters: 
\begin{eqnarray}
\{s,s\} \varphi &=& \delta_{\text{gauge}\ (-A)} \varphi, \nonumber \\
\{s_\mu, s_\nu\} \varphi &=&\delta_{\mu\nu} \delta_{\text{gauge}\ (-B)} \varphi, \nonumber \\
\{ \tilde{s},\tilde{s}\} \varphi &=& \delta_{\text{gauge}\ (-A)} \varphi, \nonumber \\
\{s,s_\mu\}\varphi &=& -i\partial_\mu \varphi+  \delta_{\text{gauge}\ (i\omega_\mu)} \varphi , \nonumber \\
\{\tilde{s},s_\mu\}\varphi&=& i\epsilon_{\mu\rho}\partial^\rho \varphi
+\delta_{\text{gauge}\ (-i\epsilon_{\mu\rho}\omega^\rho)} \varphi, \nonumber \\
\{s,\tilde{s}\}&=& 0,
\end{eqnarray} 
where $\delta_{\text{gauge}\ (\epsilon)}$ denotes gauge transformation of
parameter $\epsilon$ for any component fields 
$\{\varphi\}=\{A,\lambda_\mu,D,B,\rho,\tilde{\rho}\,E, \omega_\mu\}$.

Since we now know all the twisted $N=2$ SUSY transformation of the 
component fields of the superfields $\mathcal{F}$ and $\mathcal{W}$, 
we can reconstruct the component fields expansion of these superfields. 
Since $A$ and $B$ are parent fields of $\mathcal{F}$ and $\mathcal{W}$, 
respectively, we may define as 
\begin{eqnarray}
\mathcal{W} &=&  e^{\delta_\theta } A, \\
\mathcal{F} &=&  e^{\delta_\theta } B.
\end{eqnarray}
Explicit forms of  $\mathcal{W}$ and $\mathcal{F}$ are 
\begin{eqnarray}
\mathcal{W}(x) &=& A +\theta \theta^\mu \frac{i}{2} D_\mu A + \theta^\mu \tilde{\theta}  \frac{i}{2}\epsilon_{\mu\nu} D^\nu A + \theta^4 \frac{1}{4} D^\mu D_\mu A\nonumber \\
&+& \theta^\mu \lambda_\mu +\theta\theta^2 \frac{i}{2}\epsilon^{\mu\nu} D_\mu\lambda_\nu -\theta^2 \tilde{\theta}\frac{i}{2}D^\mu \lambda_\mu \nonumber \\
&+& \theta^2 D \nonumber \\
&+& \theta\theta^2 \frac{i}{3} [A,\tilde{\rho}] -\theta^2 \tilde{\theta} \frac{i}{3} [A,\rho] + \theta\theta^\mu \tilde{\theta} \frac{i}{6}[A,\epsilon_{\mu\nu}\lambda^\nu] \nonumber \\
&-&\theta^4 \frac{i}{6}\{\lambda^\mu, \lambda_\mu \} +\theta^4 \frac{1}{8}[A,[A,B]],
\end{eqnarray}
\begin{eqnarray}
\mathcal{F}(x) &=& B -\theta \theta^\mu \frac{i}{2} D_\mu B - \theta^\mu \tilde{\theta}  \frac{i}{2}\epsilon_{\mu\nu} D^\nu B + \theta^4 \frac{1}{4} D^\mu D_\mu B \nonumber \\
&+&\theta \rho - \theta\theta^\mu \tilde{\theta}\frac{i}{2}\epsilon_{\mu\nu}D^\nu\rho \nonumber \\
&+&\tilde{\theta} \tilde{\rho} - \theta\theta^\mu \tilde{\theta}\frac{i}{2}D_\mu\tilde{\rho} \nonumber \\
&+&\theta\tilde{\theta }(F+D) \nonumber \\
&+&\theta\theta^2 \frac{i}{6} [\tilde{\rho},B]- \theta^2 \tilde{\theta}\frac{i}{6}[\rho,B] +\theta \theta^\mu \tilde{\theta}\frac{i}{3} [\epsilon_{\mu\nu}\lambda^\nu ,B] \nonumber \\
&+&\theta^4 \{\frac{1}{8}[B,[B,A]] -\frac{i}{6}\{\rho,\rho\} -\frac{i}{6}\{\tilde{\rho},\tilde{\rho} \} \}.
\end{eqnarray}
Using the superfields $\mathcal{F}$ and $\mathcal{W}$, we can construct 
off-shell $N=2$ twisted SUSY invariant action with non-Abelian gauge symmetry: 
\begin{eqnarray}
S&=& \int d^2x d^4\theta \ \text{Tr} \mathcal{W}(x)\mathcal{F}(x) \nonumber \\
&=&\int d^2x\ \frac{1}{2} s\epsilon^{\mu\nu}s_\mu s_\nu \tilde{s}\ 
\text{Tr} (AB) \nonumber  \\
&=&  \int d^2x \ \text{Tr} \big{(} -\frac{1}{2}F_{\mu\nu}F^{\mu\nu} 
+ \frac{G^2}{4} +D^\mu D_\mu A \ B
-iD^\mu\lambda_\mu\ \rho -i\epsilon^{\mu\nu}D_\mu \lambda_\nu \ \tilde{\rho} \nonumber \\
& &\qquad\qquad -\frac{1}{4}[A,B]^2-\frac{i}{2}\{ \rho, \rho\}A -\frac{i}{2}\{\tilde{\rho}, \tilde{\rho}\}A -\frac{i}{2}\{\lambda^\mu,\lambda_\mu \}B \big{)}.
\end{eqnarray}
We identify this action as off-shell twisted $N=2$ SUSY invariant super Yang-Mills 
action. This action is equivalent to the kinetic terms of quantized topological 
Yang-Mills action which was derived previously\cite{KT,KKU}.

\subsection{$N=2$ superconnection formalism in four dimensions}

As we show in the previous subsection twisted $N=2$ SUSY invariant 
super Yang-Mills action has been derived by the superconnection 
formalim in two dimensions. Based on the philosophy of Dirac-K\"ahler 
twisting procedure we naively expect that twisted $N=4$ SUSY 
invariant super Yang-Mills action may be derived by the superconnection 
formalism parallel to the $N=D=2$ superconnection formalism. 
It turns out that it is highly nontrivial to find out twisted $N=4$ 
counterparts of the constraints (\ref{eq:curvature con}) in a systematic way. 
Instead here we consider superconnection formalism of $N=2$ truncated 
version of twisted $N=4$ SUSY algebra, which is given in subsection 4.2. 
This formulation is closely related to the $N=2$ 
formulation given by Alvarez and Labastida who used spinor formulation of 
usual four-dimensional extended $N=2$ SUSY algebra \cite{AL}.
Hereafter we omit the $+$ suffix from $N=2$ twisted SUSY counterpart decomposed 
from $N=4$ twisted SUSY for supercharges and differential operators; 
$\{s^+_I\}=\{s^+,s^+_\mu,s^+_{\mu\nu}\}$, 
$\{\mathfrak{D}^+_I\}=\{\mathfrak{D}^+,\mathfrak{D}^+_\mu,
\mathfrak{D}^+_{\mu\nu}\}$ and $\{\mathcal{Q}^+_I\}=\{\mathcal{Q}^+,
\mathcal{Q}^+_\mu,\mathcal{Q}^+_{\mu\nu}\}$. 

As in the two-dimensional formulation we first introduce supercovariant 
derivatives by introducing fermionic superconnections $\Gamma_0, 
\Gamma_\mu,\Gamma_{\mu\nu}$ and a bosonic superconnection 
$\Gamma_{\underline{\mu}}$, 
\begin{eqnarray}
\nabla_0 &=& \mathfrak{D}-i\Gamma_0, \nonumber \\
\nabla_\mu &=& \mathfrak{D}_\mu -i\Gamma_\mu, \nonumber \\
\nabla_{\mu\nu} &=& \mathfrak{D}_{\mu\nu} -i\Gamma_{\mu\nu}, \nonumber \\
\nabla_{\underline{\mu}} &=& \partial_{\underline{\mu}} 
-i\Gamma_{\underline{\mu}}, \nonumber \\
\end{eqnarray}
where the superderivative differential operators 
$\mathcal{D}^+=\mathcal{D},\mathcal{D}^+_\mu=\mathcal{D}_\mu$ and 
$\mathcal{D}^+_{\mu\nu}=\mathcal{D}_{\mu\nu}$ are given in 
(\ref{def-chirally-decomposed-superderivatives}) and satisfy the 
four-dimensional $N=2$ twisted SUSY algebra (\ref{eq:4al+-}) with 
replacements: $s^{\pm}_I \rightarrow \mathfrak{D}^\pm_I$. 
The lowest ($\theta=0$) component of the connection $\Gamma_{\underline{\mu}}$ 
is identified as the usual gauge field 
$\Gamma_{\underline{\mu}}|_{\theta=0}=\omega_\mu$.

We can introduce the following supergauge transformation: 
\begin{eqnarray}
\delta_{gauge} \Gamma_I = \nabla_I K \equiv \mathfrak{D}_I K - i [\Gamma_I,K],
\end{eqnarray}
where $K$ is an arbitrary superfield. We gauge away superfluous component 
fields by the supergauge freedom of the superfield $K$ by 
taking Wess-Zumino gauge and keep the usual gauge degrees of freedom. 

Parallel to the two-dimensional notations we summarize the commutators and 
anticommutators of these supercovariant derivatives in Table 6.
\begin{table}[htbp]
\[
 \begin{array}{|l|c|c|c||c|}
\hline
 & \nabla_0 & \nabla_{B} & \nabla_\nu & \nabla_{\underline{\nu}}\\
\hline
\nabla_0 & -i\mathcal{F}_{0,0}& -i\mathcal{F}_{0,B}   &-i\nabla_{\underline{\nu}} -i \mathcal{F}_{0,\nu} &-i\mathcal{F}_{0,\underline{\nu}}   \\
\nabla_A & &  -i\mathcal{F}_{A,B} & i\delta^+ _{A,\nu\rho}\nabla^{\underline{\rho}}-i\mathcal{F}_{\nu,A} &-i\mathcal{F}_{A,{\underline{\nu}} } \\
\nabla_{\mu} & & & -i\mathcal{F}_{\mu,\nu} & -i\mathcal{F}_{\mu,\underline{\nu}} \\
\hline
\hline
\nabla_{\underline{\mu}} & & & & -i\mathcal{F}_{\underline{\mu},\underline{\nu}}\\
\hline 
\end{array}
\]
\caption{Definition of $N=2$ Supercurvatures}
\label{def-4d-N2supercurvature}
\end{table}
Here the lowest component of 
$\mathcal{F}_{\underline{\mu},\underline{\nu}}$ is usual curvature:  
\begin{eqnarray}
\mathcal{F}_{\underline{\mu},\underline{\nu}}|=F_{\mu\nu}=
\partial_\mu \omega_\nu- \partial_\nu \omega_\mu 
-i[\omega_\mu,\omega_\nu ].
\end{eqnarray}

We impose the following constraints: 
\begin{eqnarray}
\mathcal{F}_{A,B} &=& \delta^+ _{A,B} F_{0,0}, \nonumber \\ 
\mathcal{F}_{\mu,\nu} &=& \delta_{\mu\nu} \mathcal{W}, \nonumber \\
\mathcal{F}_{0,\mu}&=& \mathcal{F}_{0,A} = \mathcal{F}_{\nu,A} =0, 
\label{4d-curvature-constrs}
\end{eqnarray}
which can be interpreted as the suppression conditions of higher spin 
fields to construct twisted $N=2$ super Yang-Mills multiplets. 
Then the Table \ref{def-4d-N2supercurvature} changes into 
Table \ref{tab:c2} after taking into account the constraints.
\begin{table}[htbp]
\[
 \begin{array}{|l|c|c|c||c|}
\hline
 & \nabla_0 & \nabla_{B} & \nabla_\nu & \nabla_{\underline{\nu}}\\
\hline
\nabla_0 & -i\mathcal{F}_{0,0}& 0  &-i\nabla_{\underline{\nu}} &-i\mathcal{F}_{0,\underline{\nu}}   \\
\nabla_A & &  -i\delta^+ _{A,B}\mathcal{F}_{0,0} & i\delta^+ _{A,\nu\rho}\nabla^{\underline{\rho}} &-i\mathcal{F}_{A,{\underline{\nu}} } \\
\nabla_{\mu} & & & -i\delta_{\mu\nu}\mathcal{W} & -i\mathcal{F}_{\mu,\underline{\nu}} \\
\hline 
\hline
\nabla_{\underline{\mu}} & & & & -i\mathcal{F}_{\underline{\mu},\underline{\nu}}\\
\hline 
\end{array}
\]
\caption{$N=2$ Supercurvatures after the constraints}
\label{tab:c2}
\end{table}

Since we have imposed the constraints (\ref{4d-curvature-constrs}), we can 
obtain series of nontrivial relations by Jacobi identities:
\begin{eqnarray}
\nabla_0 \mathcal{F}_{0,0} &\equiv& \mathfrak{D} \mathcal{F}_{0,0} -i
[\Gamma_0,\mathcal{F}_{0,0} ] = 0, \label{eq:non-chi1} \\
\nabla_A \mathcal{F}_{0,0} &\equiv& \mathfrak{D}_A \mathcal{F}_{0,0} -i
[\Gamma_A,\mathcal{F}_{0,0} ] = 0, \label{eq:non-chi2} \\
\nabla_\mu \mathcal{W} &\equiv& \mathfrak{D}_\mu \mathcal{W} -i
[\Gamma_\mu,\mathcal{W} ] = 0, \label{eq:non_achi1} \\
\mathcal{F}_{0\underline{\mu}} &=& -\frac{i}{2}\nabla_\mu \mathcal{F}_{0,0} , 
\label{4d-N2-constrs4}\\
\mathcal{F}_{A\underline{\mu}} &=& -\frac{i}{2}\delta_{A,\mu\nu}\nabla^\nu \mathcal{F}_{0,0} , \\
\mathcal{F}_{\mu\underline{\nu}} &=& -\frac{i}{2}\delta_{\mu\nu} \nabla_0 \mathcal{W} +\frac{i}{2} \nabla_{\mu\nu} \mathcal{W} ,\\
\mathcal{F}_{\underline{\mu}\underline{\nu}} &=& -\frac{1}{2} \nabla_0 \nabla_{\mu\nu} \mathcal{W} +\frac{1}{4} (\nabla_\mu\nabla_\nu - \nabla_\nu \nabla_\mu )\mathcal{F}_{0,0},
\label{Jacobi-curvature-4d-N2-ab}
\end{eqnarray}
\begin{eqnarray}
 \delta^+ _{A,\nu\rho} \nabla_\mu \nabla^\rho \mathcal{F}_{0,0} ,
&-&\frac{1}{2}\delta^+ _{A,\mu\rho}(\nabla_\nu \nabla^\rho -\nabla^\rho \nabla_\nu) \mathcal{F}_{0,0}, \nonumber \\ 
&+& \delta_{\mu\nu} \nabla_A \nabla_0 \mathcal{W} -\nabla_A \nabla_{\mu\nu} \mathcal{W} +\delta^+ _{A,\mu\rho} \nabla_0 \nabla_\nu {}^\rho \mathcal{W} =0, 
\label{4d-N2-constrs8}
\end{eqnarray}
where we assume that the gauge algebra is non-Abelian.

\subsubsection{Abelian case}

If the gauge algebra is Abelian the first three constraints 
(\ref{eq:non-chi1}), (\ref{eq:non-chi2}) and (\ref{eq:non_achi1}) turn into 
the chiral and anti-chiral constraints:
\begin{eqnarray}
\mathfrak{D}_0\mathcal{F}_{0,0} = 0, \ & & \
\mathfrak{D}_A \mathcal{F}_{0,0} = 0, 
\label{4dN2abchiralcon}\\
\mathfrak{D}_\mu \mathcal{W} &=&  0.
\label{4dN2abanti-chiralcon}
\end{eqnarray}
We can solve these chiral and anti-chiral constraints (\ref{4dN2abchiralcon}) 
and (\ref{4dN2abanti-chiralcon}), respectively, and expand these 
superfields as follows: 
\begin{eqnarray}
\mathcal{F}_{0,0}(y,\theta^{+\mu}) &=& \phi + \theta^{+\mu} C_\mu +\frac{1}{2}\theta^{+\mu}\theta^{+\nu}\phi_{\mu\nu}+ \theta^{3+} _{\mu}\tilde{\psi}^\mu +\theta^{4+} \tilde{\phi}, \nonumber \\
\mathcal{W}(w,\theta^+,\theta^{+A})&=& \overline{\phi} + \frac{1}{4}\theta^{A}\chi^+ _A + \frac{1}{4}(\theta^{2+})^{A}M^+ _A +(\theta^{3+})\tilde{\chi} \nonumber  \\
& &+\theta^+ (\chi + \frac{1}{4}\theta^{A}N^+ _A + \frac{1}{4}(\theta^{2+})^{A}\tilde{\chi}^+ _A +(\theta^{3+})L ).
\end{eqnarray}
In two-dimensional $N=2$ case we have obtained only one relation 
(\ref{DEF-relation}) between the usual curvature and the component fields of 
chiral and anti-chiral superfields. 
In four-dimensional twisted $N=2$ formulation the 
constraints (\ref{Jacobi-curvature-4d-N2-ab}) and (\ref{4d-N2-constrs8}) lead 
to the following relations:
\begin{eqnarray}
\tilde{\phi} &=& -\partial^2 \overline{\phi}, \nonumber \\
\tilde{\psi}^\mu &=& i(\partial^\mu \chi -\partial_\nu \chi^{+\mu\nu}), \nonumber \\
\phi_A &\equiv& \phi^+ _A +\phi^- _A = \phi^+ _A -2F^- _A ,
\end{eqnarray}
\begin{eqnarray}
N^+ _A &=& \phi^+ _A + 2F^+ _A, \nonumber \\
M^+ _A &=& \phi^+ _A - 2F^+ _A, \nonumber \\
\tilde{\psi} &=& -i\partial^\mu C_\mu, \nonumber \\
L &=& \partial^2 \phi, \nonumber \\
\tilde{\chi}^+ _A &=&i\delta^+ _{A,\rho\sigma}\partial^\rho \phi^\sigma  
=i(\partial \phi)^+ _A,
\end{eqnarray}
where 
we define a (anti-)self-dual part of the curvature:
\begin{eqnarray}
F^{\pm}_{\mu\nu} \equiv \frac{1}{4} \delta^{\pm} _{\mu\nu,\rho\sigma}F^{\rho\sigma}. 
\end{eqnarray}
Taking into account these relations, we obtain the explicit form of the 
superfields $\mathcal{F}_{0,0}$ and $ \mathcal{W}$ as 
\begin{eqnarray}
\mathcal{F}_{0,0} &=& \phi + \theta^{+\mu} C_\mu 
+\frac{1}{2}\theta^{+\mu}\theta^{+\nu}(\phi^+ _{\mu\nu} 
-2F^- _{\mu\nu}) \nonumber \\
& &\hspace{40mm}+ i\theta^{3+} _{\mu}(\partial^\mu \chi 
-\partial_\nu \chi^{+\mu\nu})  - \theta^{4+} \partial^2 \overline{\phi},  \\
\mathcal{W} &=& \overline{\phi} + \frac{1}{4}\theta^{A}\chi^+ _A 
+ \frac{1}{4}(\theta^{2+})^{A}(\phi^+ _A - 2F^+ _A)  
-i(\theta^{3+})\partial^\mu C_\mu  \nonumber  \\
& &+\theta^+ (\chi + \frac{1}{4}\theta^{A}(\phi^+_A + 2F^+_A) 
+ \frac{i}{4}(\theta^{2+})^{A}(\partial \phi)^+_A 
+(\theta^{3+})\partial^2 \phi ).
\end{eqnarray}
Using these superfields, we can obtain off-shell twisted $N=2$ SUSY invariant 
actions:
\begin{eqnarray}
S_1 &=& \frac{1}{2}\int d^4 y d^4 \theta \ (\mathcal{F}_{0,0})^2 \nonumber \\
 &=& \frac{1}{2}\int d^4 y\  
\frac{1}{4!} \epsilon^{\mu\nu\rho\sigma}s^+_\mu s^+_\nu s^+_\rho s^+_\sigma\  (\phi^2) 
\nonumber \\
&=&  \int d^4 x \{ -\phi \partial^2 \overline{\phi} 
-iC_\mu (\partial^\mu \chi -\partial_\nu \chi^{+\mu\nu}) 
-\frac{1}{4}(\phi^+ _{\mu\nu})^2 +(F^- _{\mu\nu})^2 \},
\label{eq:s1}
\end{eqnarray}
\begin{eqnarray}
S_2 &=&\frac{1}{2} \int d^4 w d^4 \theta \ \mathcal{W}^2 \nonumber \\
&=&\frac{1}{2} \int d^4 w \ s^+ \frac{1}{3!4^3}
\Gamma^{+A,B,C} s^+_A s^+_B s^+_C 
(\overline{\phi}^2) \  \nonumber \\
&=& -\int d^4x \{(-\phi \partial^2 \overline{\phi} 
-iC_\mu (\partial^\mu \chi -\partial_\nu \chi^{+\mu\nu}) 
-\frac{1}{4}(\phi^+ _{\mu\nu})^2 +(F^+ _{\mu\nu})^2  )\}.
\label{eq:s2}
\end{eqnarray}
We can see that these actions (\ref{eq:s1}) and (\ref{eq:s2}) are 
Abelian version of $N=2$ twisted SUSY invariant super Yang-Mills action 
{\it a la} Donaldson-Witten.  The difference of these actions are 
topologically invariant surface term. 

Here it is interesting to recognize that these actions $S_1$ and 
$S_2$, respectively, have the same forms as 
(\ref{4d-ab-N2-action-chiral1}) and 
(\ref{4d-ab-N2-action-anti-chiral1}) which don't have derivatives 
in the Lagrangian. 
It turns out that the derivatives in the actions $S_1$ and $S_2$ 
are generated by the constraints (\ref{4d-curvature-constrs}) which 
in turn relate the chiral and anti-chiral superfields. 
Eventually the corresponding actions (\ref{eq:s1}) and (\ref{eq:s2}) are related and the difference 
is only the topological surface term. 

Twisted $N=2$ SUSY transformation can be obtained by operating 
superdifferential operators to the superfields and is given in 
Table \ref{tb:trdw}.
\begin{table}[htbp]
\[
 \begin{array}{|c||c|c|c|}
\hline
 & s^+ & s^+ _\mu & s^+ _A \\
\hline
\phi & 0 & C_\mu & 0 \\
C_\nu & -i\partial_\nu \phi & -\phi^+ _{\mu\nu} +2 F^- _{\mu\nu} & i\mathcal{D}^+ _{\nu,A}\phi \\
\phi^+ _B & \frac{i}{2} \delta^+ _{B,\rho\sigma} \partial^\rho C^\sigma & 
\frac{i}{2}\delta^+ _{B,\mu\nu} (\partial^\nu \chi -\partial_\rho \chi^{+\nu\rho}) & -\frac{i}{2}\delta^+ _{B,\rho\sigma} \mathcal{D}^{+\rho} {}_{,A} C^\sigma \\
\overline{\phi} & \chi & 0 & \chi^{+}_A \\ 
\chi^+ _B & -\phi^+ _B - 2F^+ _B &  i\mathcal{D}^+ _{\mu,B} \overline{\phi} & 
-\frac{1}{4} \Gamma^+ _{ABC} (\phi^{+C}-2F^{+C}) \\
\chi &0 & -i\partial_\mu \overline{\phi} & \phi^+ _A +2 F^+ _A \\ 
\omega_\nu &-\frac{i}{2} C_\nu & -\frac{i}{2} (\delta_{\mu\nu}\chi-\chi^+ _{\mu\nu} )&
-\frac{i}{2} \delta^+ _{A,\nu\rho} C^\rho\\
\hline
 \end{array}
\]
\caption{Twisted $N=2$ SUSY transformation of component fields for 
Abelian Donaldson-Witten theory.}
\label{tb:trdw}
\end{table}
As in the two-dimensional case the twisted $N=2$ SUSY algebra (\ref{eq:4al+-}) 
is valid modulo gauge transformation.

\subsubsection{Non-Abelian case}
If the gauge algebra is non-Abelian the constrains (\ref{eq:non-chi1}), 
(\ref{eq:non-chi2}) and (\ref{eq:non_achi1}) are no longer chiral and 
anti-chiral constraints and thus $\mathcal{F}_{0,0}$ and $\mathcal{W}$ are 
not (anti-)chiral superfields. 
We cannot then obtain the explicit component fields expansion of the 
superfields from simple arguments. We then proceed to derive the twisted 
SUSY transformation of component fields from the constraint relations 
obtained from Jacobi identity as in the two-dimensional non-Abelian case.  
We first identify the following component fields: 
\begin{align}
&\mathcal{F}_{0,0}| = \phi,  &  &\nabla_\mu \mathcal{F}_{0,0}| = C_\mu, &
 &\frac{1}{2}(\nabla_\mu\nabla_\nu - \nabla_\nu\nabla_\mu)\mathcal{F}_{0,0}| 
=- \phi_{\mu\nu},
\nonumber
\end{align}
\vspace{-5mm}
\begin{align}
&\frac{1}{3!}\epsilon^{\mu\nu\rho\sigma} \nabla_\nu \nabla_\rho
 \nabla_\sigma \mathcal{F}_{0,0}| =- \tilde{\psi}^\mu, &
&\frac{1}{4!}\epsilon^{\mu\nu\rho\sigma} \nabla_\mu \nabla_\nu \nabla_\rho
\nabla_\sigma \mathcal{F}_{0,0}| = \tilde{\phi},
\end{align}

\begin{align}
&\mathcal{W}|=\overline{\phi}, &  &\nabla_A \mathcal{W}| =\chi^+ _A, 
\nonumber \\
 &\frac{1}{2!\cdot4^2}\Gamma^{+ABC}\nabla_B\nabla_C \mathcal{W}| 
=- M^{+A}, & 
&\frac{1}{3!\cdot4^3}\Gamma^{+ABC}\nabla_A\nabla_B\nabla_C \mathcal{W}| 
=-
 \tilde{\chi},\nonumber 
\end{align}
\vspace{-5mm}
\begin{align}
&\nabla_0 \mathcal{W}|=\chi, &  &\nabla_A \nabla_0 \mathcal{W}| =N^+ _A, 
\nonumber \\
 &\frac{1}{2!\cdot4^2}\Gamma^{+ABC}\nabla_B\nabla_C \nabla_0 \mathcal{W}| 
=-\tilde{ \chi}^{+A}, & &\frac{1}{3!\cdot4^3}
\Gamma^{+ABC}\nabla_A\nabla_B\nabla_C \nabla_0 \mathcal{W}| = -L,
\end{align}

We can then obtain the twisted $N=2$ SUSY transformation of these component 
fields by Jacobi identities with Wess-Zumino gauge as in the two-dimensional 
case. For example we find the transformation law of $\phi$ as follows: 
\begin{eqnarray}
s \phi = \mathcal{Q} \mathcal{F}_{0,0}| = \nabla_0  \mathcal{F}_{0,0}| 
= [\nabla_0,i\{\nabla_0,\nabla_0 \}] = 0.
\end{eqnarray}
We can similarly find the transformation laws of all the other component 
fields. We show the list of twisted $N=2$ SUSY transformation of component 
fields in Table \ref{tb:nontrdw}.
\begin{table}[htbp]
\[
 \begin{array}{|c||c|c|}
\hline
 & s^+ & s^+ _\mu  \\
\hline
\phi & 0 & C_\mu \\
C_\nu & -iD_\nu \phi & -\phi^+ _{\mu\nu} +2 F^- _{\mu\nu} +\frac{i}{2}\delta_{\mu\nu} [\phi,\overline{\phi}] \\
\phi^+ _B & \frac{i}{2} \delta^+ _{B,\rho\sigma} D^\rho C^\sigma -\frac{i}{2}[\chi^+ _{B},\phi] & 
\frac{i}{2}\delta^+ _{B,\mu\nu} (D^\nu \chi -D_\rho \chi^{+\nu\rho})+\frac{i}{2} \delta^+ _{B,\mu\nu}[\overline{\phi},C^\nu ]  \\
\overline{\phi} & \chi & 0  \\ 
\chi^+ _B & -\phi^+ _B - 2F^+ _B &  i\delta^+ _{B,\mu\nu}D^\nu \overline{\phi}  \\
\chi & -\frac{i}{2} [\phi,\overline{\phi}] & -iD_\mu \overline{\phi} \\ 
\omega_\nu &-\frac{i}{2} C_\nu & -\frac{i}{2} (\delta_{\mu\nu}\chi-\chi^+ _{\mu\nu} ) \\
\hline
 \end{array}
\]
\end{table}
\begin{table}[htbp]
\[
 \begin{array}{|c||c|}
\hline
 & s^+ _A \\
\hline
\phi  & 0 \\
C_\nu & i\delta^+ _{A,\nu\rho}D^\rho \phi \\
\phi^+ _B & -\frac{i}{2}\delta^+ _{B,\alpha\beta}\delta^+ _{A,} {}^\alpha {}_\gamma(D^\gamma C^\beta -\frac{1}{4}[\chi^{\gamma\beta},\phi] ) +\frac{i}{2} \delta^+ _{A,B} [\chi,\phi]  \\
\overline{\phi} & \chi^{+}_A \\ 
\chi^+ _B & -\frac{1}{4} \Gamma^+ _{ABC} (\phi^{+C}-2F^{+C}) -\frac{i}{2} \delta_{A,B}[\phi,\overline{\phi}]\\
\chi & \phi^+ _A +2 F^+ _A \\ 
\omega_\nu & -\frac{i}{2} \delta^+ _{A,\nu\rho} C^\rho\\
\hline
 \end{array}
\]
\caption{$N=2$ twisted SUSY transformation of component fields for non-Abelian 
Donaldson-Witten theory}
\label{tb:nontrdw}
\end{table}

The $N=2$ twisted SUSY algebra is closed for the fields up to the gauge 
transformation of the following specific gauge parameter choice: 
\begin{eqnarray}
\{s^+,s^+\}\varphi&=& \delta_{\text{gauge}\ (-\phi)} \varphi\nonumber \\
\{s^+ _\mu,s^+ _\nu\}\varphi&=& 
\delta_{\mu\nu} \delta_{\text{gauge} (-\psi)} \varphi, \nonumber \\
\{s^+ _A,s^+ _B\}\varphi&=& \delta^+ _{A,B}
\delta_{\text{gauge}\ (-\phi)} \varphi, \nonumber \\
\{s^+,s^+ _A\}\varphi&=& 0, \nonumber \\
\{s^+,s^+_\mu\}\varphi&=& -i(\partial_\mu \varphi +\delta_{\text{gauge}\ (-A_\mu)} \varphi) , \nonumber \\
\{s^+ _\mu,s^+ _A\}\varphi&=& i\delta^+ _{A,\mu\nu} (\partial^\nu \varphi +\delta_{\text{gauge}\ (-A^\nu)} \varphi)  ,
\end{eqnarray}
where $\{\varphi\} = \{\phi,\ C_\mu,\ \phi^+ _A,\ \overline{\phi},\ \chi^+ _A,\
\chi,\ \omega_\mu \}$ and $\delta_{\text{gauge}\ \phi}
\omega_\mu=\partial_\mu \phi -i[\omega_\mu,\phi]$ for the gauge field and
$\delta_{\text{gauge}\ \phi} \cdots=
-i[\cdots,\phi]$ for the other fields. We can see that $N=2$ twisted 
SUSY algebra in four dimensions is closed modulo the gauge transformation. 

We can now define the superfield $\mathcal{F}_{0,0}$ and $\mathcal{W}$ from 
the parent fields $\phi$ and $\overline{\phi}$ as follows:
\begin{eqnarray}
\mathcal{F}_{0,0}(y^\mu,\theta^{+\mu}) &=&e^{ \theta^{+\mu} s^{+}_\mu}e^{\theta^+ s^+ +\frac{1}{4} \theta^{+\mu\nu} s^+ _{\mu\nu}}\phi \nonumber  \\
&=& e^{ \theta^{+\mu} s^{+}_\mu}\phi, \\
\mathcal{W}(w^\mu,\theta^{+},\theta^{+\mu\nu}) &=&e^{\theta^+ s^+ + \frac{1}{4}\theta^{+\mu\nu} s^+ _{\mu\nu}}e^{ \theta^{+\mu} s^{+}_\mu}\overline{\phi} \nonumber  \\
&=&  e^{\theta^+ s^+ +\frac{1}{4} \theta^{+\mu\nu} s^+ _{\mu\nu}} \overline{\phi}.
\end{eqnarray}
The superfield  $\mathcal{F}_{0,0}$ and $\mathcal{W}$ are then given by 
\begin{eqnarray}
\mathcal{F}_{0,0}(y^\mu,\theta^{+\mu}) &=& \phi(y^\mu) + \theta^{+\mu} C_\mu+ \frac{1}{2}\theta^{+\mu} \theta^{+\nu}(\phi^{+}_{\mu\nu}-2F^- _{\mu\nu}) \nonumber \\
& &+i(\theta^{3+} )_\mu (D^\mu \chi -D_\nu\chi^{+\mu\nu} +\frac{1}{2}[\overline{\phi},C^\mu]) \nonumber \\
& &+\theta^{4+} (-D^\mu D_\mu \overline{\phi} +\frac{i}{8}\{\chi^+ _{\mu\nu} , \chi^{+\mu\nu}\}\nonumber \\
& &\hspace{40mm}+\frac{i}{2}\{\chi,\chi\} -\frac{1}{4}[\overline{\phi},[\overline{\phi},\phi]]),
\end{eqnarray}
\begin{eqnarray}
\mathcal{W}(w^\mu,\theta , \theta^{+\mu\nu})&=& \overline{\phi}(w^\mu)
+ \theta \chi
 +\frac{1}{4}\theta^{+A} \chi^{+}_A +\frac{1}{4}\theta^+ \theta^{+A}(\phi^+ _A +2F^+ _A) \nonumber \\
& & +\frac{1}{4}(\theta^{2+})^A(\phi^+ _A -2F^+ _A) \nonumber \\
& &+ \theta^+ (\theta^{2+})^A(\frac{i}{4} \delta^+ _{A,\mu\nu}D^\mu C^\nu -\frac{i}{8}[\chi^+ _A , \phi]) \nonumber \\ 
& &+\theta^{3+} (-iD^\mu C_\mu +\frac{i}{2}[\chi,\phi] )\nonumber \\
& &+\theta^+  \theta^{3+}(D^\mu D_\mu \phi -\frac{i}{2} \{C^\mu ,C_\mu\} -\frac{1}{4}[[\phi,\overline{\phi}], \phi]).
\end{eqnarray}
We can now define the following off-shell $N=2$ twisted SUSY invariant 
actions: 
\begin{eqnarray}
S_1 &=& \frac{1}{2}\int d^4 y d^4 \theta\ \text{Tr} {\mathcal{F}_{0,0}}^2
= \frac{1}{2}\int d^4 y \  
\frac{1}{4!} \epsilon^{\mu\nu\rho\sigma}s^+_\mu s^+_\nu s^+_\rho s^+_\sigma\ 
\text{Tr}( \phi^2) 
, \nonumber \\
S_2 &=& \frac{1}{2}\int d^4 w d^4 \theta\ \text{Tr} \mathcal{W}^2
= \frac{1}{2} \int d^4 w \  s^+\frac{1}{3!4^3}\Gamma^{+A,B,C} s^+_A s^+_B s^+_C 
\text{Tr} (\overline{\phi}^2) ,
\end{eqnarray}
where $\int d^4 \theta (\theta^{4+})=1 ,\quad\int d^4 \theta \theta(\theta^{3+})=1  $. 
We can obtain the explicit form of these actions, 
\begin{eqnarray}
S_1 &=& \int d^4 x\text{Tr}\Big{(}-\phi D^\mu D_\mu \overline{\phi} -iC^\mu(D_\mu \chi -D^\nu \chi^+ _{\mu\nu} ) -\frac{1}{4}(\phi^+ _{\mu\nu} )^2 + (F^- _{\mu\nu})^2 \nonumber \\
& &\hspace{15mm}+\frac{i}{2}\phi \{\chi,\chi\}+\frac{i}{8}\phi \{\chi^{+A},\chi^+ _A\} + \frac{i}{2} \overline{\phi} \{C^\mu,C_\mu\}+\frac{1}{4}[\phi,\overline{\phi}]^2 \Big{)},
\label{eq:s_1}
\end{eqnarray}
\begin{eqnarray}
S_2 &=& -\int d^4 x\text{Tr}\Big{(} \phi D^\mu D_\mu \overline{\phi} -iC^\mu(D_\mu \chi -D^\nu \chi^+ _{\mu\nu} ) -\frac{1}{4}(\phi^+ _{\mu\nu} )^2 + (F^+ _{\mu\nu})^2 \nonumber \\
& &\hspace{15mm}+\frac{i}{2}\phi \{\chi,\chi\}+\frac{i}{8}\phi \{\chi^{+A},\chi^+ _A\} + \frac{i}{2} \overline{\phi} 
\{ C^\mu,C_\mu\}+\frac{1}{4}[\phi,\overline{\phi}]^2\Big{)}.
\end{eqnarray}
The difference of these actions is only topological surface term. 
These actions are equivalent to non-Abelian version of Donaldson-Witten action 
which has off-shell twisted $N=2$ SUSY invariance.

We have derived the four-dimensional quantized topological Yang-Mills 
action by using superconnection formulation. Thus our twisted $N=4$ 
superspace formalism naturally includes our twisted $N=2$ superspace 
formalism as a subalgebra.

\subsubsection{Dirac-K\"ahler matter fermion}

As we have already shown that Dirac-K\"ahler(D-K) twisting procedure of supercharges generated $N=4$ twisted superalgebra which is decomposed into two 
sets of $N=2$ twisted superalgebra by (anti-)self-dual decomposition. 
On the other hand the D-K fermion mechanism transforms fermionic 
anti-symmetric tensor fields into matter fermions, we expect that 
this type of decomposition will also be generated in the matter fermion 
sector of D-K fermion. Here we explicitly construct twisted 
$N=2$ sector of D-K fermion. 

We claim that the following two "flavor" Dirac fermion action is essentially 
equivalent to $N=2$ sector of D-K fermion constructed from 
the D-K fermion mechanism:
\begin{eqnarray}
\int d^4x \sum_{i=1}^{2}\ \overline{\Psi}^{i N=2}\gamma^\mu \partial_\mu
 \Psi^{i N=2},
\label{eq:su2}
\end{eqnarray}
where $\overline{\Psi}^{i N=2} =(\Psi^{i N=2})^\dagger $ and $\Psi^{i N=
2}$ is $SU(2)$ Majorana fermion which satisfies the following condition
\cite{Kugo-T}:
\begin{eqnarray}
(\Psi^{i N=2})^* =\epsilon^{ij} B \Psi^{j N=2},
\end{eqnarray}
where $B=-\gamma^1 \gamma^3$ and $\gamma^\mu = B^{-1} \gamma^\mu B$.

The $N=2$ sector of the D-K fermion $\Psi^{i N=2}$ is the chirally 
projected sector of the original D-K fermion on the "flavor" suffix or 
$R$-symmetry related suffix as follows:
\begin{eqnarray}
(\Psi^{N=2})_{\alpha i}  = (\Psi)_{\alpha j} (P_+)_{ji},
\end{eqnarray}
where $P_{\pm}= \frac{1}{2}(1\pm \gamma_5)$ and the D-K fermion 
$\Psi$ is defined in (\ref{def-DK-fields}) as  
\begin{eqnarray}
\Psi _{\alpha i}=\frac{1}{2\sqrt{2}}\Big{(} {\bf 1} \psi +\gamma^\mu \psi_\mu + 
\frac{1}{2}\gamma^{\mu\nu} \psi_{\mu\nu} +\tilde{\gamma}^\mu \tilde{\psi
}_\mu +\gamma_5 \tilde{\psi} \Big{)}_{\alpha i},
\end{eqnarray}
where $i=1,2,3,4 $.

The action (\ref{eq:su2}) can be transformed as follows:
\begin{eqnarray}
& &\int d^4x \sum_{i=1} ^{2}\ \overline{\Psi}^{i N=2}\gamma^\mu \partial
_\mu \Psi^{i N=2} \nonumber \\
 &=& \int d^4x \mbox{Tr} \overline{\Psi} \gamma^\mu \partial_\mu \Psi P_
+ \nonumber \\
&=& \int d^4x  \frac{1}{2} \big{\{} (\psi+\tilde{\psi})\partial_{\mu} 
(\psi^\mu + \tilde{\psi}^\mu)-(\psi_\mu+ \tilde{\psi_\mu})(\psi^{\mu\nu} 
-\frac{1}{2} \epsilon^{\mu\nu\rho\sigma} \psi_{\rho\sigma}) \big{\}} 
\nonumber \\
&=& \int d^4x  \{ \chi \partial_\mu C^\mu -C^\mu \partial^\nu \chi^+ 
_{\mu\nu} \},
\end{eqnarray} 
where $ \chi =\frac{1}{\sqrt{2}}(\psi+\tilde{\psi}),\quad C_\mu
=\frac{1}{\sqrt{2}}(\psi^\mu + \tilde{\psi}^\mu), \quad \chi^+ _{\mu\nu}
= \frac{1}{\sqrt{2}}(\psi^{\mu\nu} -\frac{1}{2} 
\epsilon^{\mu\nu\rho\sigma}\psi_{\rho\sigma} )$.
These terms are included in the action (\ref{eq:s_1}). Then the action
(\ref{eq:s_1}) eventually turns into the following form:
\begin{eqnarray}
S_1 &=& \int d^4 x\text{Tr}\Big{(}-\phi D^\mu D_\mu \overline{\phi} -i 
\overline{\Psi}^{i N=2}\gamma^\mu D_\mu \Psi^{i N=2} -\frac{1}{4}(\phi^+ 
_{\mu\nu} )^2 + (F^- _{\mu\nu})^2 \nonumber \\
& &\hspace{15mm}+i \phi \ \overline{\Psi}^{i N=2}(1+\gamma_5) \Psi^{i N=
2}+i \overline{\phi}\ \overline{\Psi}^{i N=2}(1-\gamma_5) \Psi^{i N=2} 
\nonumber \\
& &\hspace{15mm}+\frac{1}{4}[\phi,\overline{\phi}]^2 \Big{)},
\end{eqnarray}
where fermions are now transformed into $SU(2)$ Majorana fermions via 
Dirac-K\"ahler fermion mechanism.

\section{Conclusions and Discussions}

We have proposed four-dimensional twisted $N=4$ superspace formalism 
based on the Dirac-K\"ahler twisting procedure. In the Dirac-K\"ahler 
twist, $N=2$ twisted superspace formalism in two dimensions and that of 
$N=4$ in four dimensions have close 
similarity in the formulation. We have examined 
the formulation in various cases to derive off-shell invariant twisted 
SUSY invariant actions. To see the basic structure in simpler 
examples we have given $D=N=2$ formulation as well. 
In order to find $N=4$ invariant action with gauge symmetry in four 
dimensions, we have introduced twisted vector superfield and found 
formal expressions of actions which have off-shell $N=4$ twisted SUSY 
invariance. It turns out, however, the actions have still many 
superfluous component fields and too higher derivatives for the gauge fields 
and we may still need to find further 
constraints to obtain $N=4$ twisted super Yang-Mills action. 

We have then proposed twisted superconnection formalism to find supergauge invariant Yang-Mills actions. In two dimensions we have 
successfully derived $N=2$ twisted SUSY invariant super Yang-Mills 
actions. In four dimensions we have not yet found out constraints to 
derive $N=4$ twisted SUSY invariant super Yang-Mills action. 
By decomposing the twisted $N=4$ algebra into two sectors of $N=2$ 
algebra and thus truncating the full algebra into $N=2$, we have 
proposed twisted $N=2$ superspace formalism in four dimensions. 
We have then examined 
the superconnection formalism on the $N=2$ sector of twisted SUSY 
algebra. We have successfully derived $N=2$ super Yang-Mills action 
from the $N=2$ twisted superspace and thus the super Yang-Mills 
action is off-shell $N=2$ twisted SUSY invariant. 
In the final form of the twisted $N=2$ SUSY invariant Yang-Mills 
action, the ghost-related fermions in the 
twisted sector turn into matter fermions by the Dirac-K\"ahler fermion 
mechanism. 

One of the main aims of establishing this twisted superspace formalism 
in four dimensions is to derive $N=4$ twisted SUSY invariant Yang-Mills 
action. As we have seen already from some examples of $N=4$ twisted 
SUSY invariant actions, there are either too many superfluous component 
fields in the action or too higher derivatives in the action. 
The strategy to find further constraints in vector superfield and 
superconnection formulations are not obvious. 
The constraints we imposed for the superconnection formalism can be 
interpreted as the suppression condition of higher spin fields 
in the supermultiplets. We may hope that this type of condition may 
lead a clue to find consistent constraints for superconnections of 
$N=4$ twisted SUSY algebra. It is known that there are some 
nontrivial examples of off-shell usual $N=4$ SUSY invariant Yang-Mills 
action\cite{GSW,Sohnius,SSW} which may have twisted superspace counterpart 
of the formulation.

\vspace{1cm}

\textbf{\Large Acknowledgements}

We would like to thank T. Tsukioka for useful discussions at the 
very early stage of this line of investigation. We also thank 
K. Nagata and J. Saito for fruitful discussions and comments. 
Thanks are also due to A. D'Adda and I. Kanamori for useful discussions. 
This work is supported in part by Japanese Ministry of Education, 
Science, Sports and Culture under the grant number 
13640250 and 13135201.

\newpage
\setcounter{equation}{0}
\setcounter{section}{0}
\renewcommand{\theequation}{\Alph{section}.\arabic{equation}}
\renewcommand{\thesection}{Appendix \Alph{section}}

\section{Four-dimensional $\gamma$-matrix and related formulae}

We use the following Euclidean four-dimensional $\gamma$-matrix throughout this paper. 
The $\gamma$-matrixes satisfy the following Clifford algebra:
\begin{eqnarray}
\{\gamma_\mu ,\gamma_\nu  \} = 2\delta_{\mu\nu}.
\end{eqnarray}
We use the following representation of $\gamma$-matrix:
\[
\gamma^\mu=
\left(
\begin{array}{cc}
0 & i\sigma^\mu \\
i\overline{\sigma}^\mu &0
\end{array}
\right),
\]
where $\sigma^\mu = (\sigma^1,\sigma^2,\sigma^3,\sigma^4)$,
$\overline{\sigma}^\mu = (-\sigma^1,-\sigma^2,-\sigma^3,\sigma^4)$ with 
$\sigma^4=i{\bf 1} _{2 \times 2}$ and $\{\sigma^i\} (i=1,2,3)$ are Pauli matrixes. 
We introduce the following charge conjugation matrix $C$ and $B$-matrix :
\begin{eqnarray}
\gamma_\mu{}^T = C \gamma_\mu C^{-1},& &\qquad C\equiv - \gamma_1  \gamma_3 , 
\qquad C^T = -C, \qquad 
\gamma_5 \equiv \gamma_1  \gamma_2 \gamma_3  \gamma_4, \nonumber \\
\gamma^{\mu\ast} &=& B\gamma^\mu B^{-1}, \qquad B^\ast B=-1 , \qquad 
B\equiv -\gamma^1\gamma^3 .
\end{eqnarray}

We then introduce the following notations:
\begin{eqnarray}
\gamma_{\mu\nu} \equiv \frac{1}{2} {[} \gamma_\mu , \gamma_\nu {]}, \qquad \tilde{\gamma}_\mu \equiv \gamma_\mu \gamma_5.
\end{eqnarray}
We list useful formulae which are used in the algebraic manipulation of $N=4$ twisted 
SUSY algebra:
\begin{eqnarray}
\gamma^\mu \gamma^\nu &=& \gamma^{\mu\nu}+\delta^{\mu\nu},\\
\gamma^\mu \gamma^\nu \gamma^\rho &=& \gamma^{\mu\nu\rho}+ \gamma^\mu \delta^{\nu\rho}-\gamma^\nu \delta^{\mu\rho}+\gamma^\rho \delta^{\mu\nu}\nonumber \\
&=&-\epsilon^{\mu\nu\rho\sigma}\tilde{\gamma}_{\sigma}+ \gamma^\mu \delta^{\nu\rho}-\gamma^\nu \delta^{\mu\rho}+\gamma^\rho \delta^{\mu\nu},\\
\gamma^\mu \gamma^\nu \gamma^\rho \gamma^\sigma &=& \epsilon^{\mu\nu\rho\sigma}\gamma_5 +\gamma^{\mu\nu} \delta^{\rho\sigma} -\gamma^{\mu\rho} \delta^{\nu\sigma}+\gamma^{\mu\sigma} \delta^{\nu\rho}+\gamma^{\nu\rho} \delta^{\mu\sigma}-\gamma^{\nu\sigma} \delta^{\mu\rho} +\gamma^{\rho\sigma} \delta^{\mu\nu}\nonumber \\
& &+\delta^{\mu\nu} \delta^{\rho\sigma} -\delta^{\mu\rho} \delta^{\nu\sigma}+\delta^{\mu\sigma} \delta^{\nu\rho},
\end{eqnarray}

\begin{eqnarray}
\gamma^{\mu\nu} \gamma^\rho &=& -\epsilon^{\mu\nu\rho\sigma} \tilde{\gamma}_\sigma +\gamma^{\mu} \delta^{\nu\rho}-\gamma^\nu \delta^{\mu\rho},\\
\gamma^{\mu\nu} \gamma^{\rho\sigma} &=&\epsilon^{\mu\nu\rho\sigma}\gamma_5
-\gamma^{\mu\rho} \delta^{\nu\sigma} +\gamma^{\mu\sigma} \delta^{\nu\rho}
+\gamma^{\nu\rho} \delta^{\mu\sigma}-\gamma^{\nu\sigma} \delta^{\mu\rho} \nonumber \\
& &\ -\delta^{\mu\rho} \delta^{\nu\sigma} +\delta^{\mu\sigma} \delta^{\nu\rho},\\
\gamma^{\mu\nu} \tilde{\gamma}^{\rho} &=&-\epsilon^{\mu\nu\rho\sigma}\gamma_\sigma +\tilde{\gamma}^{\mu} \delta^{\nu\rho}-\tilde{\gamma}^\nu \delta^{\mu\rho},\\
\gamma^{\mu\nu} \gamma_5 &=& -\frac{1}{2}\epsilon^{\mu\nu\rho\sigma} \gamma_{\rho\sigma},
\end{eqnarray}

\begin{eqnarray}
\gamma^\rho \gamma^{\mu\nu} &=& -\epsilon^{\mu\nu\rho\sigma}\tilde{\gamma}_{\sigma}-\gamma^\mu \delta^{\rho\nu} +\gamma^\nu \delta^{\rho\mu},\\
\tilde{\gamma}^\rho \gamma^{\mu\nu} &=& -\epsilon^{\mu\nu\rho\sigma} \gamma_\sigma -\tilde{\gamma}^\mu \delta^{\rho\nu} +\tilde{\gamma}^\nu \delta^{\rho\mu},\\
\gamma_5 \gamma^{\mu\nu}&=& -\frac{1}{2}\epsilon^{\mu\nu\rho\sigma} \gamma_{\rho\sigma}.
\end{eqnarray}

\setcounter{equation}{0}
\section{ Definition of $\Gamma^{ab,cd,ef,gh,ij,kl}$}

We introduce the following simbol:
\begin{eqnarray}
\theta^{ab}\theta^{cd}\theta^{ef} \theta^{gh}\theta^{ij}\theta^{kl}
\equiv \Gamma^{ab,cd,ef,gh,ij,kl} (\theta^{12}\theta^{13}\theta^{14} 
\theta^{23}\theta^{24}\theta^{34}),
\end{eqnarray}
where $\theta^{ab}$ are twisted superparameters with anti-symmetric tensor suffix. 
These tensor suffixes have six independent degrees of freedom and thus $\Gamma$ 
behaves like a six-dimensional totally anti-symmetric tensor.
Here we denote the tensor suffixes $\{ab\}$ by the upper case $A$. 
We can derive the following formulae:
\begin{eqnarray}
\Gamma^{ABCDEF} \Gamma_{ABCDEF} &=& 6!2^6,  \\
\Gamma^{ABCDEF} \Gamma_{ABCDEG} &=& 5!2^5 \delta^F { }_G , \\
\Gamma^{ABCDEF} \Gamma_{ABCDGH} &=& 4!2^4 (\delta^E { }_G \delta^F { }_H-\delta^E { }_H\delta^F { }_G), \\
\Gamma^{ABCDEF} \Gamma_{ABCGHI} &=& 3!2^3 \{ \delta^D { }_G \delta^E { }_H \delta^F { }_I-\delta^D { }_I \delta^E { }_H \delta^F { }_G+\text{cyclic in GHI }\},\nonumber  \\
& & \\
\Gamma^{ABCDEF} \Gamma_{ABGHIJ} &=&
 \left| 
\begin{array}{cccc} 
\delta_G {}^C& \delta_G {}^D &\delta_G {}^E &\delta_G {}^F\\
\delta_H {}^C& \delta_H {}^D &\delta_H {}^E &\delta_H {}^F\\
\delta_I {}^C& \delta_I {}^D &\delta_I {}^E &\delta_I {}^F\\
\delta_J {}^C& \delta_J {}^D &\delta_J {}^E &\delta_J {}^F
\end{array}
\right|, \\
\Gamma^{ABCDEF} \Gamma_{AGHIJK} &=&
 \left| 
\begin{array}{ccccc} 
\delta_G {}^B& \delta_G {}^C &\delta_G {}^D &\delta_G {}^E&\delta_G {}^F\\
\delta_H {}^B& \delta_H {}^C &\delta_H {}^D &\delta_H {}^E&\delta_H {}^F\\
\delta_I {}^B& \delta_I {}^C &\delta_I {}^D &\delta_I {}^E&\delta_I {}^F\\
\delta_J {}^B& \delta_J {}^C &\delta_J {}^D &\delta_J {}^E&\delta_J {}^F\\
\delta_K {}^B& \delta_K {}^C &\delta_K {}^D &\delta_K {}^E&\delta_K {}^F
\end{array}
\right|, \\
\Gamma^{ABCDEF} \Gamma_{GHIJKL} &=&
 \left| 
\begin{array}{cccccc} 
\delta_G {}^A& \delta_G {}^B &\delta_G {}^C &\delta_G {}^D&\delta_G {}^E&\delta_G {}^F\\
\delta_H {}^A& \delta_H {}^B &\delta_H {}^C &\delta_H {}^D&\delta_H {}^E&\delta_H {}^F\\
\delta_I {}^A& \delta_I {}^B &\delta_I {}^C &\delta_I {}^D&\delta_I {}^E&\delta_I {}^F\\
\delta_J {}^A& \delta_J {}^B &\delta_J {}^C &\delta_J {}^D&\delta_J {}^E&\delta_J {}^F\\
\delta_K {}^A& \delta_K {}^B &\delta_K {}^C &\delta_K {}^D&\delta_K {}^E&\delta_K {}^F\\
\delta_L {}^A& \delta_L {}^B &\delta_L {}^C &\delta_L {}^D&\delta_L {}^E&\delta_L {}^F
\end{array}
\right|.
\end{eqnarray}

\newpage

\renewcommand{\theequation}{C.\arabic{equation}}
\setcounter{equation}{0}
\section{$N=4$ twisted SUSY transformation of chiral multiplets }

We show the full list of the $N=4$ twisted SUSY transformation of chiral multiplets 
given in subsection 3.3.\\

\noindent
{\bf 1 Chiral Multiplets}

\vspace{5mm}

\begin{tabular}{|l||c|c|c|c|}

\hline

$\psi $ 
& $ S $ & $S_\mu$   & $\tilde{S_{\mu}}$ & $\tilde{S}$ \\

\hline

$\psi^0$ 
& $0$ 
& $\psi^1 _{\mu,}$ 

& $ \psi^0  _{,\mu}$ 
& $0$ \\

\hline  

$\psi^0 _{,a} $ 
& $0$ 
& $ -\psi^1 _{\mu,a}$

& $ -\psi^0 { }_{,\mu a}$ 
& $ -i\partial_a \psi^0$ \\

\hline

$\psi^0 _{,ab} $ 
& $0$ 
& $ \psi^1 _{\mu,ab}$

 & $\epsilon _{ab\mu \nu} \tilde{\psi^0} { }^{,\nu} $  
 & $ i(\partial_a \psi^0 { }_{,b} -\partial_b \psi^0 { }_{,a} ) $\\

\hline

$\tilde{\psi^0} _{,a} $ 
 & $0$ 
 & $-\tilde{\psi^1} _{\mu,a} $

 & $ \delta_{\mu,a} \tilde{\psi^0}$ 
 & $ \frac{i}{2} \epsilon_{a} { }^{bcd} \partial_b \psi^0 { }_{,cd}$ \\

\hline

$ \tilde{\psi^0} $ 
 & $0$ 
 & $\tilde{\psi^1} _{\mu,}$

 & $0$ 
 & $ -i \partial^a \tilde{\psi^0} { }_{,a}$ \\

\hline
\end{tabular}

\vspace{1cm}
\begin{tabular}{|l||c|c|c|c|}
\hline

$\psi$ 
& $ S $ & $S_\mu$  & $\tilde{S_{\mu}}$ & $\tilde{S}$ \\

\hline

$\psi^1 _{m,}$ 
& $-i \partial_m \psi^0 $ 
& $-\psi^2 _{\mu m,}$ 

& $ \psi^1  _{m,\mu}$ 
& $0$ \\

\hline  

$\psi^1 _{m,a} $ 
& $i\partial_m \psi^0 _{,a}$ 
& $\psi^2 _{\mu m,a}$

& $ -\psi^1 { }_{m,\mu a}$ 
& $ -i\partial_a \psi^1 _{m,}$ \\

\hline

$\psi^1 _{m,ab} $ 
& $ -i\partial_m \psi^0 _{,ab} $ 
& $- \psi^2 _{\mu m,ab}$

 & $\epsilon _{ab\mu} {}^{\nu} \tilde{\psi^1} { }_{m,\nu} $  
 & $ i (\partial_a \psi^1 { }_{m,b} -\partial_b \psi^1 { }_{m,a} ) $\\

\hline

$\tilde{\psi^1} _{m,a} $ 
 & $i\partial_m \tilde{\psi^0} _{,a}$  
 & $\tilde{\psi^2} _{\mu m,a} $
 
 & $ \delta_{\mu,a} \tilde{\psi^1} _m$ 
 & $ \frac{i}{2} \epsilon_{a} { }^{bcd} \partial_b \psi^1 { }_{m,cd}$ \\

\hline

$ \tilde{\psi^1}_{m,} $ 
 & $-i\partial_m \tilde{\psi^0}$ 
 & $-\tilde{\psi^2} _{\mu m,}$

 & $0$ 
 & $ -i \partial^a \tilde{\psi^1} { }_{m,a}$ \\

\hline
\end{tabular}

\vspace{1cm}

\begin{tabular}{|l||c|c|c|c|}

\hline

$\psi$ 
& $ S $ & $S_\mu$  & $\tilde{S_{\mu}}$ & $\tilde{S}$ \\

\hline

$\psi^2 _{mn,}$ 
& $ i ( \partial_m \psi^1 _{n,} -  \partial_n \psi^1 _{m,} )$ 
& $ \epsilon_{mn\mu} {}^{\nu} \psi^3 _{\nu,}$ 

& $ \psi^2  _{mn,\mu}$ 
& $0$ \\

\hline  

$\psi^2 _{mn,a} $ 
& $-i ( \partial_m \psi^1 _{n,a} -  \partial_n \psi^1 _{m,a} )$ 
& $-\epsilon_{mn\mu} {}^{\nu} \psi^3 _{\nu,a}$

& $ -\psi^2 { }_{mn,\mu a}$ 
& $ -i\partial_a \psi^2 _{mn,}$ \\

\hline

$\psi^2 _{mn,ab} $ 
& $i ( \partial_m \psi^1 _{n,ab} -  \partial_n \psi^1 _{m,ab} )$ 
& $ \epsilon_{mn\mu} {}^{\nu} \psi^3 _{\nu,ab}$

 & $\epsilon _{ab\mu} {}^{\nu} \tilde{\psi^2} { }_{mn,\nu} $  
 & $ i \delta_{ab,cd} \partial^c \psi^2 { }_{mn,} {}^d$
\\

\hline

$\tilde{\psi^2} _{mn,a} $ 
 &$-i ( \partial_m \tilde{\psi^1} _{n,a} -  \partial_n \tilde{\psi^1} _{m,a} )$  
 & $ -\epsilon_{mn\mu} {}^{\nu} \tilde{\psi^3} _{\nu,a}$

 & $ \delta_{\mu,a} \tilde{\psi^2} _{mn,}$ 
 & $ \frac{i}{2} \epsilon_{a} { }^{bcd} \partial_b \psi^2 { }_{mn,cd}$ \\

\hline

$ \tilde{\psi^2}_{mn,} $ 
 & $i ( \partial_m \tilde{\psi^1} _{n,} -  \partial_n \tilde{\psi^1} _{m,} )$ 
 & $\epsilon_{mn\mu} {}^{\nu} \tilde{\psi^3} _{\nu,}$
 
 & $0$ 
 & $ -i \partial^a \tilde{\psi^2} { }_{mn,a}$ \\

\hline
\end{tabular}
\vspace{1cm}

\begin{tabular}{|l||c|c|c|c|}

\hline

$\psi$ 
& $ S $ & $S_\mu$  & $\tilde{S_{\mu}}$ & $\tilde{S}$ \\

\hline

$\psi^3 _{m,}$ 
& $ \frac{i}{2}\epsilon _{m} {}^{\rho\mu\nu} \partial _\rho \psi^2 _{\mu\nu,}$ 
& $ \delta_{m,\mu} \psi^4 $ 

& $ \psi^3  _{m,\mu}$ 
& $0$ \\

\hline  

$\psi^3 _{m,a} $ 
& $-\frac{i}{2}\epsilon _{m} {}^{\rho\mu\nu} \partial _\rho \psi^2 _{\mu\nu,a}$ 
& $-\delta_{m,\mu} \psi^4 _{,a}$

& $ -\psi^3 { }_{m,\mu a}$ 
& $ -i\partial_a \psi^3 _{m,}$ \\

\hline

$\psi^3 _{m,ab} $ 
& $\frac{i}{2}\epsilon _{m} {}^{\rho\mu\nu} \partial _\rho \psi^2 _{\mu\nu,ab}$ 
& $\delta_{m,\mu} \psi^4 _{,ab}$

 & $\epsilon _{ab\mu} {}^{\nu} \tilde{\psi^3} { }_{m,\nu} $  
 & $ i (\partial_a \psi^3 { }_{m,b} -\partial_b \psi^3 { }_{m,a} ) $\\

\hline

$\tilde{\psi^3} _{m,a} $ 
 &$-\frac{i}{2}\epsilon _{m} {}^{\rho\mu\nu} \partial _\rho \tilde{\psi^2} _{\mu\nu,a}$  
 & $ -\delta_{m,\mu} \tilde{\psi^4 } _{,a}$

 & $ \delta_{\mu,a} \tilde{\psi^3} _{m,}$ 
 & $ \frac{i}{2} \epsilon_{a} { }^{bcd} \partial_b \psi^3 { }_{m,cd}$ \\

\hline

$ \tilde{\psi^3}_{m,} $ 
 & $\frac{i}{2}\epsilon _{m} {}^{\rho\mu\nu} \partial _\rho \tilde{\psi^2} _{\mu\nu,}$ 
 & $\delta_{m,\mu} \tilde{\psi^4 }$
 
 & $0$ 
 & $ -i \partial^a \tilde{\psi^3} { }_{m,a}$ \\

\hline
\end{tabular}

\vspace{5mm}

\begin{tabular}{|l||c|c|c|c|}

\hline

$\psi$ 
& $ S $ & $S_\mu$  & $\tilde{S_{\mu}}$ & $\tilde{S}$ \\

\hline

$\psi^4 $ 
& $ -i\partial^\mu \psi^3_{\mu,}$ 
& $0 $ 

& $ \psi^4  _{,\mu}$ 
& $0$ \\

\hline  

$\psi^4 _{,a} $ 
& $i\partial^\mu \psi^3_{\mu,a}$
& $0$

& $ -\psi^4 { }_{,\mu a}$ 
& $ -i\partial_a \psi^4 $ \\

\hline

$\psi^4 _{,ab} $ 
& $-i\partial^\mu \psi^3_{\mu,ab}$ 
& $0$

&$\epsilon _{ab\mu} {}^{\nu} \tilde{\psi^4} { }_{,\nu} $ 
 & $ i (\partial_a \psi^4 { }_{,b} -\partial_b \psi^4 { }_{,a} ) $\\

\hline

$\tilde{\psi^4} _{,a} $ 
 &$i\partial^\mu \tilde{\psi^3}_{\mu,a}$  
 & $0$

& $ \delta_{\mu,a} \tilde{\psi^4} $
 & $ \frac{i}{2} \epsilon_{a} { }^{bcd} \partial_b \psi^4 { }_{,cd}$ \\

\hline

$ \tilde{\psi^4} $ 
 & $-i\partial^\mu \tilde{\psi^3}_{\mu,}$ 
 & $0$

 & $0$ 
 & $ -i \partial^a \tilde{\psi^4} { }_{,a}$ \\

\hline
\end{tabular}

\vspace{10mm}

\begin{tabular}{|l||c|}

\hline

$\psi$ & $S_{\mu\nu}$ \\

\hline

$\psi^0$ & $0$ \\

\hline  

$\psi^0 _{,a} $ & $-i\epsilon_{a\mu\nu\rho} \partial^\rho \psi^0 $ \\

\hline

$\psi^0 _{,ab} $ 
& $i(\epsilon_{a\mu\nu\rho} \partial^\rho \psi^0 _{,b}-\epsilon_{b\mu\nu\rho} \partial^\rho \psi^0 _{,a})$ \\

\hline

$\tilde{\psi^0} _{,a} $ 
 & $\frac{i}{2} \epsilon_{a} { }^{bcd}\epsilon_{b\mu\nu\rho} \partial^\rho \psi^0 { }_{,cd}$ \\

\hline

$ \tilde{\psi^0} $ 

 & $i\epsilon_{\mu\nu\rho\sigma}\partial^{\rho}\tilde{\psi}^{0,\sigma}$ \\

\hline
\end{tabular}

\vspace{10mm}

\begin{tabular}{|l||c|}

\hline

$\psi$ & $S_{\mu\nu}$  \\

\hline

$\psi^1 _{m,}$  
& $i(\delta_{m,\mu} \partial_\nu -\delta_{m,\nu} \partial_\mu ) \psi^0$ \\

\hline  

$\psi^1 _{m,a} $ 
& $-i(\delta_{m,\mu} \partial_\nu -\delta_{m,\nu} \partial_\mu ) \psi^0 _{,a}
-i\epsilon_{a\mu\nu\rho} \partial^{\rho} \psi^1 _{m,}$  \\

\hline

$\psi^1 _{m,ab} $ 
& $i(\delta_{m,\mu} \partial_\nu -\delta_{m,\nu} \partial_\mu ) \psi^0 _{,ab}
+i(\epsilon_{a\mu\nu\rho} \partial^{\rho} \psi^1 _{m,b} -\epsilon_{b\mu\nu\rho} \partial^{\rho} \psi^1 _{m,a} )$ \\

\hline

$\tilde{\psi^1} _{m,a} $ 
 & $-i(\delta_{m,\mu} \partial_\nu -\delta_{m,\nu} \partial_\mu ) \tilde{\psi^0} _{,a} +\frac{i}{2} \epsilon_a {}^{\alpha\beta\gamma} \epsilon_{\alpha\mu\nu\rho} \partial^{\rho} \psi^1 _{m,\beta\gamma}$\\

\hline

$ \tilde{\psi^1}_{m,} $ 
 & $i(\delta_{m,\mu} \partial_\nu -\delta_{m,\nu} \partial_\mu ) \tilde{\psi^0}+i\epsilon_{\mu\nu\rho\sigma}\partial^{\rho}\tilde{\psi}^1 { }_{m,} { }^{\sigma} $ \\

\hline
\end{tabular}

\vspace{1cm}

\begin{tabular}{|l||c|}

\hline

$\psi$ & $S_{\mu\nu}$  \\

\hline

$\psi^2 _{mn,}$ 
& $-i(\delta_{m,\mu} \partial_\nu -\delta_{m,\nu} \partial_\mu ) \psi^1 _{n,}
+i(\delta_{n,\mu} \partial_\nu -\delta_{n,\nu} \partial_\mu ) \psi^1 _{m,}$ \\

\hline  

$\psi^2 _{mn,a} $ 
& $i(\delta_{m,\mu} \partial_\nu -\delta_{m,\nu} \partial_\mu ) \psi^1 _{n,a}
-i(\delta_{n,\mu} \partial_\nu -\delta_{n,\nu} \partial_\mu ) \psi^1 _{m,a}  
-i\epsilon_{a\mu\nu\rho}\partial^{\rho}\psi^2 _{mn,}$ \\

\hline

$\psi^2 _{mn,ab} $ 
& $-i\delta_{mk,\mu\nu}\partial^k 
\psi^1 _{n,ab}
+i\delta_{nk,\mu\nu}\partial^k 
 \psi^1 _{m,ab}
+i\epsilon_{a\mu\nu\rho}\partial^{\rho}\psi^2 { }_{mn,b} -i\epsilon_{b\mu\nu\rho}\partial^{\rho}\psi^2 { }_{mn,a}$ \\

\hline

$\tilde{\psi^2} _{mn,a} $ 
 & $i\delta_{mk,\mu\nu} \partial^k 
\tilde{\psi^1} _{n,a}
-i\delta_{nk,\mu\nu} \partial^k 
\tilde{ \psi^1} _{m,a}
+\frac{i}{2}\epsilon_a { }^{\alpha\beta\gamma} \epsilon_{\alpha\mu\nu\rho}\partial^{\rho}\psi^2 _{mn,\beta\gamma}$ \\

\hline

$ \tilde{\psi^2}_{mn,} $ 
 & $-i(\delta_{m,\mu} \partial_\nu -\delta_{m,\nu} \partial_\mu ) \tilde{\psi^1} _{n,}
+i(\delta_{n,\mu} \partial_\nu -\delta_{n,\nu} \partial_\mu )\tilde{ \psi^1} _{m,}
+i\epsilon_{\mu\nu\alpha\beta} \partial^{\alpha}\tilde{\psi}^2 _{mn,} { }^{\beta}$ \\

\hline
\end{tabular}
\vspace{1cm}

\begin{tabular}{|l||c|}

\hline

$\psi$  & $S_{\mu\nu}$ \\

\hline

$\psi^3 _{m,}$ 
& $-\frac{i}{2} (\epsilon_{m} {}^{\alpha\beta} {}_{\mu} \partial_{\nu} - \epsilon_{m} {}^{\alpha\beta} {}_{\nu} \partial_{\mu}) \psi^2 _{\alpha\beta,}$ \\

\hline  

$\psi^3 _{m,a} $ 
&  $\frac{i}{2} (\epsilon_{m} {}^{\alpha\beta} {}_{\mu} \partial_{\nu} - \epsilon_{m} {}^{\alpha\beta} {}_{\nu} \partial_{\mu}) \psi^2 _{\alpha\beta,a}-i\epsilon_{a\mu\nu\rho}\partial^{\rho}\psi^3 _{m,}$ \\

\hline

$\psi^3 _{m,ab} $ 
& $-\frac{i}{2} (\epsilon_{m} {}^{\alpha\beta} {}_{\mu} \partial_{\nu} - \epsilon_{m} {}^{\alpha\beta} {}_{\nu} \partial_{\mu}) \psi^2 _{\alpha\beta,ab}+i\epsilon_{a\mu\nu\rho}\partial^{\rho}\psi^3 { }_{m,b} -i\epsilon_{b\mu\nu\rho}\partial^{\rho}\psi^3 { }_{m,a}$ \\

\hline

$\tilde{\psi^3} _{m,a} $ 
 & $\frac{i}{2} (\epsilon_{m} {}^{\alpha\beta} {}_{\mu} \partial_{\nu} - \epsilon_{m} {}^{\alpha\beta} {}_{\nu} \partial_{\mu}) \tilde{\psi^2} _{\alpha\beta,a}+\frac{i}{2}\epsilon_a { }^{\alpha\beta\gamma} \epsilon_{\alpha\mu\nu\rho}\partial^{\rho}\psi^3 _{m,\beta\gamma}$ \\

\hline

$ \tilde{\psi^3}_{m,} $ 
 & $-\frac{i}{2} (\epsilon_{m} {}^{\alpha\beta} {}_{\mu} \partial_{\nu} - \epsilon_{m} {}^{\alpha\beta} {}_{\nu} \partial_{\mu}) \tilde{\psi^2} _{\alpha\beta,}+i\epsilon_{\mu\nu\alpha\beta} \partial^{\alpha}\tilde{\psi}^3 _{m,} { }^{\beta}$ \\

\hline
\end{tabular}

\vspace{5mm}

\begin{tabular}{|l||c|}

\hline

$\psi$ & $S_{\mu\nu}$ \\

\hline

$\psi^4 $ 
& $-i(\partial_\mu \psi^3 _{\nu,} -\partial_\nu \psi^3 _{\mu,} )$ \\

\hline  

$\psi^4 _{,a} $ 
&  $i(\partial_\mu \psi^3 _{\nu,a} -\partial_\nu \psi^3 _{\mu,a} )-i\epsilon_{a\mu\nu\rho}\partial^{\rho}\psi^4$ \\

\hline

$\psi^4 _{,ab} $ 
& $-i(\partial_\mu \psi^3 _{\nu,ab} -\partial_\nu \psi^3 _{\mu,ab} )+i\epsilon_{a\mu\nu\rho}\partial^{\rho}\psi^4 { }_{,b} -i\epsilon_{b\mu\nu\rho}\partial^{\rho}\psi^4 { }_{,a}$ \\

\hline

$\tilde{\psi^4} _{,a} $ 
 & $i(\partial_\mu \tilde{\psi^3} _{\nu,a} -\partial_\nu \tilde{\psi^3} _{\mu,a} )+\frac{i}{2}\epsilon_a { }^{\alpha\beta\gamma} \epsilon_{\alpha\mu\nu\rho}\partial^{\rho}\psi^4 _{,\beta\gamma}$ \\

\hline

$ \tilde{\psi^4} $ 
 & $-i(\partial_\mu \tilde{\psi^3} _{\nu,} -\partial_\nu \tilde{\psi^3} _{\mu,} )+i\epsilon_{\mu\nu\alpha\beta} \partial^{\alpha}\tilde{\psi}^4 { }^{\beta}$\\

\hline
\end{tabular}

\vspace{15mm}


{\bf 2. Anti-chiral Multiplet}

\vspace{10mm}

\begin{tabular}{|l||c|c|}

\hline

$\phi$ & $ S $ & $S_\mu$   \\

\hline

$\phi^0$ & $\phi^1$ & $0$  \\

\hline  

$\phi^0 _{ab} $ & $ - \phi^1 _{ab}$ & $ i\mathcal{D}_{\mu,ab} \phi^0$
 \\

\hline

$\phi^0 _{ab,cd} $ & $ \phi^1 _{ab,cd}$ & $ -i(\mathcal{D}_{\mu,ab} \phi^0 _{cd} - \mathcal{D}_{\mu,cd} \phi^0 _{ab}) $
 
  \\

\hline

$\phi^0 _{ab,cd,ef} $ 
 & $ -\phi^1 _{ab,cd,ef}$ 
 & $ i(\mathcal{D}_{\mu,ab} \phi^0 _{cd,ef} + \mathcal{D}_{\mu,ef} \phi^0 _{ab,cd}+ \mathcal{D}_{\mu,cd} \phi^0 _{ef,ab}) $
 
  \\

\hline

$ \tilde{\phi^0} _{ab,cd} $ 
 & $ \tilde{\phi^1} _{ab,cd}$ 
 & $ -i\frac{1}{3!2^4}\Gamma_{ab,cd} {   }^{ef,gh,ij,kl} \mathcal{D}_{\mu,ef} \phi^0 _{gh,ij,kl} $
  \\

\hline

$ \tilde{\phi^0} _{ab} $ 
 & $ -\tilde{\phi^1} _{ab}$ 
 & $ -\frac{i}{2}\mathcal{D}_{\mu,cd} \tilde{\phi}^0 _{ab} {  }^{cd} $
\\

\hline

$ \tilde{\phi^0} $ 
 & $ \tilde{\phi^1} $ 
 & $ \frac{i}{2}\mathcal{D}_{\mu,ab} \tilde{\phi}^{0 \ ab} $
  \\

\hline
\end{tabular}

\vspace{10mm}

\begin{tabular}{|l||c|c|}

\hline

$\phi$ & $ S $ & $S_\mu$    \\

\hline

$\phi^1$ & $0$ & $-i\partial_\mu \phi^0 $  \\

\hline  

$\phi^1 _{ab} $ & $0$ & $i\partial_\mu \phi^0 { }_{ab} +i\mathcal{D}_{\mu,ab} \phi^1$
 \\

\hline

$\phi^1 _{ab,cd} $ & $0$ & $-i\partial_\mu \phi^0 { }_{ab,cd} -i(\mathcal{D}_{\mu,ab} \phi^1 _{cd} - \mathcal{D}_{\mu,cd} \phi^0 _{ab}) $
  \\

\hline

$\phi^1 _{ab,cd,ef} $ 
 & $0$  
 & $i\partial_\mu \phi^0 { }_{ab,cd,ef} +i(\mathcal{D}_{\mu,ab} \phi^1 _{cd,ef} + \mathcal{D}_{\mu,ef} \phi^1 _{ab,cd}+ \mathcal{D}_{\mu,cd} \phi^1 _{ef,ab}) $
  \\

\hline

$ \tilde{\phi^1} _{ab,cd} $ 
 & $0$ 
 & $-i\partial_\mu \tilde{\phi^0} _{ab,cd} -i\frac{1}{3!2^4}\Gamma_{ab,cd} {   }^{ef,gh,ij,kl} \mathcal{D}_{\mu,ef} \phi^1 _{gh,ij,kl} $
 \\

\hline

$ \tilde{\phi^1} _{ab} $ 
 & $0$ 
 & $ i\partial_\mu \tilde{\phi^0}_{ab} -\frac{i}{2}\mathcal{D}_{\mu,cd} \tilde{\phi}^1 _{ab} {  }^{cd} $
  \\

\hline

$ \tilde{\phi^1} $ 
 & $0$
 & $-i\partial_\mu \tilde{\phi^0} + \frac{i}{2}\mathcal{D}_{\mu,ab} \tilde{\phi}^{1 \ ab} $
  \\

\hline

\end{tabular}

\vspace{10mm}

\begin{tabular}{|l||c|c|}

\hline

$\phi$ & $ S $ & $S_\mu$   \\

\hline

$\phi^2$ &  $-\phi^3$ & $0$  \\

\hline  

$\phi^2 _{ab} $ & $ \phi^3 _{ab}$ & $ i\mathcal{D}_{\mu,ab} \phi^2$
 \\

\hline

$\phi^2 _{ab,cd} $ & $ -\phi^3 _{ab,cd}$ & $ -i(\mathcal{D}_{\mu,ab} \phi^2 _{cd} - \mathcal{D}_{\mu,cd} \phi^2 _{ab}) $
  \\

\hline

$\phi^2 _{ab,cd,ef} $ 
 & $ \phi^3 _{ab,cd,ef}$ 
 & $ i(\mathcal{D}_{\mu,ab} \phi^2 _{cd,ef} + \mathcal{D}_{\mu,ef} \phi^2 _{ab,cd}+ \mathcal{D}_{\mu,cd} \phi^2 _{ef,ab}) $
  \\

\hline

$ \tilde{\phi^2} _{ab,cd} $ 
 & $ -\tilde{\phi^3} _{ab,cd}$ 
 & $ -i\frac{1}{3!2^4}\Gamma_{ab,cd} {   }^{ef,gh,ij,kl} \mathcal{D}_{\mu,ef} \phi^2 _{gh,ij,kl} $
  \\

\hline

$ \tilde{\phi^2} _{ab} $ 
 & $ \tilde{\phi^3} _{ab}$ 
 & $ -\frac{i}{2}\mathcal{D}_{\mu,cd} \tilde{\phi}^2 _{ab} {  }^{cd} $
  \\

\hline

$ \tilde{\phi^2} $ 
 & $ -\tilde{\phi^3} $ 
 & $ \frac{i}{2}\mathcal{D}_{\mu,ab} \tilde{\phi}^{2 \ ab} $
 \\

\hline
\end{tabular}

\vspace{10mm}

\begin{tabular}{|l||c|c|}

\hline

$\phi$ & $ S $ & $S_\mu$   \\

\hline

$\phi^3$ & $0$ & $i\partial_\mu \phi^2 $   \\

\hline  

$\phi^3 _{ab} $ & $0$ & $-i\partial_\mu \phi^2 { }_{ab}+ i\mathcal{D}_{\mu,ab} \phi^3$
  \\

\hline

$\phi^3 _{ab,cd} $ & $0$ & $i\partial_\mu \phi^2 { }_{ab,cd} -i(\mathcal{D}_{\mu,ab} \phi^3 _{cd} - \mathcal{D}_{\mu,cd} \phi^3 _{ab}) $
  \\

\hline

$\phi^3 _{ab,cd,ef} $ 
 & $0$  
 & $-i\partial_\mu \phi^2 { }_{ab,cd,ef} +i(\mathcal{D}_{\mu,ab} \phi^3 _{cd,ef} + \mathcal{D}_{\mu,ef} \phi^3 _{ab,cd}+ \mathcal{D}_{\mu,cd} \phi^3 _{ef,ab}) $
  \\

\hline

$ \tilde{\phi^3} _{ab,cd} $ 
 & $0$ 
 & $i\partial_\mu \tilde{\phi^2}_{ab,cd} -i\frac{1}{3!2^4}\Gamma_{ab,cd} {   }^{ef,gh,ij,kl} \mathcal{D}_{\mu,ef} \phi^3 _{gh,ij,kl} $
  \\

\hline

$ \tilde{\phi^3} _{ab} $ 
 & $0$ 
 & $ -i\partial_\mu \tilde{\phi^2} _{ab} -\frac{i}{2}\mathcal{D}_{\mu,cd} \tilde{\phi}^3 _{ab} {  }^{cd} $
  \\

\hline

$ \tilde{\phi^3} $ 
 & $0$
 & $i\partial_\mu \tilde{\phi^2} + \frac{i}{2}\mathcal{D}_{\mu,ab} \tilde{\phi}^{3 \ ab} $
  \\

\hline

\end{tabular}

\vspace{10mm}

\begin{tabular}{|l||c|}

\hline

$\phi$ & $S_{\mu\nu}$ \\

\hline

$\phi^0$ & $\phi^0  _{\mu\nu}$ \\

\hline  

$\phi^0 _{ab} $ & $-\phi^0  _{\mu\nu ,ab}$\\

\hline

$\phi^0 _{ab,cd} $ & $\phi^0  _{\mu\nu ,ab,cd}$
  \\

\hline

$\phi^0 _{ab,cd,ef} $ 
 & $-\frac{1}{2^3} \Gamma_{\mu\nu,ab,cd,ef,gh,ij} \tilde{\phi}^{0\ gh,ij}$
  \\

\hline

$ \tilde{\phi^0} _{ab,cd} $ 
 & $ \delta_{\mu\nu,ab} \tilde{\phi^0}_{cd} -\delta_{\mu\nu,cd} \tilde{\phi^0}_{ab}$
  \\

\hline

$ \tilde{\phi^0} _{ab} $ 
 & $ \delta_{\mu\nu,ab} \tilde{\phi^0} $
\\

\hline

$ \tilde{\phi^0} $ 
 & $ 0$
  \\

\hline
\end{tabular}
\hspace{10mm}
\begin{tabular}{|l||c|}

\hline

$\phi$ & $S_{\mu\nu}$  \\

\hline

$\phi^1$ & $\phi^1  _{\mu\nu}$ \\

\hline  

$\phi^1 _{ab} $ & $-\phi^1  _{\mu\nu ,ab}$  \\

\hline

$\phi^1 _{ab,cd} $ & $\phi^1  _{\mu\nu ,ab,cd}$
  \\

\hline

$\phi^1 _{ab,cd,ef} $ 
& $-\frac{1}{2^3} \Gamma_{\mu\nu,ab,cd,ef,gh,ij} \tilde{\phi}^{1\ gh,ij}$
  \\

\hline

$ \tilde{\phi^1} _{ab,cd} $ 
 & $ \delta_{\mu\nu,ab} \tilde{\phi^1}_{cd} -\delta_{\mu\nu,cd} \tilde{\phi^1}_{ab}$
 \\

\hline

$ \tilde{\phi^1} _{ab} $ 
 & $ \delta_{\mu\nu,ab} \tilde{\phi^1} $
  \\

\hline

$ \tilde{\phi^1} $ 
 & $ 0$
  \\

\hline

\end{tabular}

\vspace{10mm}

\begin{tabular}{|l||c|}

\hline

$\phi$ & $S_{\mu\nu}$  \\

\hline

$\phi^2$ & $\phi^2  _{\mu\nu}$  \\

\hline  

$\phi^2 _{ab} $ & $-\phi^2  _{\mu\nu ,ab}$ \\

\hline

$\phi^2 _{ab,cd} $ & $\phi^2  _{\mu\nu ,ab,cd}$
  \\

\hline

$\phi^2 _{ab,cd,ef} $ 
 & $-\frac{1}{2^3} \Gamma_{\mu\nu,ab,cd,ef,gh,ij} \tilde{\phi}^{2\ gh,ij}$
  \\

\hline

$ \tilde{\phi^2} _{ab,cd} $ 
 & $ \delta_{\mu\nu,ab} \tilde{\phi^2}_{cd} -\delta_{\mu\nu,cd} \tilde{\phi^2}_{ab}$
  \\

\hline

$ \tilde{\phi^2} _{ab} $ 
 & $ \delta_{\mu\nu,ab} \tilde{\phi^2} $
  \\

\hline

$ \tilde{\phi^2} $ 
 & $ 0$
 \\

\hline
\end{tabular}
\hspace{5mm}
\begin{tabular}{|l||c|}

\hline

$\phi$ & $S_{\mu\nu}$  \\

\hline

$\phi^3$ & $\phi^3  _{\mu\nu}$  \\

\hline  

$\phi^3 _{ab} $ & $-\phi^3  _{\mu\nu ,ab}$  \\

\hline

$\phi^3 _{ab,cd} $ & $\phi^3  _{\mu\nu ,ab,cd}$
  \\

\hline

$\phi^3 _{ab,cd,ef} $ 
 & $-\frac{1}{2^3} \Gamma_{\mu\nu,ab,cd,ef,gh,ij} \tilde{\phi}^{3\ gh,ij}$
  \\

\hline

$ \tilde{\phi^3} _{ab,cd} $ 
 & $ \delta_{\mu\nu,ab} \tilde{\phi^3}_{cd} -\delta_{\mu\nu,cd} \tilde{\phi^3}_{ab}$
  \\

\hline

$ \tilde{\phi^3} _{ab} $ 
 & $ \delta_{\mu\nu,ab} \tilde{\phi^3} $
  \\

\hline

$ \tilde{\phi^3} $ 
 & $ 0$
  \\

\hline

\end{tabular}

\vspace{10mm}

\begin{tabular}{|l||c|c|}

\hline

$\phi$ &  $\tilde{S_{\mu}}$ & $\tilde{S}$ \\

\hline

$\phi^0$ & $ 0$ & $ \phi^2$ \\

\hline  

$\phi^0 _{ab} $   & $ -i\epsilon_{ab\mu\nu} \partial^\nu \phi^0 $ 
 & $ -\phi^2 _{ab}$ \\

\hline

$\phi^0 _{ab,cd} $ 
 & $i(\epsilon_{ab\mu\nu} \partial^\nu \phi^0 _{cd} -\epsilon_{cd\mu\nu} \partial^\nu \phi^0 _{ab}) $ 
 & $ \phi^2 _{ab,cd}$ \\

\hline

$\phi^0 _{ab,cd,ef} $ 
 & $-i(\epsilon_{ab\mu\nu} \partial^\nu \phi^0 _{cd,ef}
  +\epsilon_{ef\mu\nu} \partial^\nu \phi^0 _{ab,cd}
  +\epsilon_{cd\mu\nu} \partial^\nu \phi^0 _{ef,ab}) $ 
 & $ -\phi^2 _{ab,cd,ef}$ \\

\hline

$ \tilde{\phi^0} _{ab,cd} $ 
 & $\frac{i}{3!2^4} \Gamma_{ab,cd} {   }^{ef,gh,ij,kl} \epsilon_{ef\mu\nu}\partial^\nu
 \phi^0 _{gh,ij,kl} $ 
 & $ \tilde{\phi^2} _{ab,cd}$ \\

\hline

$ \tilde{\phi^0} _{ab} $ 
 & $\frac{i}{2} \epsilon_{cd \mu\nu}\partial^\nu \tilde{ \phi }^0 _{ab} {  }^{cd} $ 
 & $ -\tilde{\phi^2} _{ab}$ \\

\hline

$ \tilde{\phi^0} $ 
 & $-\frac{i}{2} \epsilon_{ab \mu\nu}\partial^\nu \tilde{ \phi}^{0 \ ab}  $ 
 & $ \tilde{\phi^2}$ \\

\hline
\end{tabular}

\vspace{10mm}

\begin{tabular}{|l||c|c|}

\hline

$\phi$ & $\tilde{S_{\mu}}$ & $\tilde{S}$ \\

\hline

$\phi^1$ &  $ 0$ & $ \phi^3$ \\

\hline  

$\phi^1 _{ab} $  & $ -i\epsilon_{ab\mu\nu} \partial^\nu \phi^1 $ 
 & $ -\phi^3 _{ab}$ \\

\hline

$\phi^1 _{ab,cd} $ 
 & $i(\epsilon_{ab\mu\nu} \partial^\nu \phi^1 _{cd} -\epsilon_{cd\mu\nu} \partial^\nu \phi^1 _{ab}) $ 
 & $ \phi^3 _{ab,cd}$ \\

\hline

$\phi^1 _{ab,cd,ef} $ 
 
 & $-i(\epsilon_{ab\mu\nu} \partial^\nu \phi^1 _{cd,ef}
  +\epsilon_{ef\mu\nu} \partial^\nu \phi^1 _{ab,cd}
  +\epsilon_{cd\mu\nu} \partial^\nu \phi^1 _{ef,ab}) $ 
 & $ -\phi^3 _{ab,cd,ef}$ \\

\hline

$ \tilde{\phi^1} _{ab,cd} $ 

 & $\frac{i}{3!2^4} \Gamma_{ab,cd} {   }^{ef,gh,ij,kl} \epsilon_{ef\mu\nu}\partial^\nu
 \phi^1 _{gh,ij,kl} $ 
 & $ \tilde{\phi^3} _{ab,cd}$ \\

\hline

$ \tilde{\phi^1} _{ab} $ 

 & $\frac{i}{2} \epsilon_{cd \mu\nu}\partial^\nu \tilde{\phi^1} _{ab} {  }^{cd} $ 
 & $ -\tilde{\phi^3} _{ab}$ \\

\hline

$ \tilde{\phi^1} $ 

 & $-\frac{i}{2} \epsilon_{ab \mu\nu}\partial^\nu \tilde{\phi}^{1 \ ab} $ 
 & $ \tilde{\phi^3}$ \\

\hline

\end{tabular}

\vspace{10mm}

\begin{tabular}{|l||c|c|}

\hline

$\phi$  & $\tilde{S_{\mu}}$ & $\tilde{S}$ \\

\hline

$\phi^2$ & $-i\partial_\mu \phi^0$ & $0$ \\

\hline  

$\phi^2 _{ab} $ & $i\partial_\mu \phi^0 _{ab} -i\epsilon_{ab\mu\nu} \partial^\nu \phi^2 $ 
 & $0$ \\

\hline

$\phi^2 _{ab,cd} $ 
 & $-i\partial_\mu \phi^0 _{ab.cd} +i(\epsilon_{ab\mu\nu} \partial^\nu \phi^2 _{cd} -\epsilon_{cd\mu\nu} \partial^\nu \phi^2 _{ab}) $ 
 & $0$ \\

\hline

$\phi^2 _{ab,cd,ef} $ 

 & $i\partial_\mu \phi^0 _{ab,cd,ef}-i(\epsilon_{ab\mu\nu} \partial^\nu \phi^2 _{cd,ef}
  +\epsilon_{ef\mu\nu} \partial^\nu \phi^2 _{ab,cd}
  +\epsilon_{cd\mu\nu} \partial^\nu \phi^2 _{ef,ab}) $ 
 & $0$ \\

\hline

$ \tilde{\phi^2} _{ab,cd} $ 

 &$-i\partial_\mu \tilde{\phi^0}_{ab,cd} +\frac{i}{3!2^4} \Gamma_{ab,cd} {   }^{ef,gh,ij,kl} \epsilon_{ef\mu\nu}\partial^\nu \phi^2 _{gh,ij,kl} $ 
 & $0$ \\

\hline

$ \tilde{\phi^2} _{ab} $ 

 & $i\partial_\mu \tilde{\phi^0}_{ab}+\frac{i}{2} \epsilon_{cd \mu\nu}\partial^\nu \tilde{\phi}^2 _{ab} {  }^{cd} $ 
 & $0$ \\

\hline

$ \tilde{\phi^2} $ 

 & $-i\partial_\mu \tilde{\phi^0}-\frac{i}{2} \epsilon_{ab \mu\nu}\partial^\nu \tilde{\phi}^{2 \ ab} $ 
 & $0$ \\

\hline
\end{tabular}

\vspace{10mm}

\begin{tabular}{|l||c|c|}

\hline

$\phi$ & $\tilde{S_{\mu}}$ & $\tilde{S}$ \\

\hline

$\phi^3$ & $ -i\partial_\mu \phi^1 $ & $0$ \\

\hline  

$\phi^3 _{ab} $  & $ i\partial_\mu \phi^1 _{ab}-i\epsilon_{ab\mu\nu} \partial^\nu \phi^3 $ 
 & $0$ \\

\hline

$\phi^3 _{ab,cd} $ 
 & $-i\partial_\mu \phi^1 _{ab,cd}+i(\epsilon_{ab\mu\nu} \partial^\nu \phi^3 _{cd} -\epsilon_{cd\mu\nu} \partial^\nu \phi^3 _{ab}) $ 
 & $0$ \\

\hline

$\phi^3 _{ab,cd,ef} $ 

 & $i\partial_\mu \phi^1 _{ab,cd,ef}-i(\epsilon_{ab\mu\nu} \partial^\nu \phi^3 _{cd,ef}
  +\epsilon_{ef\mu\nu} \partial^\nu \phi^3 _{ab,cd}
  +\epsilon_{cd\mu\nu} \partial^\nu \phi^3 _{ef,ab}) $ 
 & $0$ \\

\hline

$ \tilde{\phi^3} _{ab,cd} $ 

 & $-i\partial_\mu \tilde{\phi^1}_{ab,cd}+\frac{i}{3!2^4} \Gamma_{ab,cd} {   }^{ef,gh,ij,kl} \epsilon_{ef\mu\nu}\partial^\nu
 \phi^3 _{gh,ij,kl} $ 
 & $0$ \\

\hline

$ \tilde{\phi^3} _{ab} $ 

 & $i\partial_\mu \tilde{\phi^1}_{ab}+\frac{i}{2} \epsilon_{cd \mu\nu}\partial^\nu \tilde{\phi}^3 _{ab} {  }^{cd} $ 
 & $0$ \\

\hline

$ \tilde{\phi^3} $ 

 & $-i\partial_\mu \tilde{\phi^1} -\frac{i}{2} 
\epsilon_{ab \mu\nu}\partial^\nu \tilde{\phi}^{3 \ ab} $ 
 & $0$ \\

\hline

\end{tabular}

\renewcommand{\theequation}{D.\arabic{equation}}
\setcounter{equation}{0}

\section{$N=4$ twisted supersymmetric action}

The full action (\ref{eq:44action}) has the following form:
\newcommand{\as}{\frac{1}{4!3!2^4}\epsilon^{\alpha\beta\gamma\delta}
\epsilon_{\alpha abc} \partial^c\epsilon_{\beta def}
\partial^f\epsilon_{\gamma ghp} \partial^p \Gamma^{ab,de,gh,ij,kl,mn}}

\newcommand{\ac}{\epsilon_{adgj}\epsilon^{abc\rho} \partial_\rho\epsilon^{def\epsilon}
\partial_\epsilon \epsilon^{ghi\delta}\partial_\delta
\epsilon^{jkl\theta}\partial_\theta \Gamma_{bc,ef,hi,kl} {}^{qr,st}\mathcal{D}_{\alpha,qr}\mathcal{D}_{\beta,st}} 
\begin{eqnarray*}
& &\mathcal{L}= s\tilde{s} \epsilon^{\mu\nu\rho\sigma}s_{\mu}s_{\nu}s_{\rho}s_{\sigma}
\epsilon^{\mu\nu\rho\sigma}\tilde{s}_{\mu}\tilde{s}_{\nu}\tilde{s}_{\rho}
\tilde{s}_{\sigma}
\Gamma^{ABCDEF}s_A s_B s_C s_D s_E s_F \phi^0 \psi^0
\\
&=&-\tilde{\phi}^3 \tilde{\psi}^4 -i \tilde{\phi}^2 \partial^\mu \tilde{\psi}^3 { }_{\mu,}
+i\tilde{\phi}^1 \partial^a \tilde{\psi}^4 { }_{,a} +\tilde{\phi}^0 \partial^\mu \partial^a \tilde{\psi}^3 {}_{\mu,a} \\ 
& &+\frac{i}{2}\mathcal{D}_{\sigma,ab}\tilde{\phi}^{3,ab}\tilde{\psi}^{3,\sigma}
+\frac{1}{4}\mathcal{D}_{\sigma,ab}\tilde{\phi}^{2,ab}\epsilon^{\sigma \alpha \beta \gamma } \partial_\alpha \tilde{\psi}^2 {}_{\beta\gamma} \nonumber \\
& &+\frac{1}{2}\mathcal{D}_{\sigma,ab}\tilde{\phi}^{1,ab}\partial_c\tilde{\psi}^{3,\sigma,c}
+\frac{i}{4}\mathcal{D}_{\sigma,ab}\tilde{\phi}^{0,ab}\partial^c \epsilon^{\sigma \alpha \beta \gamma } \partial_\alpha \tilde{\psi}^2 {}_{\beta\gamma,c}\\
& &-\frac{1}{4^2}\epsilon^{\mu\nu\rho\sigma} \mathcal{D}_{\mu,ab} \mathcal{D}_{\nu,cd}\tilde{\phi}^{3,ab,cd}\tilde{\psi}^2 {}_{\rho\sigma}
+\frac{i}{8}\epsilon^{\mu\nu\rho\sigma} \mathcal{D}_{\mu,ab} \mathcal{D}_{\nu,cd}\tilde{\phi}^{2,ab,cd}\partial_{\rho}\tilde{\psi}^1 {}_{\sigma}\\
& &+\frac{i}{4^2}\epsilon^{\mu\nu\rho\sigma} \mathcal{D}_{\mu,ab} \mathcal{D}_{\nu,cd}\tilde{\phi}^{1,ab,cd}\partial^{e}\tilde{\psi}^2 {}_{\rho\sigma,e}
-\frac{1}{8}\epsilon^{\mu\nu\rho\sigma} \mathcal{D}_{\mu,ab} \mathcal{D}_{\nu,cd}\tilde{\phi}^{0,ab,cd}\partial^e \partial_{\rho}\tilde{\psi}^1 {}_{\sigma,e}\\
& &+\frac{i}{4!3!2^4}\epsilon^{\mu\nu\rho\sigma} \mathcal{D}_{\mu,ab} \mathcal{D}_{\nu,cd} \mathcal{D}_{\rho,ef} \Gamma^{ab,cd,ef,gh,ij,kl}\phi^3 {}_{gh,ij,kl} \tilde{\psi}^1 {}_{\sigma}\\
& &-\frac{1}{4!3!2^4}\epsilon^{\mu\nu\rho\sigma} \mathcal{D}_{\mu,ab} \mathcal{D}_{\nu,cd} \mathcal{D}_{\rho,ef} \Gamma^{ab,cd,ef,gh,ij,kl}\phi^2 {}_{gh,ij,kl}\partial_{\sigma} \tilde{\psi}^0 \\
& &+\frac{1}{4!3!2^4}\epsilon^{\mu\nu\rho\sigma} \mathcal{D}_{\mu,ab} \mathcal{D}_{\nu,cd} \mathcal{D}_{\rho,ef} \Gamma^{ab,cd,ef,gh,ij,kl}\phi^1 {}_{gh,ij,kl}\partial^\alpha \tilde{\psi}^1 {}_{\sigma,\alpha}\\
& &-\frac{i}{4!3!2^4}\epsilon^{\mu\nu\rho\sigma} \mathcal{D}_{\mu,ab} \mathcal{D}_{\nu,cd} \mathcal{D}_{\rho,ef} \Gamma^{ab,cd,ef,gh,ij,kl}\phi^0 {}_{gh,ij,kl} \partial^\alpha \partial_\sigma \tilde{\psi}^0 {}_{,\alpha}\\
& &-\frac{3}{44!3!2^4}\epsilon^{\mu\nu\rho\sigma} \mathcal{D}_{\mu,ab} \mathcal{D}_{\nu,cd} \mathcal{D}_{\rho,ef} \mathcal{D}_{\sigma,gh} \Gamma^{ab,cd,ef,gh,ij,kl}\phi^3 {}_{ij,kl} \tilde{\psi}^0\\
& &+\frac{3i}{44!3!2^4}\epsilon^{\mu\nu\rho\sigma} \mathcal{D}_{\mu,ab} \mathcal{D}_{\nu,cd} \mathcal{D}_{\rho,ef} \mathcal{D}_{\sigma,gh} \Gamma^{ab,cd,ef,gh,ij,kl}\phi^1 {}_{ij,kl}\partial^{\delta} \tilde{\psi}^0 {}_{,\delta}\\
& &-\frac{i}{2} \epsilon_{abcd} \partial^d \tilde{\phi}^{3,bc}\tilde{\psi}^{4,a}
-\frac{1}{2} \epsilon_{abcd} \partial^d \tilde{\phi}^{2,bc}\partial^\delta \tilde{\psi}^3 {}_{\delta,a}\\
& &+\frac{1}{4} \epsilon_{abcd} \partial^d \tilde{\phi}^{1,bc}\epsilon^{a\alpha\beta\gamma}\partial_\alpha \psi^{4} { }_{,\beta\gamma}
+\frac{i}{4} \epsilon_{abcd} \partial^d \tilde{\phi}^{0 ,bc}\epsilon^{a\alpha\beta\gamma}\partial_\alpha \partial^\delta \psi^{3} { }_{\delta,\beta\gamma}\\
& &+\frac{1}{4}\epsilon_{abcd}\partial^d \mathcal{D}_{\mu,ef} \tilde{\phi}^{3,bc,ef}\tilde{\psi}^{3,\mu,a}
+\frac{i}{8}\epsilon_{abcd}\partial^d \mathcal{D}_{\mu,ef} \tilde{\phi}^{2,bc,ef}\epsilon^{\mu\alpha\beta\gamma}\partial_{\alpha} \tilde{\psi}^2 {}_{\beta\gamma} {}^{,a} \\
& &+\frac{i}{8}\epsilon_{abcd}\partial^d \mathcal{D}_{\mu,ef} \tilde{\phi}^{1,bc,ef}\epsilon^{a\alpha\beta\gamma}\partial_{\alpha} \psi^{3\mu} 
{}_{,\beta\gamma} \\
& &+\frac{1}{16}\epsilon_{abcd}\partial^d \mathcal{D}_{\mu,ef} \tilde{\phi}^{0,bc,ef}\epsilon^{a\alpha\beta\gamma}\partial_{\alpha} \epsilon^{\mu xyz} \partial_x\psi^2 { }_{yz,\beta\gamma}\\
& & +\frac{3i}{2\cdot4!3!2^4}\epsilon^{\mu\nu\rho\sigma}\epsilon_{abcd}\partial^d\mathcal{D}_{\mu ,ef} \mathcal{D}_{\nu ,gh}\Gamma^{bc,ef,gh,ij,kl,mn}
\phi^3 {}_{ij,kl,mn} \tilde{\psi}^2 {}_{\rho\sigma} {}^{,a}\\
& & -\frac{3i}{4!3!2^4}\epsilon^{\mu\nu\rho\sigma}\epsilon_{abcd}\partial^d\mathcal{D}_{\mu ,ef} \mathcal{D}_{\nu ,gh}\Gamma^{bc,ef,gh,ij,kl,mn}
\phi^2 {}_{ij,kl,mn}\partial_\rho \tilde{\psi}^1 {}_{\sigma} {}^{,a}\\
& & -\frac{3i}{4\cdot4!3!2^4}\epsilon^{\mu\nu\rho\sigma}\epsilon_{abcd}\partial^d\mathcal{D}_{\mu ,ef} \mathcal{D}_{\nu ,gh}\Gamma^{bc,ef,gh,ij,kl,mn}
\phi^1 {}_{ij,kl,mn}\epsilon^{a\alpha\beta\gamma}\partial_\alpha \psi^2 {}_{\rho\sigma,\beta\gamma}\\
& & +\frac{3i}{2\cdot4!3!2^4}\epsilon^{\mu\nu\rho\sigma}\epsilon_{abcd}\partial^d\mathcal{D}_{\mu ,ef} \mathcal{D}_{\nu ,gh}\Gamma^{bc,ef,gh,ij,kl,mn}
\phi^0 {}_{ij,kl,mn}\epsilon^{a\alpha\beta\gamma}\partial_\alpha \partial_\rho\psi^1 {}_{\sigma,\beta\gamma}\\
& &-\frac{3}{4!3!2^4} \epsilon^{\mu\nu\rho\sigma}\epsilon_{abcd}\partial^d\mathcal{D}_{\mu ,ef} \mathcal{D}_{\nu ,gh} \mathcal{D}_{\rho ,ij}\Gamma^{bc,ef,gh,ij,kl,mn}\phi^3 {}_{kl,mn} \tilde{\psi}^1 {}_{\sigma} {}^{,a} \\
& &+\frac{3i}{4!3!2^4} \epsilon^{\mu\nu\rho\sigma}\epsilon_{abcd}\partial^d\mathcal{D}_{\mu ,ef} \mathcal{D}_{\nu ,gh} \mathcal{D}_{\rho ,ij}\Gamma^{bc,ef,gh,ij,kl,mn}\phi^2 {}_{kl,mn}\partial_\sigma \tilde{\psi}^{0,a}\\
& &-\frac{3i}{2\cdot4!3!2^4} \epsilon^{\mu\nu\rho\sigma}\epsilon_{abcd}\partial^d\mathcal{D}_{\mu ,ef} \mathcal{D}_{\nu ,gh} \mathcal{D}_{\rho ,ij}\Gamma^{bc,ef,gh,ij,kl,mn}\phi^1 {}_{kl,mn}\epsilon^{a\alpha\beta\gamma}\partial_\alpha \tilde{\psi}^1 {}_{\sigma,\beta\gamma}\\
& &+\frac{3}{2\cdot4!3!2^4} \epsilon^{\mu\nu\rho\sigma}\epsilon_{abcd}\partial^d\mathcal{D}_{\mu ,ef} \mathcal{D}_{\nu ,gh} \mathcal{D}_{\rho ,ij}\Gamma^{bc,ef,gh,ij,kl,mn}\phi^0 {}_{kl,mn}\epsilon^{a\alpha\beta\gamma}\partial_\alpha \partial_\sigma \tilde{\psi}^0 {}_{,\beta\gamma} \nonumber \\
& &+\frac{3i}{2\cdot4!3!2^4} \epsilon^{\mu\nu\rho\sigma}\epsilon_{abcd}\partial^d\mathcal{D}_{\mu ,ef} \mathcal{D}_{\nu ,gh} \mathcal{D}_{\rho ,ij}\mathcal{D}_{\sigma ,kl}\Gamma^{bc,ef,gh,ij,kl,mn}\phi^3 {}_{mn} \tilde{\psi}^{0,a}\\
& &+\frac{3}{4\cdot4!3!2^4} \epsilon^{\mu\nu\rho\sigma}\epsilon_{abcd}\partial^d\mathcal{D}_{\mu ,ef} \mathcal{D}_{\nu ,gh} \mathcal{D}_{\rho ,ij}\mathcal{D}_{\sigma ,kl}\Gamma^{bc,ef,gh,ij,kl,mn}\phi^1 {}_{mn} \epsilon^{a\alpha\beta\gamma} \partial_\alpha\psi^{0} {}_{,\beta\gamma}\\
& &+\frac{1}{4}\epsilon_{abcd}\partial^d \partial^f \tilde{\phi}^{3,ea,bc}\psi^4 {}_{,ef}+\frac{i}{4}\epsilon_{abcd}\partial^d \partial^f \tilde{\phi}^{2,ea,bc}\partial^\delta \psi^3 {}_{\delta,ef}\\
& &+\frac{i}{4}\epsilon_{abcd}\partial^d \partial^f \tilde{\phi}^{1,ea,bc}
(\partial_e \psi^4 {}_{,f} -\partial_f \psi^4 {}_{,e})+\frac{1}{4}\epsilon_{abcd}\partial^d \partial^f \tilde{\phi}^{0,ea,bc}
(\partial_e \partial^\delta \psi^3 {}_{\delta,f} -\partial_f \partial^\delta \psi^3 {}_{\delta,e})\\
& &+\frac{i}{3!2^6}\epsilon_{abcd}\partial^d \partial^f \Gamma^{ea,bc,gh,ij,kl,mn}\mathcal{D} _{\mu,gh} \phi^3 {}_{ij,kl,mn} \psi^{3\mu} { }_{,ef}\\
& &+\frac{1}{3!2^7}\epsilon_{abcd}\partial^d \partial^f \Gamma^{ca,bc,gh,ij,kl,mn}\mathcal{D} _{\mu,gh} \phi^2 {}_{ij,kl,mn}\epsilon^{\mu\alpha\beta\gamma} \partial_{\alpha}\psi^{2\mu} { }_{\beta\gamma,ef}\\
& &-\frac{1}{3!2^6}\epsilon_{abcd}\partial^d \partial^f \Gamma^{ca,bc,gh,ij,kl,mn}\mathcal{D} _{\mu,gh} \phi^1 {}_{ij,kl,mn}( \partial_e\psi^{3\mu} { }_{,f}- \partial_f\psi^{3\mu} { }_{,e})\\
& &-\frac{i}{3!2^7}\epsilon_{abcd}\partial^d \partial^f \Gamma^{ca,bc,gh,ij,kl,mn}\mathcal{D} _{\mu,gh} \phi^0 {}_{ij,kl,mn}( \partial_e\epsilon^{\mu\alpha\beta\gamma} \partial_{\alpha}\psi^{2} { }_{\beta\gamma,f}- \partial_f\epsilon^{\mu\alpha\beta\gamma} \partial_{\alpha}\psi^{2} { }_{\beta\gamma,e})\\
& &+\frac{1}{2^9}\epsilon^{\mu\nu\rho\sigma}\epsilon_{abcd}\partial^d \partial^f \Gamma^{ea,bc,gh,ij,kl,mn}\mathcal{D} _{\mu,gh} \mathcal{D} _{\nu,ij}\phi^3 {}_{kl,mn} \psi^{2} { }_{\rho\sigma,ef}\\
& &+\frac{i}{2^9}\epsilon^{\mu\nu\rho\sigma}\epsilon_{abcd}\partial^d \partial^f \Gamma^{ea,bc,gh,ij,kl,mn}\mathcal{D} _{\mu,gh} \mathcal{D} _{\nu,ij}\phi^2 {}_{kl,mn}( \partial_\rho\psi^{1} { }_{\sigma,ef} - \partial_\sigma \psi^{1} { }_{\rho,ef} )\\
& &-\frac{i}{2^9}\epsilon^{\mu\nu\rho\sigma}\epsilon_{abcd}\partial^d \partial^f \Gamma^{ea,bc,gh,ij,kl,mn}\mathcal{D} _{\mu,gh} \mathcal{D} _{\nu,ij}\phi^1 {}_{kl,mn}( \partial_e\psi^{2} { }_{\rho\sigma,f} - \partial_f \psi^{2} { }_{\rho\sigma,e} )\\
& &-\frac{1}{2^8}\epsilon^{\mu\nu\rho\sigma}\epsilon_{abcd}\partial^d \partial^f \Gamma^{ea,bc,gh,ij,kl,mn}\mathcal{D} _{\mu,gh} \mathcal{D} _{\nu,ij}\phi^0 {}_{kl,mn}( \partial_e\partial_\rho \psi^{1} { }_{\sigma,f} - \partial_f\partial_\rho \psi^{1} { }_{\sigma,e} )\\
& &-\frac{i}{4\cdot3!2^4}\epsilon^{\mu\nu\rho\sigma}\epsilon_{abcd}\partial^d \partial^f \Gamma^{ea,bc,gh,ij,kl,mn}\mathcal{D} _{\mu,gh}\mathcal{D} _{\nu,ij}
\mathcal{D} _{\rho,kl} \phi^3 {}_{mn} \psi^1 {}_{\sigma , ef} \\
& &+\frac{1}{4\cdot3!2^4}\epsilon^{\mu\nu\rho\sigma}\epsilon_{abcd}\partial^d \partial^f \Gamma^{ea,bc,gh,ij,kl,mn}\mathcal{D} _{\mu,gh}\mathcal{D} _{\nu,ij}\mathcal{D} _{\rho,kl} \phi^2 {}_{mn}\partial_\sigma \psi^0 {}_{, ef} \\
& &+\frac{1}{4\cdot3!2^4}\epsilon^{\mu\nu\rho\sigma}\epsilon_{abcd}\partial^d \partial^f \Gamma^{ea,bc,gh,ij,kl,mn}\mathcal{D} _{\mu,gh}\mathcal{D} _{\nu,ij}\mathcal{D} _{\rho,kl} \phi^1 {}_{mn} ( \partial_e \psi^1 {}_{\sigma , f} -\partial_f \psi^1 {}_{\sigma , e})\\
& &-\frac{i}{4\cdot3!2^4}\epsilon^{\mu\nu\rho\sigma}\epsilon_{abcd}\partial^d \partial^f \Gamma^{ea,bc,gh,ij,kl,mn}\mathcal{D} _{\mu,gh}\mathcal{D} _{\nu,ij}
\mathcal{D} _{\rho,kl} \phi^0 {}_{mn}\partial_\sigma ( \partial_e \psi^0 {}_{, f} -\partial_f \psi^0 {}_{, e})\\
& &-\frac{1}{4!2^6}\epsilon^{\mu\nu\rho\sigma}\epsilon_{abcd}\partial^d \partial^f \Gamma^{ea,bc,gh,ij,kl,mn}\mathcal{D} _{\mu,gh}\mathcal{D} _{\nu,ij}\mathcal{D} _{\rho,kl}\mathcal{D} _{\sigma,mn} \phi^3 \psi^0 _{,ef} \\
& &+\frac{i}{4!2^6}\epsilon^{\mu\nu\rho\sigma}\epsilon_{abcd}\partial^d \partial^f \Gamma^{ea,bc,gh,ij,kl,mn}\mathcal{D} _{\mu,gh}\mathcal{D} _{\nu,ij}\mathcal{D} _{\rho,kl}\mathcal{D} _{\sigma,mn} \phi^1 ( \partial_e\psi^0 _{,f} -\partial_f \psi^0 _{,e}) \\
%
%
%
& - &i\as \phi^3 {}_{ij,kl,mn} \psi^4 {}_{\delta} \\
& -&\as\phi^2 {}_{ij,kl,mn}\partial^\epsilon \psi^3 {}_{\epsilon,\delta} \\
& -&\as\phi^1 {}_{ij,kl,mn}\partial_\delta \psi^4  \\
& -&i\as\phi^0 {}_{ij,kl,mn}\partial_\delta\partial^\epsilon \psi^3 {}_{\epsilon,} \\
& -&3\as \mathcal{D}_{\mu,ij} \phi^{3} {}_{kl,mn}\psi^{3,\mu} {}_{,\delta}\\
& -&\frac{3i}{2}\as\mathcal{D}_{\mu,ij}\phi^{2} {}_{kl,mn} \epsilon^{\mu\nu\rho\sigma} \partial_\nu \psi^2 {}_{\rho\sigma,\delta}\\
& +&3i\as\mathcal{D}_{\mu,ij}\phi^{1} {}_{kl,mn}  \partial_\delta \psi^3 {}^{\mu}\\
& +&\frac{3}{2} \as \mathcal{D}_{\mu,ij} \phi^0 {}_{kl,mn} \partial_\delta \epsilon^{\mu\nu\rho\sigma} \partial_\nu \psi^2 {}_{\rho\sigma}\\
&-&i\frac{3!3!}{4!} \as \epsilon^{\mu\nu\rho\sigma} \mathcal{D}_{\mu,ij}\mathcal{D}_{\nu,kl} \phi^3 {}_{mn} \psi^2 {}_{\rho\sigma,\delta} \\
& +&\frac{3!3!}{4!} \as \\
& &\hspace{50mm} \times  \epsilon^{\mu\nu\rho\sigma} \mathcal{D}_{\mu,ij}\mathcal{D}_{\nu,kl} \phi^2 {}_{mn} \delta_{\rho\sigma,rs} \partial^s \psi^{1s} {}_{,\delta}
 \\
&-&\frac{3!3!}{4!} \as  \\
& & \hspace{50mm} \times \epsilon^{\mu\nu\rho\sigma} \mathcal{D}_{\mu,ij}\mathcal{D}_{\nu,kl} \phi^1 {}_{mn} \partial_\delta \psi^2 {}_{\rho\sigma,} \\
& +&i\frac{3!3!}{4!} \as \\
& & \hspace{50mm} \times \epsilon^{\mu\nu\rho\sigma} \mathcal{D}_{\mu,ij}\mathcal{D}_{\nu,kl} \phi^0 {}_{mn} \partial_\delta \delta_{\rho\sigma,rs}\partial^r \psi^{1s}
\\
& +& \as \epsilon^{\mu\nu\rho\sigma} \mathcal{D}_{\mu,ij}\mathcal{D}_{\nu,kl} \mathcal{D}_{\rho,mn} \phi^3 \psi^1 {}_{\sigma,\delta} \\
& -& \as \epsilon^{\mu\nu\rho\sigma} \mathcal{D}_{\mu,ij}\mathcal{D}_{\nu,kl} \mathcal{D}_{\rho,mn} \phi^2 \partial_\sigma \psi^0 {}_{\delta,} \\
& -& \as \epsilon^{\mu\nu\rho\sigma} \mathcal{D}_{\mu,ij}\mathcal{D}_{\nu,kl} \mathcal{D}_{\rho,mn} \phi^1 \partial_\delta \psi^1 {}_{\sigma,} \\
& +& \as \epsilon^{\mu\nu\rho\sigma} \mathcal{D}_{\mu,ij}\mathcal{D}_{\nu,kl} \mathcal{D}_{\rho,mn} \phi^0 \partial_\delta\partial_\sigma \psi^0 \\
& +&\frac{i}{4!2^7} \ac \phi^2 \epsilon^{\alpha\beta uv}\partial_u \psi^2 {}_{v,}\\
& -&\frac{1}{4!2^8} \ac \phi^2 \epsilon^{\alpha\beta uv} \psi^2_{uv,} \\
& -&\frac{1}{4!2^7} \epsilon_{adgj}\epsilon^{abc\rho} \partial_\rho\epsilon^{def\epsilon}\partial_\epsilon \epsilon^{ghi\delta}\partial_\delta
\epsilon^{jkl\theta}\partial_\theta \Gamma_{bc,ef,hi,kl} {}^{qr,st}\mathcal{D}_{\alpha,qr} \phi^2 {}_{st} \epsilon^{\alpha uvw} \partial_u \psi^2 {}_{vw}\\
& -&\frac{i}{4!2^6} \epsilon_{adgj}\epsilon^{abc\rho} \partial_\rho\epsilon^{def\epsilon}\partial_\epsilon \epsilon^{ghi\delta}\partial_\delta
\epsilon^{jkl\theta}\partial_\theta \Gamma_{bc,ef,hi,kl} {}^{qr,st}\mathcal{D}_{\alpha,qr}\psi^3 {}_{st} \psi^{3 \alpha}\\
& -& \frac{i}{4!2^7}\epsilon_{adgj}\epsilon^{abc\rho} \partial_\rho\epsilon^{def\epsilon}\partial_\epsilon \epsilon^{ghi\delta}\partial_\delta
\epsilon^{jkl\theta}\partial_\theta \Gamma_{bc,ef,hi,kl} {}^{qr,st} \phi^2 {}_{qr,st} \partial_v \psi^{3 v} \\
& -&\frac{1} {4!2^7} \epsilon_{adgj}\epsilon^{abc\rho} \partial_\rho\epsilon^{def\epsilon}\partial_\epsilon \epsilon^{ghi\delta}\partial_\delta
\epsilon^{jkl\theta}\partial_\theta \Gamma_{bc,ef,hi,kl} {}^{qr,st}\phi^3 {}_{qr,st} \psi^4.
\end{eqnarray*}

\renewcommand{\theequation}{E.\arabic{equation}}
\setcounter{equation}{0}
\section{Useful formulae in four dimensions}

Definition of 
$\delta^+ _{\mu\nu,\rho\sigma}$ 
is 
\begin{eqnarray}
& &\delta^+ _{\mu \nu , \rho \sigma}= \delta_{\mu \rho} \delta_{\nu \sigma}
-\delta_{\mu \sigma} \delta_{\nu \rho} - \varepsilon_{\mu \nu \rho \sigma},\\
& &\delta^{+A}, {}_A = 4\cdot 3,
\end{eqnarray}
where the suffix $A$ denotes tensor suffix $\mu\nu$. Thus $\delta^{+A},{}_A$ 
stands for $\delta^{+\mu\nu},{}_{\mu\nu}$. 
We introduce self-dual field $\phi^{+\mu\nu}$\ ($\phi^{+\mu\nu} =
-\frac{1}{2}\epsilon^{\mu\nu\rho\sigma}\phi^+ _{\rho\sigma}$) 
then 
\begin{eqnarray}
\delta^+ _{\mu\nu,\rho\sigma} \phi^{+\rho\sigma} &=& 
(\delta_{\mu \rho} \delta_{\nu \sigma}
-\delta_{\mu \sigma} \delta_{\nu \rho} - 
\varepsilon_{\mu \nu \rho \sigma})\phi^{+\rho\sigma}\nonumber  \\
&=& 2\phi^{+} _{\mu\nu} -\varepsilon_{\mu \nu \rho \sigma}\phi^{+\rho\sigma} 
\nonumber \\
&=& 4\phi^{+} _{\mu\nu}. \nonumber 
\end{eqnarray}

\noindent
Variants of the definition of $\Gamma^{+ABC}$ are  
\begin{eqnarray}
\Gamma^{+ \mu \alpha , \nu \beta , \rho \gamma}
&=& \delta^{\alpha \nu} \delta^{\beta \rho} \delta^{\gamma \mu} 
+ \delta^{\mu \nu} \delta^{\beta \gamma } \delta^{\rho \alpha} 
+ \delta^{\alpha \beta} \delta^{\nu \gamma} \delta^{\rho \mu} 
+ \delta^{\mu \beta} \delta^{\nu \rho} \delta^{\gamma \alpha} \nonumber \\
& &-( \delta^{\alpha \nu} \delta^{\beta \gamma} \delta^{\rho \mu} 
+ \delta^{\mu \nu} \delta^{\beta \rho } \delta^{\gamma \alpha} 
+ \delta^{\alpha \beta} \delta^{\nu \rho} \delta^{\gamma \mu} 
+ \delta^{\mu \beta} \delta^{\nu \gamma } \delta^{\rho \alpha} ) \nonumber \\
& & -\varepsilon^{\mu \alpha \beta \gamma } \delta^{\nu \rho}
- \varepsilon^{\mu \alpha \nu \rho } \delta^{\beta \gamma}
+ \varepsilon^{\mu \alpha \nu \gamma } \delta^{\beta \rho}
+ \varepsilon^{\mu \alpha \beta \rho } \delta^{\nu \gamma} \nonumber \\ 
&=& \delta^{+\mu\alpha,\nu\rho}\delta^{\beta\gamma}
+ \delta^{+\mu\alpha,\beta\gamma}\delta^{\nu\rho}
- \delta^{+\mu\alpha,\nu\gamma}\delta^{\beta\rho}
- \delta^{+\mu\alpha,\beta\rho}\delta^{\nu\gamma},
\end{eqnarray}
\begin{eqnarray}
\Gamma^{+AB \rho \sigma} &=& \frac{1}{2}(\delta^{+A,\nu\rho}\delta^{+B}, {}_\nu {}^\sigma - \delta^{+A,\nu\sigma}\delta^{+B}, {}_\nu {}^\rho),\\ 
\Gamma^{+ABC} &=& \frac{1}{4}\delta^{+A,\nu\rho} \delta^{+B}, {}_\nu {}^\sigma \delta^{+C}, {}_{\rho\sigma}.
\end{eqnarray}

Useful formulae are in order.
\begin{eqnarray}
\varepsilon^{\mu \alpha \gamma \nu} \delta^{\beta \rho}
&=& \delta^{\mu \beta} \varepsilon^{\rho \alpha \gamma \nu}
+ \delta^{\gamma \beta} \varepsilon^{\rho \mu \alpha \nu}
- \delta^{\alpha \beta} \varepsilon^{\rho \mu \gamma \nu}
- \delta^{\nu \beta} \varepsilon^{\rho \mu \alpha \gamma } \nonumber \\
&=& \delta^{\mu \rho} \varepsilon^{\beta \alpha \gamma \nu}
+ \delta^{\gamma \rho} \varepsilon^{\beta \mu \alpha \nu}
- \delta^{\alpha \rho} \varepsilon^{\beta \mu \gamma \nu}
- \delta^{\nu \rho} \varepsilon^{\beta \mu \alpha \gamma } \nonumber \\
&=& \frac{1}{2}(\delta^{\mu \beta} \varepsilon^{\rho \alpha \gamma \nu}
+ \delta^{\gamma \beta} \varepsilon^{\rho \mu \alpha \nu}
- \delta^{\alpha \beta} \varepsilon^{\rho \mu \gamma \nu}
- \delta^{\nu \beta} \varepsilon^{\rho \mu \alpha \gamma }  \nonumber \\
& & + \delta^{\mu \rho} \varepsilon^{\beta \alpha \gamma \nu}
+ \delta^{\gamma \rho} \varepsilon^{\beta \mu \alpha \nu}
- \delta^{\alpha \rho} \varepsilon^{\beta \mu \gamma \nu}
- \delta^{\nu \rho} \varepsilon^{\beta \mu \alpha \gamma }),
\end{eqnarray}

\begin{eqnarray}
& &-\varepsilon^{\mu \alpha \gamma \nu} \delta^{\beta \rho}
-\varepsilon^{\mu \alpha \rho \beta} \delta^{\nu \gamma}
+\varepsilon^{\mu \alpha \rho \nu} \delta^{\beta \gamma}
+\varepsilon^{\mu \alpha \gamma \beta} \delta^{\nu \rho} \nonumber \\
&=& \frac{1}{2} ( -\delta^{\mu \beta} \varepsilon^{\rho \alpha \gamma \nu}
-\delta^{\mu \nu} \varepsilon^{\gamma \alpha \rho \beta}
+\delta^{\mu \gamma} \varepsilon^{\beta \alpha \rho \nu}
+\delta^{\mu \rho} \varepsilon^{\nu \alpha \gamma \beta} \nonumber \\
& &\hspace{24mm} +\delta^{\alpha \beta} \varepsilon^{\rho \mu \gamma \nu}
+\delta^{\alpha \nu} \varepsilon^{\gamma \mu \rho \beta}
-\delta^{\alpha \gamma} \varepsilon^{\beta \mu \rho \nu}
-\delta^{\alpha \rho} \varepsilon^{\nu \mu \gamma \beta}),
\end{eqnarray}

\begin{eqnarray}
& &\Gamma^{+ABC} \cdot\Gamma^+_{ABC}=3!\cdot4^3, \\
& &\Gamma^{+ABC} \cdot\Gamma^+_{ABD}=2!\cdot4^2 \delta^{+C} {}_D,\\
& &\Gamma^{+ABC} \cdot\Gamma^+_{ADE}
=4(\delta^{+B} {}_D \delta^{+C} {}_E -\delta^{+B} {}_E \delta^{+C} {}_D),\\
& &\Gamma^{+ABC} \cdot\Gamma^+_{DEF}
=\Bigr( \delta^{+A} {}_D \delta^{+B} {}_E \delta^{+C} {}_F +(\mbox{ cyclic in DEF}) \Bigr) \nonumber \\
& &\hspace{4cm}-\Bigr( \delta^{+A} {}_F \delta^{+B} {}_E \delta^{+C} {}_D +(\mbox{ cyclic in DEF}) \Bigr),
\end{eqnarray}

\begin{eqnarray}
\mathcal{D}^+ _{\mu , \rho \sigma} = \delta_{\mu \rho}\partial_\sigma
-\delta_{\mu \sigma}\partial_\rho 
- \varepsilon_{\mu \rho \sigma \nu}\partial^\nu,
\end{eqnarray}

\begin{eqnarray}
& &\Gamma^{+\mu\nu,\rho\alpha,\sigma} {}_\alpha=2\delta^{+\mu\nu,\rho\sigma},\\
& &\delta^{+\mu\rho,\nu} {}_\rho =3 \delta^{\mu\nu},
\end{eqnarray}

\begin{eqnarray}
& &\Gamma^{+ABC}\mathcal{D}_{\mu,A}\mathcal{D}_{\nu,B}\mathcal{D}_{\rho,C}
=-4^3\epsilon_{\sigma\mu\nu\rho}\partial^2\partial^\sigma, \\
& &\Gamma^{+ABC}\mathcal{D}_{\mu,A}\mathcal{D}_{\nu,B}
= 4^2 \Bigr( \partial_\mu \mathcal{D}^+_{\nu,} { }^C-\partial_\nu \mathcal{D}^+_{\mu,} { }^C - \partial^2 \delta^+_{\mu\nu,} {}^C\Bigr),
\end{eqnarray}

\begin{eqnarray}
\delta^{+}_{\mu\nu,\alpha\gamma}\delta^{+}_{\rho\sigma,\beta} {}^{\gamma}
=\delta_{\alpha\beta} \delta^{+}_{\mu\nu,\rho\sigma} 
+\delta_{\beta\rho} \delta^{+}_{\mu\nu,\alpha\sigma}
-\delta_{\beta\sigma} \delta^{+}_{\mu\nu,\alpha\rho}
-\delta_{\alpha\rho} \delta^{+}_{\mu\nu,\beta\sigma}
+\delta_{\alpha\sigma} \delta^{+}_{\mu\nu,\beta\rho},
\end{eqnarray}

\begin{eqnarray}
\delta^{+}_{\mu\nu,\alpha\gamma}\delta^{+}_{\rho\sigma,\beta} {}^{\gamma}
+\delta^{+}_{\mu\nu,\beta\gamma}\delta^{+}_{\rho\sigma,\alpha} {}^{\gamma}
=2\delta_{\alpha\beta} \delta^{+}_{\mu\nu,\rho\sigma},
\end{eqnarray}

\begin{eqnarray}
\delta^{+}_{\mu\nu,\alpha\gamma}\delta^{+}_{\rho\sigma,\beta} {}^{\gamma}
-\delta^{+}_{\mu\nu,\beta\gamma}\delta^{+}_{\rho\sigma,\alpha} {}^{\gamma}
=2(\delta_{\beta\rho} \delta^{+}_{\mu\nu,\alpha\sigma}
-\delta_{\beta\sigma} \delta^{+}_{\mu\nu,\alpha\rho}
-\delta_{\alpha\rho} \delta^{+}_{\mu\nu,\beta\sigma}
+\delta_{\alpha\sigma} \delta^{+}_{\mu\nu,\beta\rho} ).
\end{eqnarray}

\renewcommand{\theequation}{F.\arabic{equation}}
\setcounter{equation}{0}
\section{Wess-Zumino gauge in two dimensions}

In this Appendix we explicitly show how the superfluous component 
fields in the superconnection can be gauged away consistently with 
the constraints by Wess-Zumino gauge. We consider twisted $N=D=2$ Abelian 
case for simplicity. We have introduced the fermionic superconnection $\{\Gamma_I\}$ 
as in (\ref{def_2d-super-covariant-derivative}). Here we first show the explicit 
component expansion of the superconnection as: 
\begin{eqnarray}
\Gamma &=& \Gamma^0 \nonumber \\
&+& \theta \Gamma^1 + \theta^\mu \Gamma^1_\mu
+ \tilde{\theta} \tilde{\Gamma}^1 \nonumber \\
&+& \theta\tilde{\theta} \Gamma^2 +\theta^2 \tilde{\Gamma}^2 + \theta\theta^\mu\Gamma^2_\mu +\theta^\mu \tilde{\theta} \tilde{\Gamma}^2_\mu+\cdots,
\end{eqnarray}
\begin{eqnarray}
\tilde{\Gamma} &=& H^0 \nonumber \\
&+& \theta H^1 + \theta^\mu H^1_\mu+ \tilde{\theta} \tilde{H}^1 \nonumber \\
&+& \theta\tilde{\theta} H^2+\theta^2 \tilde{H}^2 + \theta\theta^\mu H^2_\mu +\theta^\mu \tilde{\theta} \tilde{H}^2_\mu +  \cdots,
\end{eqnarray}
\begin{eqnarray}
\Gamma_\mu &=& M^0_\mu \nonumber \\
&+& \theta M^1_\mu+ \theta^\rho M^1_{\mu\rho} + \tilde{\theta}\tilde{M}^1_\mu  \nonumber \\
&+& \theta\tilde{\theta} M^2_\mu +\theta^2 \tilde{M}^2_\mu + \theta\theta^\rho M^2_{\mu\rho} +\theta^\rho \tilde{\theta} \tilde{M}^2_{\mu\rho}+\cdots,
\end{eqnarray}
\begin{eqnarray}
\Gamma_{\underline{\mu}} &=& \omega_\mu \nonumber \\
&+& \theta A^1_\mu  + \theta^\rho A^1_{\mu\rho}+ \tilde{\theta} \tilde{A}^1_\mu \nonumber \\
&+& \theta\tilde{\theta} A^2_\mu +\theta^2 \tilde{A}^2_\mu + \theta\theta^\rho A^2_{\mu\rho} +\theta^\rho \tilde{\theta} \tilde{A}^2_{\mu\rho} +\cdots.
\end{eqnarray}

According to the introduction of the superconnection we introduce the following 
gauge transformation:
\begin{eqnarray}
\delta \Gamma &=& \mathfrak{D} K , \label{eq:g1} \\
\delta \tilde{\Gamma} &=& \tilde{ \mathfrak{D}} K ,\label{eq:g2} \\
\delta \Gamma_\mu &=&  \mathfrak{D}_\mu K, \\
\delta \Gamma_{\underline{\mu}} &=&  \mathfrak{D}_{\underline{\mu}} K,
\end{eqnarray}
where $K$ is the supergauge parameter defined by 
\begin{eqnarray}
K &=& K^0 \nonumber \\
&+& \theta K^1 + \theta^\mu K^1_\mu
+ \tilde{\theta} \tilde{K}^1 \nonumber \\
&+& \theta\tilde{\theta} K^2 +\theta^2 \tilde{K}^2 + \theta\theta^\mu K^2_\mu
 +\theta^\mu \tilde{\theta} \tilde{K}^2_\mu \nonumber \\
&+& \theta\theta^2 K^3 +\theta^2 \tilde{\theta} \tilde{K}^3+\theta\theta^\mu \tilde{\theta} K^3_\mu \nonumber \\
&+&\theta^4 K^4 .
\end{eqnarray}
All the component fields in the supergauge parameter $K$ except for the
$K^0$ which is identified as an ordinary gauge parameter can be used to gauge 
away the component fields of the superconnection $\{\Gamma_I\}$. 

We show some of gauge transformations in (\ref{eq:g1}) and (\ref{eq:g2})  
\begin{eqnarray}
\delta \Gamma^{0} &=& K^{1}, \ \ \ 
\delta\Gamma^{1} = 0, \\
\delta\Gamma^{1}_\mu &=& K^{2 }_\mu  -\frac{i}{2}\partial_\mu K^0,\ \ \ 
\delta\tilde{\Gamma}^{1} = K^{2} ,
\end{eqnarray}
\begin{eqnarray*}
\delta H^{0} &=& \tilde{K}^{1} , \ \ \ 
\delta H^{1} = -K^2,\\
\delta H^{1}_\mu &=&  -\tilde{K}^{2 }_\mu  +
\frac{i}{2}\epsilon_{\mu\nu}\partial^\nu K^0, \ \ \ 
\delta\tilde{H}^{1} = 0.
\end{eqnarray*}

For example we can gauge away $\Gamma^0$ by using the gauge parameter $K^1$, 
similarly $H^0$ can be gauged away by $\tilde{K}^1$ and so on. 
In order to gauge away $\tilde{\Gamma}^1$ and $H^1$, we have to use the same 
gauge parameter $K^2$ and thus it is {\it a priori} not obvious if this 
gauging away procedure is consistent or not. It turns out that the consistency 
comes out from the constraints in (\ref{eq:curvature con}). 
Here we explicitly show curvature constraints (\ref{eq:curvature con}), 
\begin{eqnarray}
\mathcal{W}_1 &=& 2\mathfrak{D}\Gamma = \mathcal{W},\\
\mathcal{W}_2 &=& 2\tilde{\mathfrak{D}}\tilde{\Gamma}= \mathcal{W},\\
\mathcal{W}_3 &=& \mathfrak{D} \tilde{\Gamma} + \tilde{\mathfrak{D}} \Gamma =0,
\label{eq:w3}\\
\mathcal{F}_{\mu\nu} &=& \mathfrak{D}_\mu \Gamma_\nu - \mathfrak{D}_\nu \Gamma_\mu =\delta_{\mu\nu} \mathcal{F},
\end{eqnarray}
\begin{eqnarray}
\mathfrak{D} \Gamma_\mu + \mathfrak{D} \Gamma_\mu &=&
- i\Gamma_{\underline{\mu}}, \\
\tilde{\mathfrak{D}} \Gamma_\mu + \mathfrak{D} \tilde{\Gamma}_\mu &=&
 i\epsilon_{\mu\nu} \Gamma^{\underline{\nu}}.
\end{eqnarray}
The constraint (\ref{eq:w3}) leads to the following relation: 
\begin{eqnarray}
0=& &H^1+\tilde{\Gamma}^1 \nonumber \\
&+&\theta(-\Gamma^2) \nonumber \\
&+&\theta^\rho (H^2_\mu -\tilde{\Gamma}^2_\mu ) \nonumber \\
&+&\tilde{\theta} (H^2   )\nonumber \\
&+&\theta\tilde{\theta}(0 ) \nonumber \\
&+&\theta^2 (H^3-\frac{i}{2}\epsilon^{\mu\nu}\partial_\mu H^1_\nu +\tilde{\Gamma}^3 +\frac{i}{2}\partial^\mu \Gamma^1_\mu  )\nonumber \\
&+&\theta\theta^\mu (\frac{i}{2}\partial_\mu H^1 +\Gamma^3_\mu -\frac{i}{2}\epsilon_{\mu\nu}\partial^\nu\Gamma^1         )\nonumber \\
&+&\theta^\mu\tilde{\theta}( H^3_\mu -\frac{i}{2}\partial_\mu \tilde{H}^1  +\frac{i}{2}\epsilon_{\mu\nu}\partial^\nu \tilde{\Gamma}^1  ) +\cdots,
\end{eqnarray}
This constraint thus requires, $H^1+\tilde{\Gamma}^1 =0$,\quad $\Gamma^2=0 $,\quad 
$H^2_\mu-\tilde{\Gamma}^2_\mu=0 $, and so on. 
Thus $\tilde{\Gamma}^1$ and $H^1$ can be gauge away by the same gauge parameter $K^2$.

We can then gauge away most of the component fields in the superconnection 
except for the following fields: $\Gamma^1=\frac{1}{2}A,\ \Gamma^2 _\mu =
\frac{1}{2}\lambda_\mu,\ \Gamma^3 =
\frac{1}{2}D+\frac{1}{8}\epsilon^{\mu\nu}F_{\mu\nu}  ,\  M^1 _{\mu\nu}=
\frac{1}{2}\delta_{\mu\nu}B , \  M^2 _{\mu\nu}= -\frac{1}{2}(\delta_{\mu\nu}\rho+
\epsilon_{\mu\nu}\tilde{\rho} )$ and $\omega_\mu $. 
As we can see these are the basic component fields of the chiral supermultiplets. 

We show the most explicit form of the superconnection with non-Abelian gauge 
algebra in the following: 
\begin{eqnarray}
\Gamma &=& \Gamma^0(0) \nonumber \\
&+& \theta \Gamma^1(\frac{1}{2}A) + \theta^\mu \Gamma^1_\mu
(-\frac{i}{2}\omega_\mu)+ \tilde{\theta} \tilde{\Gamma}^1(0) \nonumber \\
&+& \theta\tilde{\theta} \Gamma^2(0) +\theta^2 \tilde{\Gamma}^2(-\frac{1}{3}\tilde{\rho} ) + \theta\theta^\mu\Gamma^2_\mu(\frac{1}{2}\lambda_\mu) +\theta^\mu \tilde{\theta} \tilde{\Gamma}^2_\mu(-\frac{i}{6}\epsilon_{\mu\nu}\lambda^\nu) \nonumber \\
&+& \theta\theta^2 \Gamma^3(\frac{1}{2}(D+\frac{1}{4}\epsilon^{\rho\sigma}F_{\rho\sigma})) +\theta^2 \tilde{\theta} \tilde{\Gamma}^3( \frac{i}{8}[A,B])+\theta\theta^\mu \tilde{\theta} \Gamma^3_\mu(\frac{i}{4}\epsilon_{\mu\nu}D^\nu A) \nonumber \\
&+&\theta^4 \Gamma^4 (-\frac{i}{6}D^\mu \lambda_\mu-\frac{i}{6}[A,\rho] ),
\end{eqnarray}
\begin{eqnarray}
\Gamma_\mu &=& M^0_\mu(0) \nonumber \\
&+& \theta M^1_\mu(-\frac{i}{2}\omega_\mu) + \theta^\rho M^1_{\mu\rho}(\frac{1}{2}\delta_{\mu\rho} B ) + \tilde{\theta}\tilde{M}^1_\mu (\frac{i}{2}\epsilon_{\mu\nu}\omega^{\nu}) \nonumber \\
&+& \theta\tilde{\theta} M^2_\mu(-\frac{1}{3}\epsilon_{\mu\nu}\lambda^{\nu}) +\theta^2 \tilde{M}^2_\mu(0) + \theta\theta^\rho M^2_{\mu\rho}(-\frac{1}{2}\delta_{\mu\rho}\rho  - \frac{1}{6}\epsilon_{\mu\rho} \tilde{\rho}  )\nonumber  \\
& & +\theta^\rho \tilde{\theta} \tilde{M}^2_{\mu\rho} (\frac{1}{2}\delta_{\mu\rho} \tilde{\rho} -\frac{1}{6}\epsilon_{\mu\rho} \rho)   )\nonumber \\
&+& \theta\theta^2 M^3_\mu(\frac{i}{4}\epsilon_{\mu\rho}D^\rho B) +\theta^2 \tilde{\theta} \tilde{M}^3_\mu(\frac{i}{4}D_\mu B) +\theta\theta^\rho \tilde{\theta} M^3_{\mu\rho}(-\frac{1}{2}\delta_{\mu\rho}(D +\frac{3}{4} \epsilon^{\nu\sigma} F_{\nu\sigma})  + \frac{i}{8} \epsilon_{\mu\rho}[A,B]) \nonumber \\
&+&\theta^4 M^4_\mu (-\frac{i}{6} D_\mu \rho -\frac{i}{6} \epsilon_{\mu\nu} D^\nu \tilde{\rho} -\frac{i}{6} [B,\lambda_\mu ]).
\end{eqnarray}
Using these superconnections, we can explicitly construct supercurvatures.


\newpage

\begin{thebibliography}{30}


\bibitem{W}
 E. Witten, 
 Commun. Math. Phys. \textbf{117} (1988) 353; 
 Commun. Math. Phys. {\bf 118} (1988) 411. 


 \bibitem{bs}
  L. Baulieu and I. Singer,
  Nucl. Phys. Proc. Suppl. {\bf B5} (1988) 12.

 \bibitem{bms}
  R. Brooks, D. Montano and J. Sonnenschein, 
  Phys. Lett. {\bf B214} (1988) 91.

 \bibitem{lp}
  J.M.F. Labastida and M. Pernici, 
  Phys. Lett. {\bf B212} (1988) 56; 
  Phys. Lett. {\bf B213} (1988) 319.

\bibitem{BRT1}
 D. Birmingham, M. Rakowski and G. Thompson, 
 Phys. Lett. {\bf B214} (1988) 381; 
 Phys. Lett. {\bf B212} (1988) 187; 
 Nucl. Phys. {\bf B315} (1989) 577.


\bibitem{BBRT}
D. Birmingham, M. Blau, M. Rakowski and G. Thompson, 
Phys. Rep. {\bf 209} (1991) 129.

\bibitem{BG}
 L. Baulieu and B. Grossman, 
 Phys. Lett. {\bf B212} (1988) 351; 
 Phys. Lett. {\bf B214} (1988) 223.


\bibitem{Kanno}
  H. Kanno, 
  Z. Phys. {\bf C43} (1989) 477.

\bibitem{BK}
 R. Brooks and D. Kastor, 
 Phys. Lett. {\bf B246} (1990) 99.

\bibitem{EY}
 T. Eguchi and S.-K. Yang, 
 Mod. Phys. Lett. {\bf A5} (1990) 1693.

\bibitem{BBT}
 D. Birmingham, M. Blau and G. Thompson, 
 Int. J. Mod. Phys. {\bf A5} (1990) 4721.


\bibitem{GM}
B. Geyer and D. Mulsch,
Nucl. Phys. {\bf B616} (2001) 476. 

\bibitem{LSSTV}
 V.E.R. Lemes, M.S. Sarandy, S.P. Sorella, A. Tnazini and O.S. Ventura, 
 JHEP {\bf 0101} (2001) 016 .


\bibitem{BRT}
 D. Birmingham, M. Rakowski, and G. Thompson, Nucl. Phys. \textbf{B329},
(1990) 83.

\bibitem{BR}
 D. Birmingham and M. Rakowski, 
 Mod. Phys. Lett. \textbf{A4}(1989) 1753; Phys. Lett. \textbf{B269} 
 (1991) 103; Phys. Rev. \textbf{B272} (1991) 217. 

\bibitem{DGS}
 F. Delduc, F. Gieres, and S.P.Sorella,
	Phys. Lett. {\bf B225} (1989) 367.

\bibitem{DLPS}
 F. Delduc, C. Lucchesi, O. Piguet and S.P. Sorella, 
 Nucl. Phys. {\bf B346} (1990) 313.


\bibitem{MS}
 N. Maggiore and S.P. Sorella, 
 Nucl. Phys. {\bf B377} (1992) 236.

\bibitem{SSVV} 
 C.A.G. Sasaki, A.P. Sorella, L.C.Q. Vilar and
 O.S. Ventura, 
 Journ. Math. Phys. {\bf 39} (1998) 848.

\bibitem{Pi}   
O. Piguet and S.P. Sorella, 
"{\it Algebraic Renormalization}"Lecture Notes in Physics, Vol. m28, 
Springer Verlag, 1995. 


\bibitem{CLS}

 O. M. Del Cima, K. Landsteiner, and M. Schweda, 
Phys.Lett. {\bf B439} (1998) 289.


\bibitem{GMS}
 E. Guadagnini, N. Maggiore and S.P. Sorella, 
 Phys. Lett. {\bf B247} (1990) 543;
 Phys. Lett. {\bf B255} (1991) 65.

\bibitem{BM}
A. Blasi and N. Maggiore, 
Class. Quantum Grav. {\bf 10} (1993) 37.

\bibitem{MSore}
 N. Maggiore and S.P. Sorella,
 Int. Journ. Mod. Phys. {\bf A8} (1993) 325.

\bibitem{LPS}
 C. Lucchesi, O. Piguet and S. P. Sorella,
 Nucl. Phys. {\bf B395} (1993) 325.

\bibitem{LSZ}
 R. Leitgeb, M. Schweda, and H. Zerrouki, 
 Nucl. Phys \textbf{B542} (1999) 425.

\bibitem{BSSV}
J. L. Bold, C. A. G. Sasaki, S. P. Sorella, and L. C. Q. Vilar, 
J. Phys. \textbf{A34} (2001) 2743. 

\bibitem{GGPS}
 F. Gieres, J. Grimstrup. T. Pisar and M. Schweda, JHEP {\bf 0006} (2000) 018.

\bibitem{GGNPS}
F. Gieres, J. Grimstrup, H. Nieder, T. Pisar and M. Schweda,
Phys. Rev. {\bf D66} (2002) 025027

\bibitem{FTVVSS}
 F. Fucito, A. Tnazini, L.C.Q. Vilar, O.S. Ventura, C.A.G. Sasaki, 
 S.P. Sorella, 
 hep-th/9707209.


\bibitem{BT}
 M. Blau and G. Thompson, 
 Ann. Phys. (N.Y.) {\bf 205} (1991) 130.

\bibitem{Soda}
 J. Soda, 
 Phys. Lett. {\bf B267} (1991) 214.

\bibitem{GMR}
  S. Cordes, G. Moore and S. Rangoolam, 
  Nucl. Phy. Proc. Suppl. {\bf 41} (1995) 184.
   

\bibitem{Pa}
P. Pasanen, J. Phys. {\bf A29} (1996) 8123.


\bibitem{DR}
 P.H. Damgaard and V.O. Rivelles, 
 Phys. Lett. {\bf B245} (1990) 48. 

\bibitem{BDL}
 R. Brooks, J-G. Demers and C. Lucchesi, 
 Nucl. Phys. {\bf B415} (1994) 353.


\bibitem{KW}
 N. Kawamoto and Y. Watabiki, Commun. Math. Phys. \textbf{144}, 641(1992);\\
 Mod. Phys. Lett. {\bf A7} (1992) 1137.

\bibitem{KW2}
 N. Kawamoto and Y. Watabiki, Phys.Rev.\textbf{D45} (1992) 605;\\
 Nucl. Phys. {\bf B396} (1993) 326.


 \bibitem{KOS}
  N. Kawamoto, E. Ozawa and K. Suehiro,
  Mod. Phys. Lett. {\bf A12} (1997) 219.

\bibitem{KSTU}
  N. Kawamoto, K. Suehiro, T. Tsukioka and H. Umetsu, 
  Commun. Math. Phys. {\bf 195} (1998) 233, 
  Nucl. Phys. {\bf B532} (1998) 429.


\bibitem{KT}
 N. Kawamoto and T. Tsukioka, Phys. Rev. \textbf{D61}(2000)105009.


\bibitem{IL} 
D. Ivanenko and L. Landau,  Z. Phys. \textbf{48} (1928) 340.

\bibitem{Kahler}
 E. K\"ahler, Rend. Math. Appl. \textbf{21} (1962) 425.


 \bibitem{G}
  W. Graf, 
  Ann. Inst. Henri Poincare {\bf A29} (1978) 85.

 \bibitem{BJ}
  P. Becher and H. Joos,
  Z. Phys. {\bf C15} (1982) 343. 

 \bibitem{Rabin}
  J.M. Rabin, 
  Nucl. Phys. {\bf B201} (1982) 315.

 \bibitem{BDH}
  T. Banks, Y. Dothan and D. Horn,
  Phys. Lett. {\bf B117} (1982) 413. 

 \bibitem{BennT}
  I.M. Benn and R.W. Tucker, 
  Commun. Math. Phys. {\bf 89} (1983) 341. 

 \bibitem{Bull}
  J.A. Bullinaria, 
  Ann. Phys. {\bf 168} (1986) 301. 


\bibitem{KKU}
J. Kato, N. Kawamoto and Y. Uchida,
 Int. Jour. Mod. Phys. {\bf A 19} (2004) 2149.


\bibitem{DVF}
  P. Di Vecchia and S. Ferrara, 
  Nucl. Phys. {\bf B130} (1977) 93. 

 \bibitem{SchapT}
  F.S. Schaposnik and G. Thompson, 
  Phys. Lett. {\bf B224} (1989) 379.


\bibitem{LL}
J.M.F. Labastida and P.M. Llatas, 
Nucl. Phys. {\bf B379} (1992) 220. 

\bibitem{AL}
M. Alvarez and J.M.F. Labastida, 
Nucl. Phys. {\bf B437} (1995) 356.


\bibitem{Yam}
 J.P. Yamron, 
 Phys. Lett. {\bf B213} (1988) 325.

\bibitem{Mac}
N. Marcus,
Nucl. Phys. {\bf B452} (1995) 331.

\bibitem{VW}
C. Vafa and E. Witten,
Nucl. Phys. {\bf B431} (1994) 3.

\bibitem{DM}
R. Dijkgraaf and G. W. Moore,
Commun. Math. Phys. {\bf 185} (1997) 411.

\bibitem{LCL}
J. M. F. Labastida and C. Lozano,
Nucl. Phys. {\bf B502} (1997) 741.


\bibitem{GSW}
 R. Grim, M. Sohnius and J. Wess, 
 Nucl. Phys. {\bf B 133} (1978) 275. 

\bibitem{Sohnius}
 M. Sohnius, 
 Nucl. Phys. {\bf B 136} (1978) 461.


\bibitem{SSW}
 M. Sohnius, K.S. Stelle and P.C. West, 
 Phys. Lett {\bf 92 B} (1980) 123; 
 Nucl. Phys. {\bf B 173} (1980) 127. 


\bibitem{DKKN} 
 A. D'Adda, I. Kanamori, N. Kawamoto and K. Nagata, 
 Nucl. Phys. {\bf B 707} (2005) 100, hep-lat/0406029.


\bibitem{KSuss}
  J. Kogut and L. Susskind, 
  Phys. Rev. {\bf D11} (1975) 395.

\bibitem{Suss}
  L. Susskind, Phys. Rev. {\bf D16} (1977) 3031.

\bibitem{KS} 
  N. Kawamoto and J. Smit, 
  Nucl. Phys. {\bf B192} (1981) 100. 


\bibitem{Gliozzi:1982ib}
F.~Gliozzi, 
Nucl. Phys. \textbf{B204} (1982), 419. 

\bibitem{Kluberg-Stern:1983dg}
H.~Kluberg-Stern, A.~Morel, O.~Napoly, and B.~Petersson, 
Nucl. Phys. \textbf{B220} (1983), 447.


\bibitem{Woronowicz:1989rt}
S.~L. Woronowicz, 
Commun. Math. Phys. \textbf{122} (1989), 125.

\bibitem{Dimakis:1992pk}
A.~Dimakis, F.~M{\"u}ller-Hoissen, and T.~Striker, 
J. Phys. \textbf{A26} (1993), 1927; 
Phys. Lett. \textbf{B300} (1993), 141.

\bibitem{Aschieri:1993wg}
P. Aschieri and L. Castellani, 
Int. J. Mod. Phys. \textbf{A8} (1993), 1667,  \texttt{hep-th/9207084}.


\bibitem{Dai:2000vf}
Jian Dai and Xing-Chang Song, 
 Phys. Lett. \textbf{B508} (2001), 385,
  \texttt{hep-th/0101130}, Commun. Theor. Phys. {\bf 39} (2003) 519.

\bibitem{Vaz1}
J. Vaz, 
Advances in Applied Clifford algebras, \textbf{7} (1997) 37.

\bibitem{KK} 
 I. Kanamori and N. Kawamoto, 
 Int. Jour. Mod. Phys. {\bf A 19} (2004) 695,  
 hep-th/0305094.


\bibitem{JMM} 
J. M. Maldacena,
Adv. Theor. Math. Phys. {\bf 2} (1998) 231, \texttt{hep-th/9711200}

\bibitem{AGMOO} 
O. Aharony, S. S. Gubser, J. M. Maldacena, H. Ooguri and Y. Oz,
 Phys. Rept. {\bf 323} (2000) 183. 

\bibitem{Kugo-T}
T. Kugo and  P. Townsend, 
 Nucl. Phs. {\bf B221} (1983) 357.

\end{thebibliography}
\end{document}